\documentclass[twocolumn]{aastex631}

\newcommand{\ha}{H$\alpha$}
\newcommand{\hb}{H$\beta$}
\newcommand{\oiii}{\mbox{[\ion{O}{3}]}}
\newcommand{\nii}{\mbox{[\ion{N}{2}]}}
\newcommand\smpy{\mbox{$M_\odot\ \textrm{yr}^{-1}$}}
\usepackage{multirow}
\usepackage{color}

\begin{document}

\title{Medium-band Astrophysics with the Grism of NIRCam In Frontier fields (MAGNIF): \\ Spectroscopic Census of \ha\ Luminosity Functions and Cosmic Star Formation at $z\sim$\,4.5 and 6.3}

\suppressAffiliations

\author[0000-0003-0964-7188]{Shuqi Fu}
\affiliation{Department of Astronomy, School of Physics, Peking University, Beijing 100871, People's Republic of China}
\affiliation{Kavli Institute for Astronomy and Astrophysics, Peking University, Beijing 100871, People's Republic of China}

\author[0000-0002-4622-6617]{Fengwu Sun}
\affiliation{Center for Astrophysics $|$ Harvard \& Smithsonian, 60 Garden St., Cambridge, MA 02138, USA}

\author[0000-0003-4176-6486]{Linhua Jiang}
\affiliation{Department of Astronomy, School of Physics, Peking University, Beijing 100871, People's Republic of China}
\affiliation{Kavli Institute for Astronomy and Astrophysics, Peking University, Beijing 100871, People's Republic of China}

\author[0000-0001-6052-4234]{Xiaojing Lin}
\affiliation{Department of Astronomy, Tsinghua University, Beijing 100084, China}
\affiliation{Steward Observatory, University of Arizona, 933 N Cherry Avenue, Tucson, AZ 85721, USA}

\author[0000-0001-9065-3926]{Jose M. Diego}
\affiliation{Instituto de F\'{i}sica de Cantabria (CSIC-UC).\ Avda. Los Castros s/n.\ 39005 Santander, Spain}

\author[0000-0001-6278-032X]{Lukas J. Furtak}
\affiliation{Department of Physics, Ben-Gurion University of the Negev, P.O. Box 653, Be'er-Sheva 84105, Israel}

\author[0000-0003-1974-8732]{Mathilde Jauzac}
\affiliation{Centre for Extragalactic Astronomy, Department of Physics, Durham University, South Road, Durham DH1 3LE, UK}
\affiliation{Institute for Computational Cosmology, Durham University, South Road, Durham DH1 3LE, UK}
\affiliation{Astrophysics Research Centre, University of KwaZulu-Natal, Westville Campus, Durban 4041, South Africa}
\affiliation{School of Mathematics, Statistics \& Computer Science, University of KwaZulu-Natal, Westville Campus, Durban 4041,
South Africa}

\author[0000-0002-6610-2048]{Anton M. Koekemoer}
\affiliation{Space Telescope Science Institute, 3700 San Martin Drive, Baltimore, MD 21218, USA}

\author[0000-0001-6251-649X]{Mingyu Li}
\affiliation{Department of Astronomy, Tsinghua University, Beijing 100084, People's Republic of China}

\author[0000-0003-3484-399X]{Masamune Oguri}
\affiliation{Center for Frontier Science, Chiba University, 1-33 Yayoicho, Inage, Chiba 263-8522, Japan}
\affiliation{Department of Physics, Graduate School of Science, Chiba University, 1-33 Yayoicho, Inage, Chiba 263-8522, Japan}

\author[0000-0001-6804-0621]{Nency R. Patel}
\affiliation{Centre for Extragalactic Astronomy, Department of Physics, Durham University, South Road, Durham DH1 3LE, UK}
\affiliation{Institute for Computational Cosmology, Durham University, South Road, Durham DH1 3LE, UK}

\author[0000-0001-9262-9997]{Christopher N. A. Willmer}
\affiliation{Steward Observatory, University of Arizona, 933 N Cherry Avenue, Tucson, AZ 85721, USA}

\author[0000-0001-8156-6281]{Rogier A. Windhorst} 
\affiliation{School of Earth and Space Exploration, Arizona State University, Tempe, AZ 85287-6004, USA}

\author[0000-0002-0350-4488]{Adi Zitrin}
\affiliation{Department of Physics, Ben-Gurion University of the Negev, P.O. Box 653, Be'er-Sheva 84105, Israel}

\author[0000-0002-8686-8737]{Franz E. Bauer}
\affiliation{Instituto de Astrof{\'{\i}}sica and Centro de Astroingenier{\'{\i}}a, Facultad de F{\'{i}}sica, Pontificia Universidad Cat{\'{o}}lica de Chile, Campus San Joaquín, Av. Vicuña Mackenna 4860, Macul Santiago, Chile, 7820436} 
\affiliation{Millennium Institute of Astrophysics, Monse{\~{n}}or Nuncio S{\'{o}}tero Sanz 100, Of 104, Providencia, Santiago, 7500011, Chile} 
\affiliation{Space Science Institute, 4750 Walnut Street, Suite 205, Boulder, Colorado 80301} 

\author[0000-0002-3805-0789]{Chian-Chou Chen}
\affiliation{Academia Sinica Institute of Astronomy and Astrophysics (ASIAA), No. 1, Sec. 4, Roosevelt Road, Taipei 106319, Taiwan}

\author[0000-0003-1060-0723]{Wenlei Chen}
\affiliation{Department of Physics, Oklahoma State University, 145 Physical Sciences Bldg, Stillwater, OK 74078, USA}

\author[0000-0003-0202-0534]{Cheng Cheng}
\affiliation{Chinese Academy of Sciences South America Center for Astronomy, National Astronomical Observatories, CAS, Beijing 100101, People’s Republic of China}

\author[0000-0003-1949-7638]{Christopher J. Conselice}
\affiliation{Jodrell Bank Centre for Astrophysics, Alan Turing Building, University of Manchester, Oxford Road, Manchester M13 9PL, UK}

\author[0000-0002-2929-3121]{Daniel J.\ Eisenstein}
\affiliation{Center for Astrophysics $|$ Harvard \& Smithsonian, 60 Garden St., Cambridge, MA 02138, USA}

\author[0000-0003-1344-9475]{Eiichi Egami}
\affiliation{Steward Observatory, University of Arizona, 933 N Cherry Avenue, Tucson, AZ 85721, USA}

\author[0000-0002-8726-7685]{Daniel Espada}
\affiliation{Departamento de F\'{i}sica Te\'{o}rica y del Cosmos,
Campus de Fuentenueva, Edificio Mecenas, Universidad de Granada,
E-18071, Granada, Spain}
\affiliation{Instituto Carlos I de F\'{i}sica Te\'{o}rica y
Computacional, Facultad de Ciencias, E-18071, Granada, Spain}

\author[0000-0003-3310-0131]{Xiaohui Fan}
\affiliation{Steward Observatory, University of Arizona, 933 N Cherry Avenue, Tucson, AZ 85721, USA}

\author[0000-0001-7201-5066]{Seiji Fujimoto}
\affiliation{Department of Astronomy, University of Texas at Austin, Austin, TX, USA}

\author[0000-0003-4512-8705]{Tiger Yu-Yang Hsiao}
\affiliation{Center for Astrophysics $|$ Harvard \& Smithsonian, 60 Garden St., Cambridge, MA 02138, USA}
\affiliation{Space Telescope Science Institute, 3700 San Martin Drive, Baltimore, MD 21218, USA}
\affiliation{Center for Astrophysical Sciences, Department of Physics and Astronomy, The Johns Hopkins University, 3400 N Charles St. Baltimore, MD 21218, USA}

\author[0000-0002-5768-738X]{Xiangyu Jin}
\affiliation{Steward Observatory, University of Arizona, 933 N Cherry Avenue, Tucson, AZ 85721, USA}

\author[0000-0002-4052-2394]{Kotaro Kohno}
\affiliation{Institute of Astronomy, Graduate School of Science, The University of Tokyo, 2-21-1 Osawa, Mitaka, Tokyo 181-0015, Japan}
\affiliation{Research Center for the Early Universe, Graduate School of Science, The University of Tokyo, 7-3-1 Hongo, Bunkyo-ku, Tokyo 113-0033, Japan}

\author[0000-0002-7633-2883]{David J. Lagattuta}
\affiliation{Centre for Extragalactic Astronomy, Department of Physics, Durham University, South Road, Durham DH1 3LE, UK}
\affiliation{Institute for Computational Cosmology, Durham University, South Road, Durham DH1 3LE, UK}

\author[0000-0001-5951-459X]{Zihao Li}
\affiliation{Cosmic Dawn Center (DAWN), Denmark}
\affiliation{Niels Bohr Institute, University of Copenhagen, Jagtvej 128, DK2200 Copenhagen N, Denmark}

\author[0000-0003-3762-7344]{Weizhe Liu}
\affiliation{Steward Observatory, University of Arizona, 933 N Cherry Avenue, Tucson, AZ 85721, USA}

\author[0000-0002-2316-8370]{Jordi Miralda-Escud\'{e}}
\affiliation{Institut de Ci\`{e}ncies del Cosmos, Universitat de Barcelona, 08028 Barcelona, Spain}
\affiliation{Instituci\'{o} Catalana de Recerca i Estudis Avan\c{c}ats, 08010 Barcelona, Spain}
\affiliation{Institut d’Estudis Espacials de Catalunya, 08860 Barcelona, Spain}

\author[0000-0001-9442-1217]{Yuanhang Ning}
\affiliation{Department of Scientific Research, Beijing Planetarium, Beijing 100044, China}

\author[0000-0002-8224-4505]{Sandro Tacchella}
\affiliation{Kavli Institute for Cosmology, University of Cambridge, Madingley Road, Cambridge, CB3 0HA, UK}
\affiliation{Cavendish Laboratory - Astrophysics Group, University of Cambridge, 19 JJ Thomson Avenue, Cambridge, CB3 0HE, UK}

\author[0000-0003-0747-1780]{Wei Leong Tee}
\affiliation{Steward Observatory, University of Arizona, 933 N Cherry Avenue, Tucson, AZ 85721, USA}

\author[0000-0003-1937-0573]{Hideki Umehata}
\affiliation{Institute for Advanced Research, Nagoya University, Furocho, Chikusa, Nagoya 464-8602, Japan}
\affiliation{Department of Physics, Graduate School of Science, Nagoya University, Furocho, Chikusa, Nagoya 464-8602, Japan}

\author[0000-0002-7633-431X]{Feige Wang}
\affiliation{Department of Astronomy, University of Michigan, 1085 S. University Ave., Ann Arbor, MI 48109, USA}

\author[0000-0001-7592-7714]{Haojing Yan}
\affiliation{Department of Physics and Astronomy, University of Missouri, Columbia, MO 65211, USA }

\author[0000-0003-3307-7525]{Yongda Zhu}
\affiliation{Steward Observatory, University of Arizona, 933 N Cherry Avenue, Tucson, AZ 85721, USA}

\begin{abstract}
We measure \ha\ luminosity functions (LFs) at redshifts $z \sim 4.5$ and 6.3 using the JWST MAGNIF (Medium-band Astrophysics with the Grism of NIRCam In Frontier fields) survey. MAGNIF obtained NIRCam grism spectra with the F360M and F480M filters in four Frontier Fields. We identify 248 \ha\ emitters based on the grism spectra and photometric redshifts from combined HST and JWST imaging data. The numbers of the \ha\ emitters show a large field-to-field variation, highlighting the necessity of multiple fields to mitigate cosmic variance. We calculate both observed and dust-corrected \ha\ LFs in the two redshift bins. Thanks to the gravitational lensing, the measured \ha\ LFs span three orders of magnitude in luminosity, and the faint-end luminosity reaches $L_{\mathrm{H}\alpha} \sim 10^{40.3}\,\mathrm{erg}\,\mathrm{s}^{-1}$ at $z \sim 4.5$ and $10^{41.5}\,\mathrm{erg}\,\mathrm{s}^{-1}$ at $z \sim 6.3$. They correspond to star-formation rates (SFRs) of $\sim$\,0.1 and 1.7\,$\mathrm{M}_\odot\,\mathrm{yr}^{-1}$, respectively.
We conclude no or weak redshift evolution of the faint-end slope of \ha\ LF across $z\simeq0.4-6.3$, and the comparison with the faint-end slopes of UV LF indicates stochastic star formation history among low-mass \ha\ emitters.
The derived cosmic SFR densities are $0.058^{+0.008}_{-0.006}\ \ M_\odot\ \mathrm{yr}^{-1}\ \mathrm{Mpc}^{-3}$ at $z \sim 4.5$ and $0.025^{+0.009}_{-0.007}\ \ M_\odot\ \mathrm{yr}^{-1}\ \mathrm{Mpc}^{-3}$ at $z \sim 6.3$. These are approximately 2.2 times higher than previous estimates based on dust-corrected UV LFs, but consistent with recent measurements from infrared surveys. We discuss uncertainties in the \ha\ LF measurements, including those propagate from the lens models, cosmic variance, and AGN contribution, and we find that they have negligible impact on the above results.

\end{abstract}

\keywords{High-redshift galaxies (734) --- Luminosity function (942) --- Star formation (1569) --- James Webb Space Telescope (2291) --- Strong gravitational lensing (1643)}

\section{Introduction} \label{sec:intro}
The star formation history of the Universe remains a key question in observational cosmology.
To map the formation and evolution history of galaxies across cosmic time, multiple star formation rate (SFR) indicators have been developed and adopted \citep[e.g., see reviews by][]{Kennicutt_2012,Madau_2014}.
At high redshifts ($z > 4$), the SFR function (SFRF) and cosmic SFR density (CSFRD) have traditionally been investigated using rest-frame UV observations of Lyman-break galaxies \citep[e.g.,][]{Schenker_2013, Bouwens_2015,Finkelstein_2015}. 
While these studies have provided valuable insights into the evolution of CSFRD using rest-frame, dust-corrected UV observations,  dust extinction correction is challenging and often introduces systematic uncertainties in SFR estimates. As a result, there is a strong motivation to employ alternative SFR tracers at high redshifts.

The \ha\ emission line has been widely regarded as one of the most reliable tracers among the easily accessible nebular-line SFR indicators. Compared with rest-frame UV luminosity, \ha\ emission is less attenuated by dust and directly traces the ionizing luminosity from massive stars with lifetimes shorter than 10\,Myr \citep[e.g.,][]{Tacchella_2022a}. This makes \ha\ a robust and near-instantaneous measure of star formation \citep{Kennicutt_2012}. Many surveys have leveraged the \ha\ line to study SFRs at  redshift $z\lesssim2.5$, providing critical benchmarks for galaxy evolution studies \citep[e.g.,][]{Hayashi_2012, Sobral_2013, Kashino_2013, Steidel_2014,  Kriek_2015, Stroe_2015}. However, at higher redshifts where \ha\ moves out of the observed-frame $K$ band, observations of \ha-emitting galaxies become extremely challenging.
As a consequence, all previous studies at $z\gtrsim4$ relied on Spitzer/IRAC \citep{Fazio_2004}  broadband photometry to estimate the luminosity of \ha\ and other emission lines through spectral energy distribution (SED) modeling \citep[e.g.,][]{Schaerer_2010, Labbe_2013, Smit_2016, Faisst_2019, Roberts-Borsani_2020, Bollo_2023}.

With the advent of JWST, \ha\ and other optical emission lines from high-redshift galaxies are now accessible for high-resolution imaging and spectroscopy. The wide-field slitless spectroscopy (WFSS) mode with NIRCam \citep{Greene_2017,Rieke_2023} grism observations is particularly efficient for identifying \ha\ emitters out to $z\sim6.6$ \citep[e.g.,][]{Sun_2022,Sun_2023}. Slitless spectroscopy provides flux-limited selections of emission-line galaxies across contiguous redshift windows without the limitations of atmospheric interference or slit losses that affect ground-based spectroscopy \citep[e.g.,][]{Matharu_2024}. By capturing a complete census of \ha\ emitters across the field of view (FoV),  grism observations enable reliable measurements of star formation activities in the high-redshift Universe.

Based on shallow flux-calibration observations taken during the NIRCam WFSS commissioning, \citet{Sun_2023} discovered four \ha\ emitters at $z>6$ and found weak evolution in the \ha\ luminosity function (LF) from $z=2$ to 6.
 \citet{Covelo-Paz_2024} conducted a blind search of \ha\ emitters at $3.8 < z < 6.6$ in the GOODS fields using Cycle-1 FRESCO \citep{Oesch_2023} and Cycle-2 CONGRESS (\citealt{Egami_2023}; Sun, F. et al., in preparation) programs.
 They computed observed \ha\ LFs down to $L_{\mathrm{H}\alpha}\sim10^{42}\,\mathrm{erg \,s}^{-1}$ in three redshift bins centered at $z \sim 4.45$, 5.30, and 6.15. These observations successfully measured the CSFRD evolution beyond the peak at $z\sim2$.
The \ha-based CSFRDs have been found to be generally consistent with UV-based measurements from Lyman-break galaxies, with differences largely attributed to potential selection biases and dust correction assumptions.

So far, \ha\ LFs at $z\gtrsim4$ have only been studied for galaxies with SFRs $\gtrsim3$\,\smpy  \citep[e.g.,][]{Bollo_2023,Sun_2023,Covelo-Paz_2024}.
To properly investigate the contribution of less massive or lower-luminosity galaxies to the CSFRD, gravitational lensing by massive galaxy clusters has been widely used.
The Hubble Frontier Fields \citep{Lotz_2017} campaign obtained deep optical/near-infrared images of six massive galaxy clusters with the Hubble Space Telescope (HST), enabling the detection of intrinsically faint galaxies that probe the faint end LFs of various galaxy populations \citep[e.g.,][]{Atek_2015,Bouwens_2015,Bouwens_2022c,Livermore_2017,Ishigaki_2018,Kawamata_2018,YueB_2018}. 
Leveraging the highly magnified regions, we can explore the properties of low-luminosity galaxies and their contributions to the CSFRD in the early Universe. 

In this work, we present the spectroscopic determination of \ha\ LFs and CSFRDs at $z\sim 4.5$ and $z\sim6.3$ through the JWST program \textsl{``Medium-band Astrophysics with the Grism of NIRCam in Frontier Fields''} (MAGNIF; Cycle-2 GO-2883, PI: F.\ Sun).
Combining the sensitivity of JWST with gravitational lensing from four of the Frontier Field clusters, we detect \ha\ emitters down to $L_{\mathrm{H}\alpha}\sim 10^{40.3}\,\mathrm{erg}\,\mathrm{s}^{-1}$ at $z\sim 4.5$ and $L_{\mathrm{H}\alpha}\sim 10^{41.5}\,\mathrm{erg}\,\mathrm{s}^{-1}$ at $z\sim 6.3$, corresponding to $\mathrm{SFR}\sim0.1 \ \ M_\odot\ \mathrm{yr}^{-1}$ and $1.7\ \ M_\odot\ \mathrm{yr}^{-1}$ at these two redshifts, respectively. We then derive the SFRFs and CSFRDs at these two redshifts based on the dust-corrected \ha\ LFs.

This paper is organized as follows: In Section~\ref{sec:data}, we present the observations and data reduction. Section~\ref{sec:method} details the selection of the \ha\ emitters, quantifies their properties, selection completeness and the effective survey volume. Our main results are presented in Section~\ref{sec:result}, followed by further discussion in Section~\ref{sec:discuss}. Finally, our conclusions are summarized in Section~\ref{sec:summary}. 
Throughout this paper, we assume a flat $\Lambda$CDM cosmology with $H_0=70\ \mathrm{km}\ \mathrm{s}^{-1}\ \mathrm{Mpc}^{-1}$ and $\Omega_m = 0.3$. The AB magnitude system \citep{Oke_1983} is used. We also assume a \citet{Chabrier_2003} initial mass function.

\section{Observations and data reduction} \label{sec:data}

Our JWST program MAGNIF observed four Frontier-Field lensing clusters \citep{Lotz_2017}, namely Abell 2744, Abell 370, MACS\,J0416--2403 and MACS\,J1149+2223 (hereafter A2744, A370, M0416 and M1149, respectively).
These four cluster fields have been observed by HST and JWST through many programs. 
In this work, we primarily use JWST NIRCam WFSS data from MAGNIF, as well as available JWST Cycle-1 and Cycle-2 and archival HST imaging data. 
In this section, we present the data collection and our reduction procedure.
More details of JWST imaging and WFSS data reduction will be presented in a forthcoming paper from the collaboration (Sun et al. in preparation).

\subsection{Direct imaging} \label{subsec:image}
All of these cluster fields were observed with NIRCam imaging at $1-5\  \mu$m with various JWST programs. 
We make use of all NIRCam and NIRISS imaging data publicly released prior to September 2024, including Cycle-1 GTO-1176 PEARLS \citep{Windhorst_2023}, GTO-1199 \citep{Stiavelli_2023}, GTO-1208 CANUCS \citep{Willott_2022}, ERS-1324 GLASS \citep{Treu_2022}, Pure Parallel GO-2514 PANORAMIC \citep{Williams_2024}, GO-2561 UNCOVER \citep{Bezanson_2024}, DDT-2756 \citep{ChenW_2022jwst.prop.2756C}; Cycle-2 GO-2883 MAGNIF (Sun et al.\ in preparation), GO-3516 \citep{Naidu_2024}, GO-3362 Technicolor \citep{Muzzin_2023}, GO-3538 \citep{Iani_2023} and GO-4111 MegaScience \citep{Suess_2024}. 

\begin{figure*}[!t]
\centering
\includegraphics[width=\textwidth]{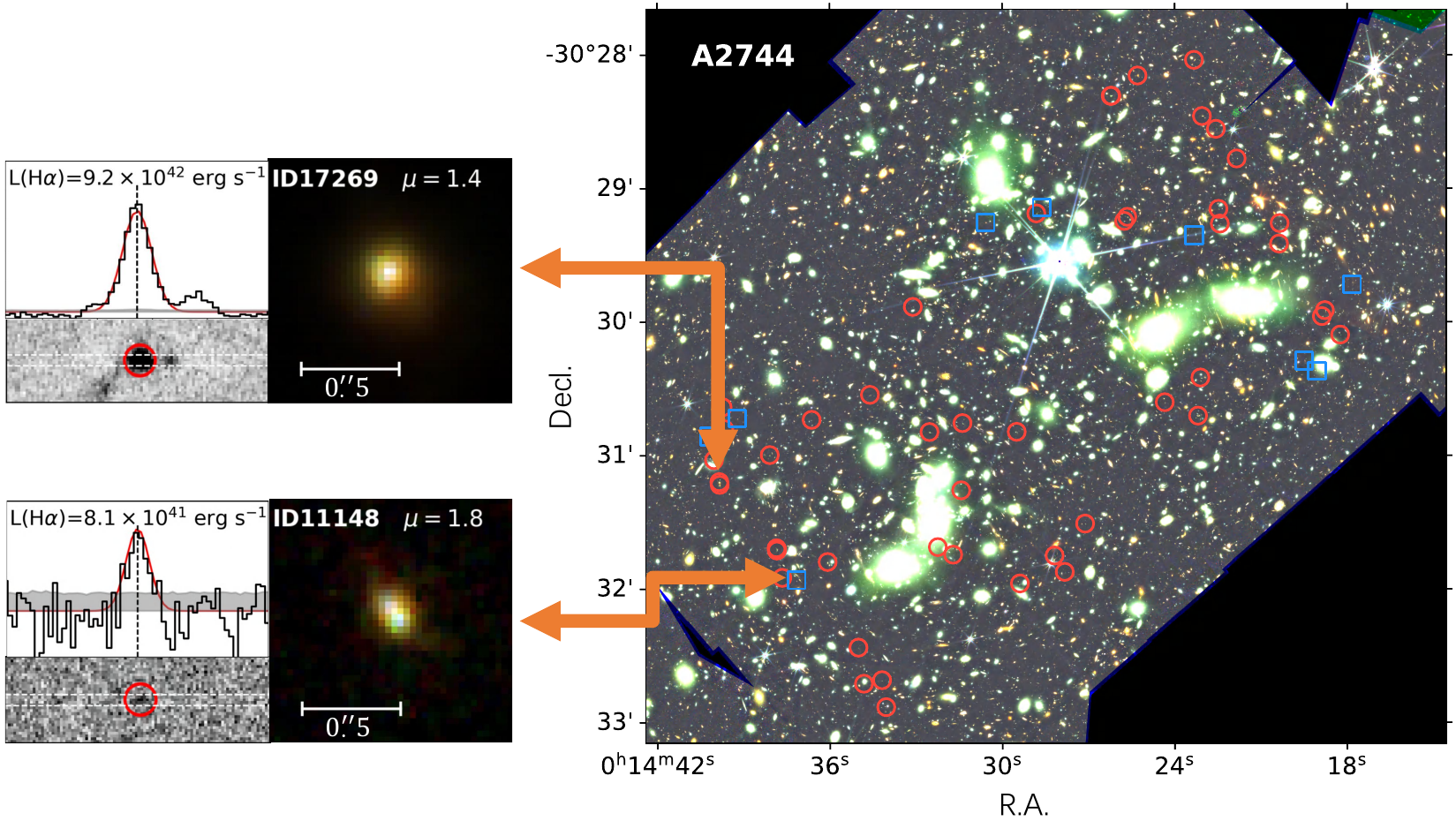}
\caption{Spatial distribution for $z\sim 4.5$ (red circles) and 6.3 (blue squares) \ha\ emitters in the A2744 field (right panel). The background is the NIRCam RGB image (R: F444W, G: F277W, B: F115W). Left panel: Cutout images and \ha\ spectra of two galaxies in our sample. ID17269 is the intrinsically brightest \ha\ emitter at $z\sim 4.5$ in this field, with the \nii\,$\lambda$6585 line also visible. ID11148 is the intrinsically faintest \ha\ emitter at $z\sim 6.3$ in this field, which would not be detected without a lensing magnification.  }
\label{fig:a2744_sample}
\end{figure*}

The JWST images were reduced and mosaicked using a modified stage-1/2/3 JWST pipeline \citep{jwst_bushouse_2023} version \verb|1.11.2| and CRDS calibration reference file context \verb|jwst_1188.pmap|. The $1/f$ noise was modeled and removed from the stage-2 products (i.e., the \texttt{\_cal.fits} files). We also applied bad-pixel masks constructed from the median stacks of images taken with the same detector. 
We subtracted wisp templates in NIRCam short-wavelength (SW) images. We also generate smoothed median-stacked images with sources masked to mimic other scattering light, and subtract them from the images. Individual images were astrometrically registered to DESI Legacy Imaging Survey \citep{Dey_2019}. We did not use GAIA stars because of their low spatial density. Specifically, we first processed the F444W images to produce reference source catalogs and registered the astrometry of images taken with other filters. A global sky background was subtracted prior to stage-3 mosaicing, and the image products were resampled to a pixel size of 0\farcs03 with \texttt{pixfrac=1}.
Finally, we used the F277W mosaics as the source mask images to subtract remaining diffuse scattering light in other bands (mostly seen at $\lesssim2$\,\micron). Figure \ref{fig:a2744_sample} illustrates an example of image products.

The HST/ACS and WFC3 images were taken in the F435W, F606W, F814W, F105W, F125W, F140W, and F160W bands by the Hubble Frontier Field program \citep{Lotz_2017} and the Beyond Ultradeep Frontier Fields And Legacy Observations (BUFFALO) program \citep{Steinhardt_2020}. We directly made use of the HST/ACS imaging data products by the BUFFALO team\footnote{\url{https://archive.stsci.edu/hlsp/buffalo}} but registered their astrometry to that of the NIRCam images.

We stacked the F182M and F444W images (inverse-variance weighed) to produce a detection image. This detection image was iteratively median-filtered to suppress the diffuse light from intracluster light and bright central galaxies, and to enhance the detection of faint blended sources. Sources were then extracted and deblended using a \textsc{photutils} pipeline \citep{photutils}. Photometry in all bands was measured within circular apertures of $r = 0\farcs1, 0\farcs15, 0\farcs25$ and $0\farcs30$, and within Kron apertures with Kron factors of 2.5 and 1.2. Local background was subtracted before the photometry. 
All photometry was aperture-corrected using a point-source aperture correction factor based on \texttt{webbpsf} models \citep{webbpsf}.
We use the photometry within $r = 0\farcs1$ for the photometric redshift determination to ensure the highest signal-to-noise ratio (SNR) since a flux scaling does not affect photometric redshifts. For SED modeling, we adopt Kron photometry  with the Kron factor of 2.5 to include most of the light for $z>4$ galaxies \citep{Sunwen_2024}.
Flux uncertainties were measured in two approaches, one from the error images and the other one from random apertures following a similar method detailed by \citet{Rieke_2023}. The larger uncertainty was used in our analysis. Usually, the random-aperture method results in a larger uncertainty because of the more realistic treatment of correlated noise. 

\subsection{Grism spectroscopy} \label{subsec:grism}
MAGNIF obtained NIRCam WFSS observations of the four clusters using the column-direction grism (Grism C) with medium-band filters F360M and F480M. These filters are optimized to detect \ha\ emission from galaxies at $z \sim 4$ and both H$\alpha$ and [\ion{O}{3}] emission from galaxies at $z \sim 6$. Compared with the commonly used wide-band filter, the use of medium-band filters mitigates source confusion and reduces sky background.

The WFSS data were reduced following the routine outlined by \citet{Sun_2023}, and the codes and calibration data are publicly available\footnote{\url{https://github.com/fengwusun/nircam_grism}}. We first processed NIRCam WFSS data through the standard JWST stage-1 calibration pipeline with version \verb|1.11.4| and calibration reference data file \verb|jwst_1263.pmap|. We applied a flat-field correction using flat-field data taken with the same filter and detector. We constructed and subtracted a super sky background for each exposure using all flat-fielded exposures taken with the same filter+pupil+module combination.
We found that there were still some background residuals, and thus performed a further 2D sky background subtraction using the Source-Extractor algorithm \citep{Bertin_1996}: the background was calculated on the $2.5\sigma$ clipped image using a mode estimator of the form $(2.5 \times \mathrm{median}) - (1.5 \times \mathrm{mean})$, which is considerably less affected by source crowding than a simple clipped mean. If $(\mathrm{mean} - \mathrm{median}) / \mathrm{STD} > 0.3$ (STD is the standard deviation), then the mode estimator does not work, and a simple median is used instead. The astrometry of the grism data was corrected by matching the simultaneously taken SW images to the long-wavelength (LW) source catalog that was previously registered to the DESI Legacy Imaging Survey catalog \citep{Dey_2019}.

\begin{figure*}[!t]
\centering
\includegraphics[width=\textwidth]{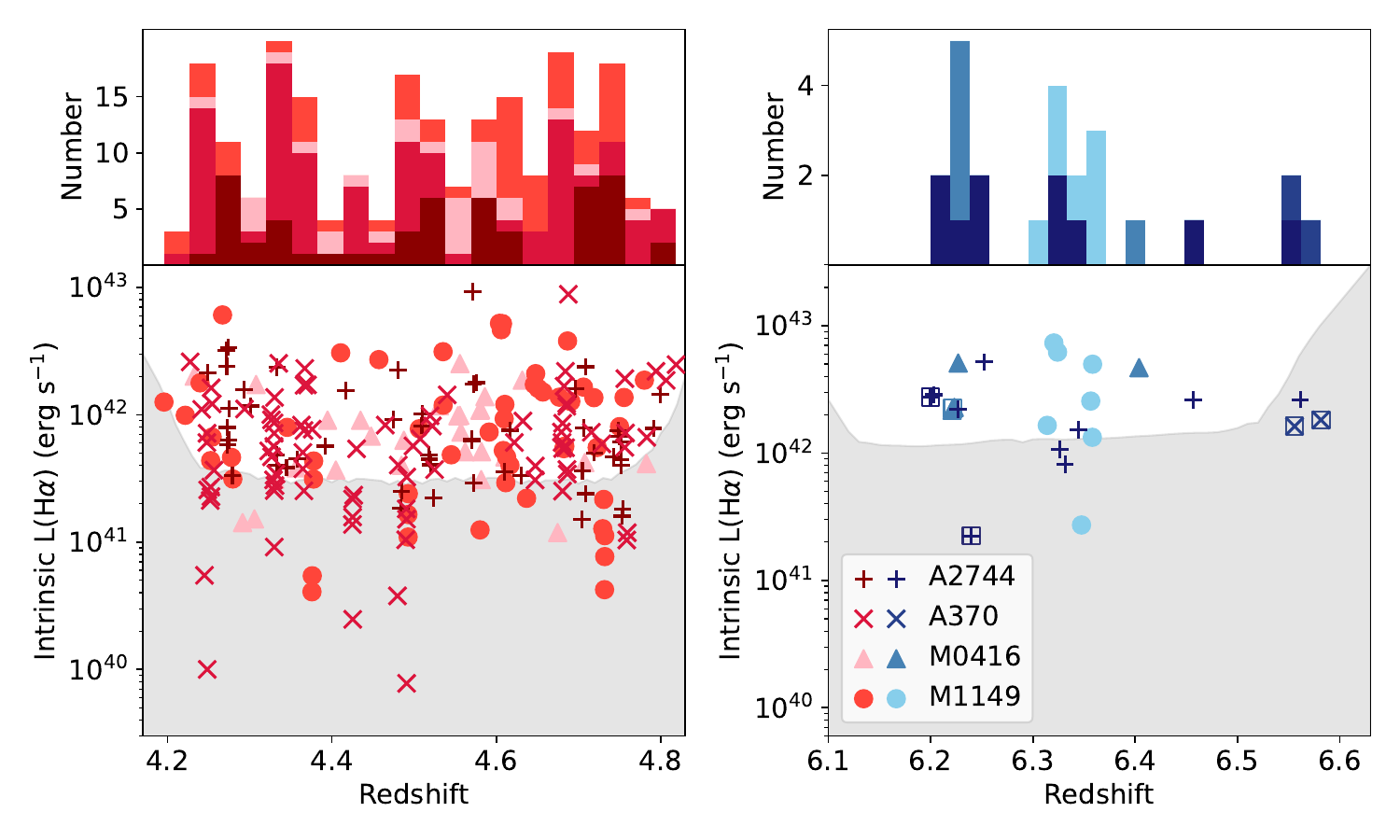}
\caption{Distribution of the redshifts and intrinsic \ha\ luminosities (de-magnified, but not corrected for dust) for our \ha\ emitter sample at $z=4.2-4.8$ (left) and $z=6.2-6.6$ (right). The upper boundaries of the gray-shaded regions represent the median $5\sigma$ detection limits in the absence of gravitational lensing. The sources with reliable \oiii\,$\lambda\lambda$4960,5008 detections but with faint \ha\ lines  ($\mathrm{SNR}<5$) are encircled by squares.   \label{fig:LHa_z}}
\end{figure*}

\section{\ha\ emitter samples} \label{sec:method}

In this section, we construct our \ha\ emitter samples at $z\sim 4.5$ and $z \sim 6.3$ (see Figure \ref{fig:LHa_z} and Table \ref{tab:sample}). We measure \ha\ line emission flux and derive survey volumes. We also estimate our sample completeness and discuss lens models.

\subsection{Construction of Samples at $z\sim 4.5$ and $z \sim 6.3$}

We first select \ha-emitting galaxy candidates from the photometric catalogs constructed from all publicly available HST and JWST imaging data listed in Section~\ref{subsec:image}. We estimate photometric redshifts using EAZY \citep{Brammer_2008} with templates from \citet{Hainline_2023}, and allow a wide redshift range ($z = [0.01,20]$) for model fitting. We select all sources with EAZY photometric redshifts within the range $[4.18, 4.95]$ and $[6.00, 6.62]$, corresponding to the wavelength coverage of the grism filters F360M and F480M. 
We argue that the photometric selection is highly complete because \ha\ emitters detected with the grism spectroscopy are also strong line emitters revealed with the medium-band imaging data, and the photometric redshifts of these emitters can be easily constrained with $\sim$\,20-band deep HST and JWST data. 
We extract the 2D spectra of these candidates from individual exposures and stacked the 2D spectra after registering them to a common wavelength and spatial grid. We then extract these 1D spectra from the stacked 2D spectra using the boxcar method. Finally, we run a line detection algorithm to search for possible emission lines in all 1D and 2D spectra. 

Following previous works that have performed emission line detections on NIRCam slitless spectra \citep[e.g.,][]{Wang_ASPIRE_2023,Kashino_2023}, we have implemented the following line-detection procedure. We first apply a 1D median filter with a size of 51 pixels and a central gap of 5 pixels (spectral resolution $\mathrm{FWHM}_\mathrm{res} \sim 24$~\AA\ for F360M and $\mathrm{FWHM}_\mathrm{res} \sim 29$~\AA\ for F480M) to the 1D and 2D spectra to remove the continuum from all sources, and then smooth the line spectra using a 1D/2D Gaussian kernel with FWHM equal to the spectral resolution. For the 1D spectra, we detect all peaks with SNR larger than 3 in the smoothed 1D spectra, then fit a Gaussian model to all peaks to measure the FWHM, and combine the flux and error within $\pm \mathrm{FWHM}$ to measure the 1D line SNR. We rejected lines with 1D $\mathrm{SNR} < 5$ and an FWHM wider than seven times $\mathrm{FWHM}_\mathrm{res}$ or narrower than half of $\mathrm{FWHM}_\mathrm{res}$. These criteria allow a broad \ha\ line with $\mathrm{FWHM}<1400\ \mathrm{km}\ \mathrm{s}^{-1}$ at $z\sim 4.5$ or $\mathrm{FWHM}<1300\ \mathrm{km}\ \mathrm{s}^{-1}$ at $z\sim 6.3$ (if presents) to be detected. We discuss the possible contribution from broad \ha\ emitters in Section~\ref{subsec:broad_Ha}. For the 2D spectra, we run \textsc{photutils} to do the source detection using $\mathrm{DETECT\_MINAREA} = 5$ and $\mathrm{DETECT\_THRESH} = 1.2$. We require the detected 2D emission lines to have less than 2 pixels offset from the spatial center, and the wavelength difference between 1D and 2D emission lines should be less than $\mathrm{FWHM}_\mathrm{res}$.
Finally, we visually inspect all emission lines and remove spurious detections. 

To further assess the completeness of our photometric selection, we extract spectra for all sources with a photometric $\mathrm{SNR} > 3$ in the F360M or F480M bands in the A370 and M1149 fields, regardless of their photometric redshifts. No additional, high-confidence \ha\ line emitters are found within the redshift range of interest. In regions with high noise levels (such as those close to bright central galaxies), large photometric errors can severely affect our SED measurement. This potential source of uncertainty has been incorporated into our completeness analysis (see \ref{subsec:complt}).

The slitless spectra may largely overlap. In other words, an emission line emitted by another source that is aligned with our target in the dispersion direction may be mistaken as the \ha\ line of the target. To alleviate this problem, we search and determine contamination lines. In short, for each target, we search for nearby sources with spectral traces overlapping with the target, and determine their possible emission lines falling in the corresponding wavelength range based on its photometric redshift.
We then perform a simple forward modeling based on the fully reduced LW images to fit the emission lines, and determine the sources for individual lines using the minimum reduced chi-square ($\chi_{\nu}^2$) method. During this process,  lines brighter than the maximum brightness allowed by the broad-band photometry are not considered. We then visually inspect the fitting results, and find that we are not able to determine the sources of a small fraction of the emission lines. For example, one emission line may either be the \ha\ line of a $z_{\mathrm{phot}}\sim 4.5$ candidate or a Paschen\,$\beta$ line of an adjacent $z_{\mathrm{phot}}\sim 1.8$ source. The line morphology does not provide enough information to produce different $\chi_{\nu}^2$ values. In this case, we assign this \ha\ emitter candidate with a `possible' quality flag. 

We eventually obtain a sample of 2000 `good' and 22 `possible' \ha\ emitters at redshift 4.18 to 4.95, and a sample of 15 `good' and 4 `possible' \ha\ emitters at redshift 6.00 to 6.62. Table~\ref{tab:sample} shows the sample size in each field. We visually inspect all emission lines and do not find broad components that may be evidence of AGN contribution. Two galaxies in our \ha\ emitter sample are shown in Figure~\ref{fig:a2744_sample}. 
The first source at $z=4.57$ is a luminous \ha\ emitter, and \nii\,$\lambda$6585 line is also detected in the spectrum. The second source at $z=6.33$ has a low intrinsic (de-magnified) \ha\ luminosity ($8.1\times10^{41}\ \mathrm{erg}\ \mathrm{s}^{-1}$). Without the gravitational lensing magnification ($\mu=1.8$), this source would have not been detected. 

The distributions of spectroscopic redshifts and intrinsic \ha\ luminosities for our $z\sim4.5$ and $z\sim6.3$ \ha\ emitter sample are shown in Figure~\ref{fig:LHa_z}. 
We also found another five $z\sim6.3$ sources with reliable \oiii\,$\lambda\lambda$4959,5007 detections and corresponding \ha\ lying in the redshift range from 6.00 to 6.62. However, their \ha\ line emisson does not satisfy our SNR criteria. We thus slightly lower the detection limit to measure their \ha\ flux, but we do not include them for the LF measurement.

\begin{deluxetable}{ccccccc}[!t]
\tabletypesize{\footnotesize}
\tablewidth{\textwidth} 
\tablecaption{Summary of \ha\ emitters in the four fields \label{tab:sample}}
\tablehead{
\colhead{Redshift} & \colhead{Flag} & \colhead{A2744} & \colhead{A370} & \colhead{M0416} & \colhead{M1149} & \colhead{Total} 
}
\startdata{
\multirow{2}{*}{$z\sim4.5$} & good & 52 & 78 & 24 & 46 & 200 \\
                      & possible & 2 & 8 & 4 & 8 & 22 \\ \hline
\multirow{3}{*}{$z\sim6.3$} & good & 6 & 0 & 3 & 6 & 15 \\
                      & possible & 2 & 0 & 1 & 1 & 4 \\
                      & ${\rm SNR}<5$ & 4 & 3 & 1 & 0 & 7 \\
}\enddata
\tablecomments{\textsf{Flag `good'} means that we confirm this source as an \ha\ emitter with high confidence. \textsf{Flag `possible'} means that a line may come from a nearby source, or that the line SNR is not high enough to determine through visual inspection (e.g., a possible line falls within the spectrum of a bright continuum contaminant). 
\textsf{Flag `${\rm SNR}<5$'} means that a source is detected with \oiii\ lines and thus its redshift is confirmed, but its \ha\ emission (within our wavelength coverage) is fainter than our SNR threshold.
}
\end{deluxetable}

\subsection{\ha\ Flux Measurement} \label{subsec:flux}

\begin{figure*}[!t] 
    \centering
    \includegraphics[width=\textwidth]{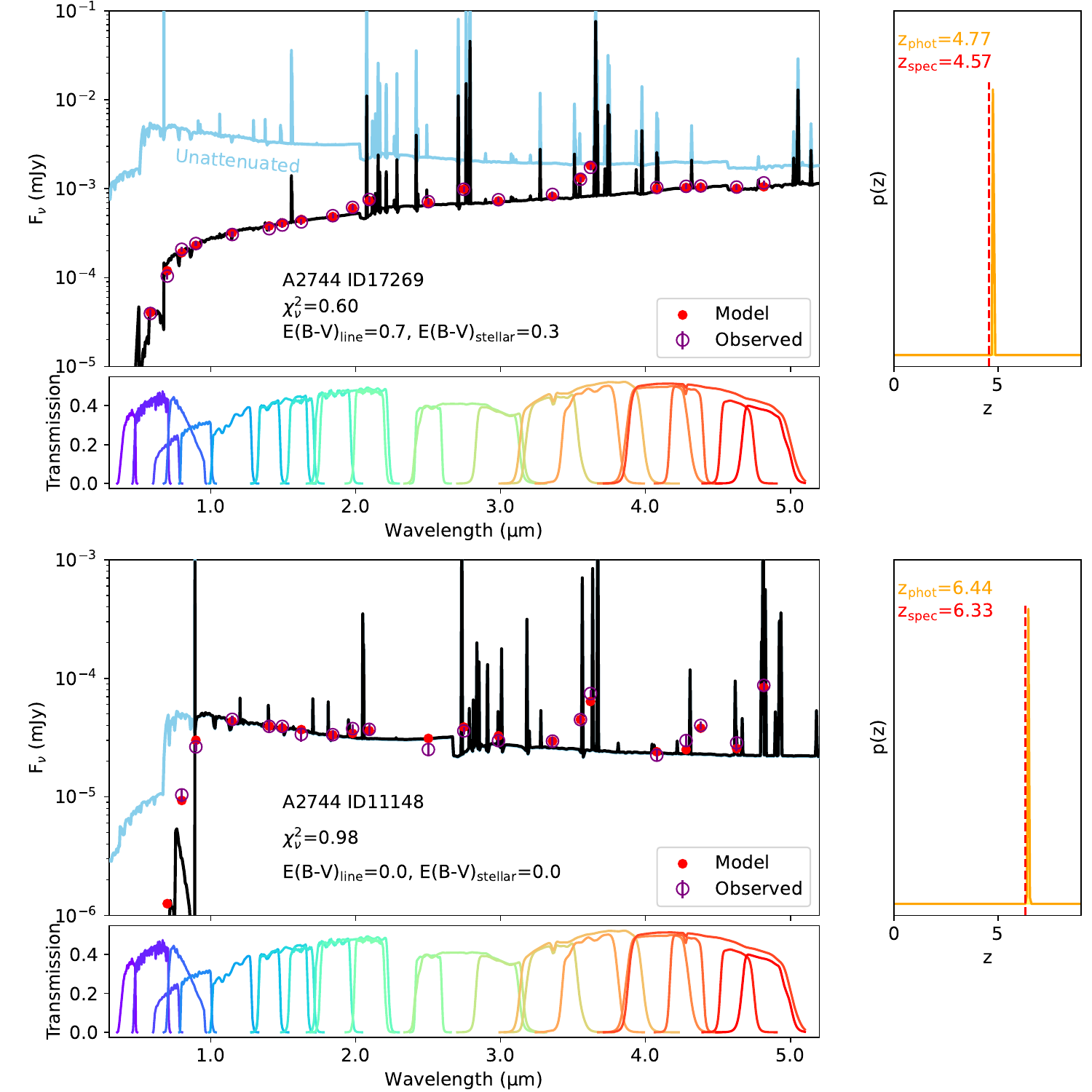} 
    \caption{Best-fit SED models and dust-attenuation correction for two \ha\ emitters. Upper panels: A2744 ID17269 is a heavily dust-obscured galaxy at $z=4.57$. It is also detected at 1.2\,mm by ALMA DUALZ survey \citep{Fujimoto_2023}. Lower panels: A2744 ID11148 is a galaxy at $z=6.33$ with negligible dust obscuration. The black lines show the best-fit SED models and the blue lines represent the dust-unattenuated SEDs. The transmission curves for all filters used in the SED modeling are shown in the lower attached panels. The probability density functions of the photometric redshifts derived from EAZY and the spectroscopic redshift calculated using \ha\ line are displayed on the right-hand side.}
    \label{fig:sed_pz}
\end{figure*}

\begin{figure*}[!t] 
    \centering
    \includegraphics[width=\textwidth]{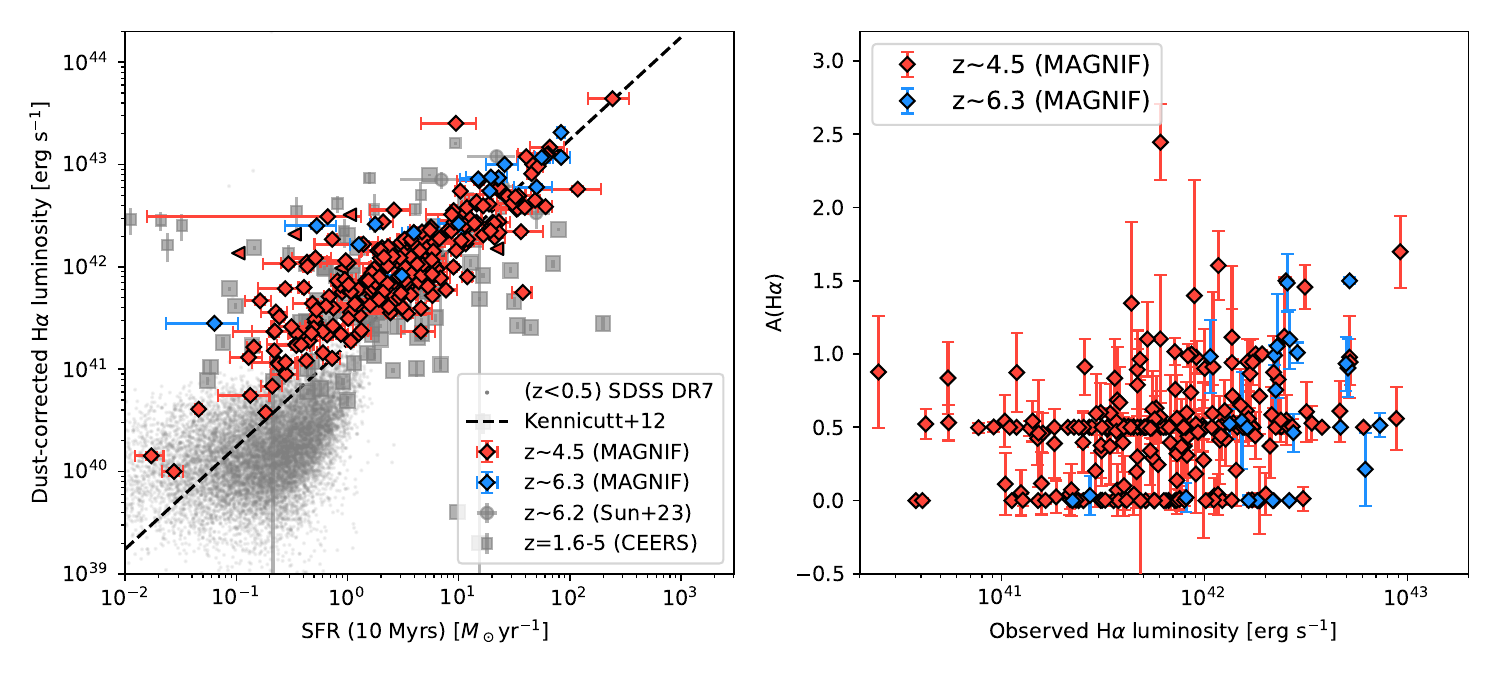} 
    \caption{Left: Dust-corrected \ha\ luminosities measured from the grism spectroscopy versus the average SFRs in the most recent 10\,Myr modeled using \textsc{cigale}. The left-pointing triangles represent the upper limit of the SFR. Results from low-redshift SDSS galaxies \citep[small grey dots;][]{Brinchmann_2004} and some other high-redshift samples are also shown for comparison \citep[e.g.,][]{Sun_2023,Backhaus_2024}. The relation between \ha\ luminosities and average SFRs in most recent 10\,Myr is consistent with the \citet{Kennicutt_2012} relation (dashed black line). Right: \ha\ attenuation modeled using \textsc{cigale} versus the observed \ha\ luminosities corrected for lensing magnification (but without dust correction). Galaxies with higher \ha\ luminosities are subject to higher dust attenuation. }
    \label{fig:LHa_SFR_AHa}
\end{figure*}

We run \textsc{photutils} \citep{photutils} on the 2D spectra to obtain the source detection and flux measurement for the \ha\ lines. To account for both the spatial distribution and the morphology broadening effect of the slitless spectra, we obtain Kron-aperture photometry \citep{Kron_1980} with a Kron factor of 2.5, which is large enough to include nearly all flux while maintaining a high SNR. The nearby \nii\,$\lambda$6585 line is masked if it presents. 
Although the median-filter continuum-subtraction method is helpful for emission-line detection, it can cause a flux loss of about 10 percent due to an oversubtraction of continuum near the line region.
Therefore, we fit a cubic polynomial continuum on each row of the 2D spectra around the emission line region, and measure the line flux on the continuum-subtracted spectra.   

The flux measurements of low-SNR emission lines are often artificially boosted due to random noise. If not corrected, this can overestimate the volume density of more luminous sources, which is also known as the Eddington bias \citep{eddington1913}. To quantify this effect, we use a Monte Carlo (MC) method (see Section~\ref{subsec:complt} for details) and compare injected and output line flux for a range of line SNRs from 3 to 50. For each input SNR, we apply a correction to our sources using the median value of the output-to-input line flux ratios.

We perform SED modeling with \textsc{cigale} \citep{Boquien_2019} to estimate dust extinction. An example is shown in Figure \ref{fig:sed_pz}. For each galaxy, its redshift is fixed at the spectroscopic redshift, and its photometric data points blueward of Lyman limit are not used for modeling. We assume a commonly-used, delayed star formation history (SFH; \texttt{sfhdelayed}), where $\mathrm{SFR} \propto te^{-t/\tau}$ and $\tau$ can vary in a wide range from 10\,Myr to 2\,Gyr. An optional late starburst is allowed in the latest 1–5\,Myr, which can contribute to $0\%–80\%$ of the total stellar mass. We use \citet{Bruzual_2003} stellar population synthesis models and adopt the \citet{Chabrier_2003} initial mass function (IMF). We also allow a metallicity range of 0.2 $Z_\odot$ to $Z_\odot$ and a broad ionization parameter ($\log U$) range of $-1.0$ to $-3.5$. 
We adopt the SMC extinction curve \citep{Pei_1992} for nebular emission and the modified \citet{Calzetti_2000} attenuation curve for the UV continuum, with the slope of the power-law modification to the attenuation curve varying in the range of -0.6 to 0.2 \citep[e.g.,][]{Buat_2018}.
The color excess of the nebular lines is allowed between $E(B-V)_{\mathrm{line}}=0-1.4$ mag. We allow the ratio between $E(B-V)_{\mathrm{line}}$ and the color excess of the stellar continuum $E(B-V)_{\mathrm{stellar}}$ to be 0.5 or 1 instead of the local measurement of 0.44 (\citealt{Calzetti_2000}; recently \citealt{sun_2025} measured $0.60\pm0.13$ for $z\sim5$ dusty galaxies). This allows that stellar and nebular emission can originate from the same birth clouds. 
The $E(B-V)_{\mathrm{line}}$ values are derived using a Bayesian method and span a range of $0-1.15$ mag. At both $z\sim4.5$ and 6.3, more than $80\%$ of the sources have best-fit models showing $E(B-V)_{\mathrm{line}}/E(B-V)_{\mathrm{stellar}}$ close to 0.5, and the remaining ratios are close to 1 (i.e., consistent with \citealt{sun_2025}). 
We then compute the extinction of \ha\ line $A_{\mathrm{H}\alpha}$ and dust-corrected \ha\ luminosity using $L_{\mathrm{H}\alpha, \mathrm{corr}} = L_{\mathrm{H}\alpha, \mathrm{obs}} \times 10^{0.4A_{\mathrm{H}\alpha}}$. Figure~\ref{fig:sed_pz} shows examples of SED modeling and dust-attenuation correction.

For the $z\sim 6.3$ \ha\ emitters, their \hb\ lines are observable in the F360M band, so we can calculate the dust attenuation based on the Balmer decrement, assuming the theoretical ratio $H_{\alpha}/H_{\beta}=2.86$ with an electron temperature $T_e=10^4$~K and a density $n_e=10^2$. For sources with reliable \hb\ detections ($\textrm{SNR}>2.5$), we find that the dust attenuation measurements based on the SED fitting and Balmer decrements are consistent within $1\sigma$.

Given that the \ha\ luminosity is a good indicator of SFR within $\sim 10$~Myr, we also compare the SFR averaged over the most recent 10 Myr derived using the SED modeling and the SFR estimated from the dust-corrected \ha\ luminosities using the \citet{Kennicutt_2012} relation (Figure~\ref{fig:LHa_SFR_AHa}).
We observe an excellent agreement among the two sets of the measurements, indicating that the results of the SED modeling and spectroscopic measurements are self-consistent. 
We also find that star-forming galaxies with higher \ha\ luminosities generally have higher \ha\ attenuation values and thus larger fractions of dust-obscured star formation. The Spearman's rank correlation analysis between the observed \ha\ luminosities and \ha\ attenuation values yields a moderate positive correlation coefficient of 0.30, with a highly significant $p$-value of $1.5 \times 10^{-6}$, indicating a statistically meaningful relationship between \ha\ luminosities and \ha\ attenuation.
Using data from the Deep UNCOVER-ALMA Legacy High-Z Survey (DUALZ, \citealt{Fujimoto_2023}), one of our sources (UNCOVER ID=17269; see also Figure~\ref{fig:a2744_sample} and \ref{fig:sed_pz}) is detected at 1.2-mm (continuum flux density $=0.28\pm0.06$\,mJy), and the dust extinction obtained from our SED modeling is also high ($E(B-V)_{\mathrm{line}}=0.68\pm0.10$).

\subsection{Lens Models} \label{subsec:lensmod}

\begin{figure*}[!t] 
    \centering
    \includegraphics[width=\textwidth]{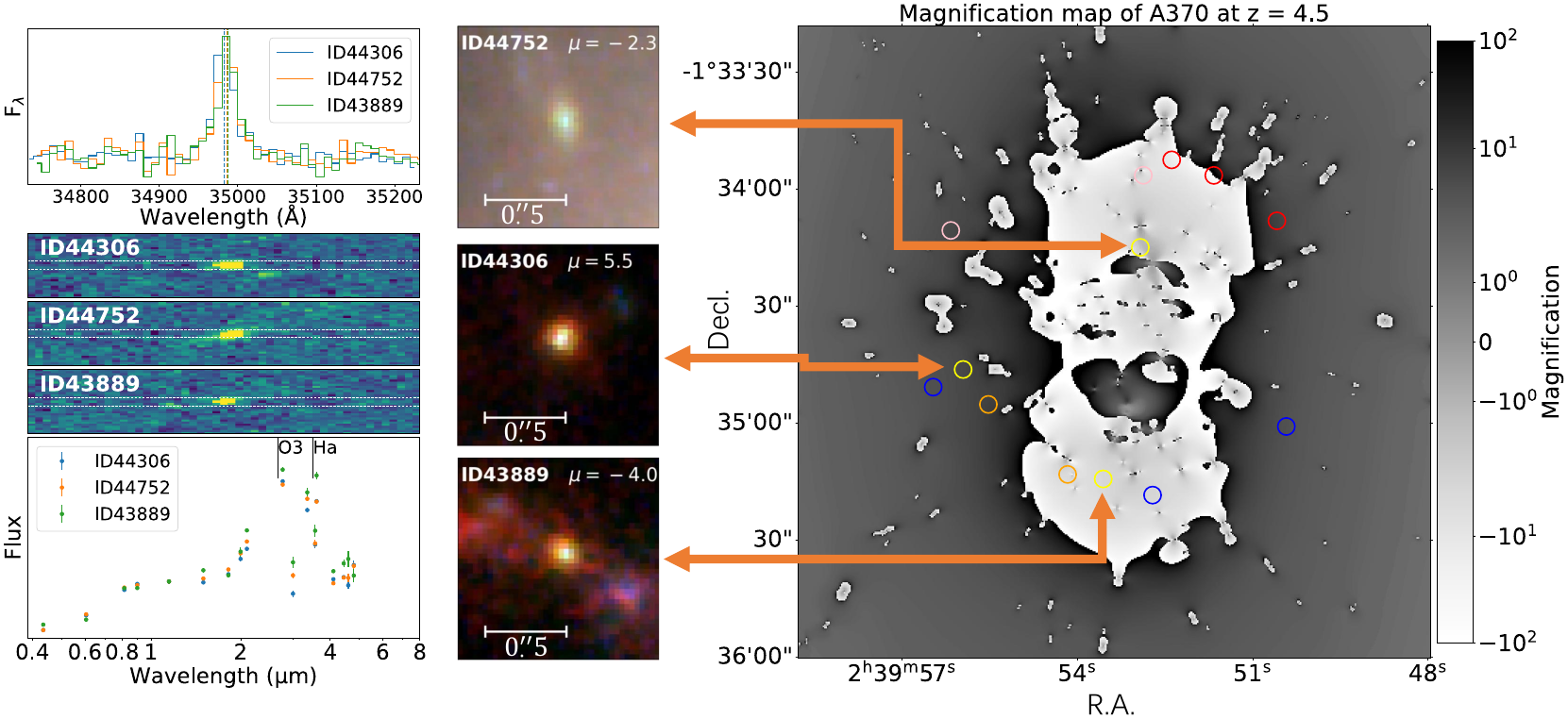} 
    \caption{An example of three multiple lens images for one source in A370. The panels on the left show their 1D and 2D \ha\ spectra and their optical-to-MIR SEDs.The dashed lines in the 1D spectra panel indicate the line centers. The 1D spectra have been scaled to the same maximum value, and the SEDs have been scaled to the same 1.15-\micron\ brightness. The middle panels show their cutout RGB images. On the right panel, the locations of multiple images in the A370 field are shown on the magnification map at $z=4.5$ derived from the \citet{Niemiec_2023} model. The multiple images of the same source are shown in the same color. }
    \label{fig:multi_image}
\end{figure*}

Observations of highly magnified regions in galaxy clusters can probe galaxy populations at very low luminosities that would otherwise not be undetected in blank field surveys. On the other hand, this benefit comes with significant challenges, particularly the need for precise lens models to calculate magnification factors and identify multiply imaged sources. 
The accuracy of these lens models depends on the number of confirmed multiple images with robust redshifts and how they spatially probe the lensing mass field, as well as a comprehensive understanding of the member galaxies that contribute to the cluster mass distribution.

In this work, for the A370 cluster we adopt the publicly available lens model constructed by \citet{Niemiec_2023} using the \textsc{Lenstool} \citep{Jullo_2007} public software and constrained using data from the Beyond the Ultra-deep Frontier Fields And Legacy Observations (BUFFALO) programme. For the M0416 field we adopt the lens model constructed by \citet{Richard_2021}. For the M1149 field, we apply the lens model produced by the CATS Team \citep{Jauzac_2016}. These models were constructed using a large number of secure multiple images, including a large number of spectroscopically confirmed multiple systems with VLT/MUSE that provide accurate mass distributions around the cluster center. The lens model of A2744 presented by \citet{Bergamini_2023} prior to the JWST observations included more spectroscopically confirmed multiple images than previous works. 
Several massive structures residing at large distances from the main cluster core were modeled, which have a non-negligible impact on the reconstruction of the cluster mass profile \citep[][e.g.]{Steinhardt_2020}. More recently, A2744 has been observed by JWST NIRSpec \citep[e.g.,][]{Bezanson_2024,Treu_2022}, and \citet{Bergamini_2023A} presents a new JWST-based lens model for A2744. 
This new model, adopted in our following analyses for A2744, uses 149 multiple images from 50 background sources, compared to, e.g., the model from \citet{Richard_2021} with 83 multiple images from 29 background sources.

Throughout this work, our results are based on these \textsc{Lenstool}-based models, which we refer to as fiducial results. Figure \ref{fig:multi_image} shows an example of magnification map for A370.
In Section~\ref{subsec:difflensmod}, we compare results among different lens models for the four fields, to quantify the uncertainties from the choice of lens models. Our analysis shows that the use of different lensing models does not lead to significantly different conclusions.

\subsection{Survey Volume} \label{subsec:vmax}

\begin{figure*}[!t] 
    \centering
    \includegraphics[width=\textwidth]{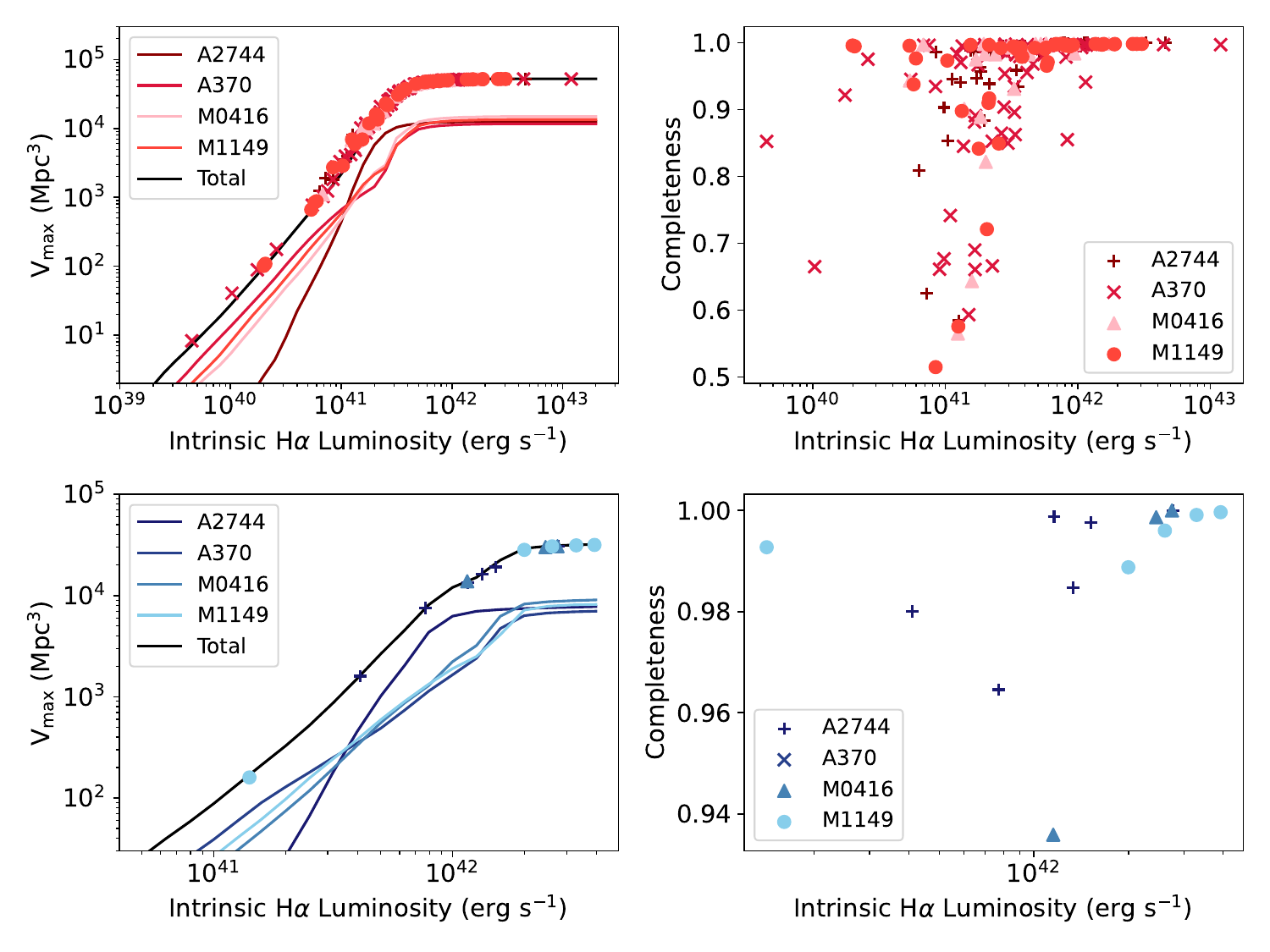} 
    \caption{Maximum survey volume (left panels) and completeness (right panels) versus intrinsic luminosity for each source at $z\sim4.5$ (top) and $z\sim6.3$ (bottom). Under a certain observed \ha\ flux limit, the maximum survey volume is a monotonic function of the intrinsic \ha\ luminosity, flattening at the intrinsic \ha\ luminosity limit corresponding to a magnification factor of $\mu = 1$. Meanwhile, the completeness is a function of the observed \ha\ SNR, leading to a scattered relation with the intrinsic \ha\ luminosity.} 
    \label{fig:c_vmax}
\end{figure*}

We compute the maximum survey volume ($V_\mathrm{max}$) in which a galaxy at a given intrinsic \ha\ luminosity can be detected at $\mathrm{SNR} > 5$ using the following method.
Over the entire survey footprint, we force extraction of 2D and 1D grism spectra over a $ 1\farcs5 \times 1\farcs5$ grid. 
The extraction method is the same as that for actual sources.
We compute the noise for a pseudo emission line located in the spectrum by co-adding the 1D error in $\pm$~FWHM, which is the median FWHM values of our \ha\ lines ($\mathrm{FWHM}=38.4$~\AA\ for $z\sim 4.5$ \ha\ lines and $\mathrm{FWHM}=54.8$~\AA\ for $z\sim 6.3$ \ha\ lines). If the extracted spectrum is contaminated by bright sources, the derived noise will be large, and therefore it would be difficult to detect faint line emitters around this region. 
Through this experiment, we obtain a $5\sigma$ depth map at a resolution of $1\farcs5 \times 1\farcs5$ for lines between $z=4.18$ to $z=4.95$ and $z=6.00$ to $z=6.62$) with a step size of $\Delta z=0.01$. 
For each line luminosity, we identify the area in the image plane in which their magnified flux would be larger than the local $5\sigma$ detection limit at each redshift. The survey area on the source plane is calculated by dividing the area on the image plane by the corresponding magnification factor, and the survey volume is integrated from the survey area across the two redshift bins.
We also examine whether $V_\mathrm{max}$ is sensitive to the choice of the grid size in the magnification maps. 
We find that binning the magnification maps to a grid size of 1\farcs5 would only cause a minor difference from the original pixel size (0\farcs15) in the maximum area on the source plane across all redshift, except for the regions with extremely large magnifications $(\mu>100)$ that no source in our sample fall into.
The survey volumes versus intrinsic luminosities (i.e., corrected for magnification) for each source are shown in the left panels of Figure~\ref{fig:c_vmax}. The survey volume is a monotonic function of the intrinsic \ha\ luminosity, flattening at the intrinsic \ha\ luminosity limit corresponding to a magnification factor of $\mu = 1$.

\subsection{Sample Completeness} \label{subsec:complt}

To determine the LF, it is essential to consider the completeness of our source detection. Our photometric detection limit by design is sufficient to detect emission-line galaxies selected with grism spectroscopy ($\geq5\sigma$), and noise fluctuations over the entire survey area are considered in the calculation of the survey volume in Section \ref{subsec:vmax}. 
In terms of completeness in the selection of photometric redshifts, the abundant medium- and wide-band photometry data presented in Section \ref{subsec:image} place excellent constraints on the redshifts. 
For verification, we also extract and examine the spectra of a subsample of sources with F360M or F480M photometry $\mathrm{SNR}>3$ regardless of their photometric redshifts, and we do not find sources with \ha\ emission lines with high confidence.

The completeness of \ha\ line detections in the 1D and 2D spectra is evaluated by the MC method. Similar to \citet{Sun_2023}, we first select faint continuum sources ($> 24$ mag) observed with the same numbers of integrations with each module/grism combination (AC and BC) as those of each line emitter, respectively. These faint continuum sources do not have line emission detected.
We then extract and co-add their 2D spectra, and inject mock line emissions with a SNR range from 3 to 50 randomly across the same wavelength range of \ha\ lines at $z=4.18-4.95$ and $6.00-6.62$. Line emission is injected as a 2D Gaussian profile with the same FWHM as the median value of our sample. The line flux and errors are measured using the same method as that for the real line detections. Based on the median output line flux and error, we derive the fraction of $5\sigma$ detections for each of the input line SNR, which is the completeness of lines at this SNR. The completeness versus intrinsic luminosities (corrected for magnification) for each source are shown in the right panels of Figure~\ref{fig:c_vmax}. Unlike the survey volume, the completeness versus intrinsic luminosity does not follow a simple function because of the different observing conditions for different sources (e.g., exposure time and contaminant brightness).

\subsection{Multiple image systems} 
\label{subsec:multimage}

In order to avoid double counting sources in the measurement of \ha\ LFs, we examine whether our \ha\ emitter sample contains multiple images from the same source.
For each \ha\ emitter with observed position $\vec{\theta}_0$, we run \textsc{Lenstool} to calculate its actual position in the sky $\vec{\beta}$ and the possible positions of all observed images $\vec{\theta}$, which we match against other \ha\ emitters in the sample to identify multiple images. 
As the precision of the lens model is typically $\lesssim 1\arcsec$ and the density of objects in a given redshift plane is low, we choose a conservative matching radius $3\arcsec$ to find all possible multiple images. 

We obtain 14 multiple image systems (see Table~\ref{tab:multi_image}). In each system, the differences of the \ha\ line centers are less than $\mathrm{FWHM}_\mathrm{res}$ ($\sim200\ \mathrm{km}\ \mathrm{s}^{-1}$). We visually inspect the spectra and SEDs of the multiple images to ensure that they have similar spectral features and SED shapes. We also examine their image morphology, and find that most of them are consistent and the remaining images exhibit elongated lensing features. The left panel of Figure~\ref{fig:multi_image} shows the spectra, SEDs, and images of one multiply imaged system in the A370 field. The figure also shows the locations of five multiply imaged systems in the field (another system in this field, No.~1 in Table~\ref{tab:multi_image}, is far away from the cluster center and is not shown in this figure). Some of these systems have been identified by VLT/MUSE observations \citep{Lagattuta_2019}.

\begin{deluxetable}{ccccc}[!t]
\tabletypesize{\footnotesize}
\tablewidth{\textwidth} 
\tablecaption{Multiple image systems in the four fields \label{tab:multi_image}}
\tablehead{
\colhead{No.} & \colhead{Field} & \colhead{Redshift} & \colhead{R.A.} & \colhead{Decl.} }
\startdata{
\multirow{2}{*}{1}  & \multirow{2}{*}{A370}  & \multirow{2}{*}{4.7564} & 39.98051  & --1.632664  \\
                    &                        &                         & 39.98103  & --1.633273  \\ \hline
\multirow{3}{*}{2}  & \multirow{3}{*}{A370}  & \multirow{3}{*}{4.3294} & 39.98325  & --1.579611  \\
                    &                        &                         & 39.97061  & --1.570912  \\
                    &                        &                         & 39.97326  & --1.587409  \\ \hline
\multirow{3}{*}{3}  & \multirow{3}{*}{A370}  & \multirow{3}{*}{4.4904} & 39.96838  & --1.564676  \\
                    &                        &                         & 39.96088  & --1.569020  \\
                    &                        &                         & 39.96538  & --1.565781  \\ \hline
\multirow{2}{*}{4}  & \multirow{2}{*}{A370}  & \multirow{2}{*}{4.425}  & 39.98145  & --1.582099  \\
                    &                        &                         & 39.97580  & --1.587086  \\ \hline
\multirow{3}{*}{5}  & \multirow{3}{*}{A370}  & \multirow{3}{*}{4.2524} & 39.96974  & --1.588559  \\
                    &                        &                         & 39.98538  & --1.580864  \\
                    &                        &                         & 39.96021  & --1.583670  \\ \hline
\multirow{2}{*}{6}  & \multirow{2}{*}{A370}  & \multirow{2}{*}{4.7591} & 39.97040  & --1.565774  \\
                    &                        &                         & 39.98412  & --1.569705  \\ \hline
\multirow{2}{*}{7}  & \multirow{2}{*}{M1149} & \multirow{2}{*}{4.6081} & 177.39476 & 22.364505  \\
                    &                        &                         & 177.39397 & 22.363669  \\ \hline
\multirow{2}{*}{8}  & \multirow{2}{*}{M1149} & \multirow{2}{*}{4.3759} & 177.40014 & 22.404139  \\
                    &                        &                         & 177.39834 & 22.403769  \\ \hline
\multirow{3}{*}{9}  & \multirow{3}{*}{M1149} & \multirow{3}{*}{4.4927} & 177.39457 & 22.394305  \\
                    &                        &                         & 177.40032 & 22.396415  \\
                    &                        &                         & 177.39582 & 22.400015  \\ \hline
\multirow{3}{*}{10} & \multirow{3}{*}{M1149} & \multirow{3}{*}{4.7319} & 177.41122 & 22.388449  \\
                    &                        &                         & 177.40994 & 22.387228  \\
                    &                        &                         & 177.40656 & 22.384497  \\ \hline
\multirow{2}{*}{11} & \multirow{2}{*}{M1149} & \multirow{2}{*}{6.3573} & 177.40522 & 22.376684  \\
                    &                        &                         & 177.40525 & 22.376619  \\ \hline
\multirow{2}{*}{12} & \multirow{2}{*}{A2744} & \multirow{2}{*}{4.5735} & 3.61756   & --30.395003 \\
                    &                        &                         & 3.61751   & --30.395353 \\ \hline
\multirow{2}{*}{13} & \multirow{2}{*}{A2744} & \multirow{2}{*}{4.2732} & 3.55848   & --30.361914 \\
                    &                        &                         & 3.55892   & --30.362427 \\ \hline
\multirow{2}{*}{14} & \multirow{2}{*}{A2744} & \multirow{2}{*}{4.7535} & 3.58593   & --30.403133 \\
                    &                        &                         & 3.58371   & --30.404080
}\enddata
\end{deluxetable}

\begin{deluxetable}{ccccc}
\tablewidth{\textwidth} 
\tablecaption{\ha\ luminosity functions \label{tab:HaLF}}

\tablehead{
\colhead{$\log{L_{\mathrm{H}\alpha}}$} & \colhead{N} & \colhead{$<c>$} & \colhead{$<V_{\mathrm{max}}>$} & \colhead{$\log{\Phi_{H\alpha}}$} \\
\colhead{($\mathrm{erg}\ \mathrm{s}^{-1}$)} & \colhead{} & \colhead{} & \colhead{($/V_{\mathrm{max}}(L_{\mathrm{max}})$)} & \colhead{($\mathrm{Mpc}^{-3}\ \mathrm{dex}^{-1}$)}}
\decimalcolnumbers
\startdata
\multicolumn{5}{c}{$z\sim 4.5$ Observed} \\ \hline
$40.80 \pm 0.50$ & $21$ & $0.88$ & $0.118$ & $-1.63\pm0.24$ \\
$41.55 \pm 0.25$ & $72$ & $0.90$ & $0.713$ & $-2.31\pm0.12$ \\
$42.05 \pm 0.25$ & $71$ & $0.98$ & $0.968$ & $-2.54\pm0.11$ \\
$42.55 \pm 0.25$ & $22$ & $1.00$ & $0.998$ & $-3.07\pm0.14$ \\
$43.05 \pm 0.25$ & $3$ & $1.00$ & $1.000$ & $-3.94\pm0.27$ \\
$43.55 \pm 0.25$ & $1$ & $1.00$ & $1.000$ & $-4.42\pm0.45$ \\
\hline
\multicolumn{5}{c}{$z\sim 4.5$ Dust-corrected} \\ \hline
$41.00 \pm 0.50$ & $28$ & $0.87$ & $0.196$ & $-1.62\pm0.18$ \\
$41.75 \pm 0.25$ & $72$ & $0.91$ & $0.768$ & $-2.37\pm0.11$ \\
$42.25 \pm 0.25$ & $56$ & $0.99$ & $0.964$ & $-2.64\pm0.12$ \\
$42.75 \pm 0.25$ & $26$ & $0.99$ & $0.991$ & $-2.99\pm0.13$ \\
$43.25 \pm 0.25$ & $7$ & $1.00$ & $1.000$ & $-3.57\pm0.19$ \\
$43.75 \pm 0.25$ & $1$ & $1.00$ & $1.000$ & $-4.42\pm0.45$ \\ \hline
\multicolumn{5}{c}{$z\sim 6.3$ Observed} \\ \hline
$41.70 \pm 0.25$ & $2$ & $0.99$ & $0.144$ & $-2.46\pm0.35$ \\
$42.20 \pm 0.25$ & $6$ & $0.95$ & $0.742$ & $-3.22\pm0.22$ \\
$42.70 \pm 0.25$ & $7$ & $1.00$ & $0.997$ & $-3.35\pm0.21$ \\
\hline
\multicolumn{5}{c}{$z\sim 6.3$ Dust-corrected} \\ \hline
$41.68 \pm 0.25$ & $2$ & $0.99$ & $0.144$ & $-2.46\pm0.36$ \\
$42.18 \pm 0.25$ & $4$ & $0.95$ & $0.643$ & $-3.33\pm0.25$ \\
$42.68 \pm 0.25$ & $5$ & $0.99$ & $0.971$ & $-3.48\pm0.23$ \\
$43.18 \pm 0.25$ & $4$ & $1.00$ & $1.000$ & $-3.60\pm0.25$ \\
\enddata
\tablecomments{Column (1): logarithm \ha\ luminosity bins. (2): number of galaxies in each bin. (3): mean completeness of \ha\ emitters in each bin. (4): mean volume $V_\mathrm{max}$ per bin in which an \ha\ emitter with the given intrinsic luminosity can be detected, compared with the total survey volume. (5): number density $\log{\Phi_{H\alpha}}$, with the uncertainty including both Poisson error and cosmic variance (Section~\ref{subsec:HaLF}). 
}
\end{deluxetable}

\begin{deluxetable}{cccc}
\tablewidth{\textwidth} 
\tablecaption{Schechter fit parameters of the \ha\ luminosity functions \label{tab:HaLF_para}}
\tablehead{
\colhead{$z$} & \colhead{$\log{L^*_{\mathrm{H}\alpha}}$} & \colhead{$\log \Phi^*_{\mathrm{H}\alpha}$} & \colhead{$\alpha_{\mathrm{H}\alpha}$} \\
\colhead{} & \colhead{($\mathrm{erg}\ \mathrm{s}^{-1}$)} & \colhead{($\mathrm{Mpc}^{-3}\ \mathrm{dex}^{-1}$)} & \colhead{} 
} 
\decimalcolnumbers
\startdata
\multicolumn{4}{c}{Observed} \\ \hline
$4.5$ & $43.07^{+0.23}_{-0.16}$ & $-3.73^{+0.25}_{-0.37}$ &  $-1.83^{+0.13}_{-0.13}$ \\
$6.3$ & $43.42^{+0.88}_{-0.68}$ & $-4.57^{+0.95}_{-0.93}$ & $-1.85^{+0.33}_{-0.19}$ \\ \hline
\multicolumn{4}{c}{Dust-corrected} \\ \hline
$4.5$ & $43.45^{+0.32}_{-0.18}$ & $-3.89^{+0.27}_{-0.40}$ &  $-1.76^{+0.11}_{-0.10}$\\  
$6.3$ & $44.11^{+0.53}_{-0.58}$ & $-5.05^{+0.65}_{-0.51}$ &  $-1.79^{+0.20}_{-0.19}$ \\
\enddata
\end{deluxetable}

\section{Results} \label{sec:result}

\begin{figure*}[!t]
\gridline{\fig{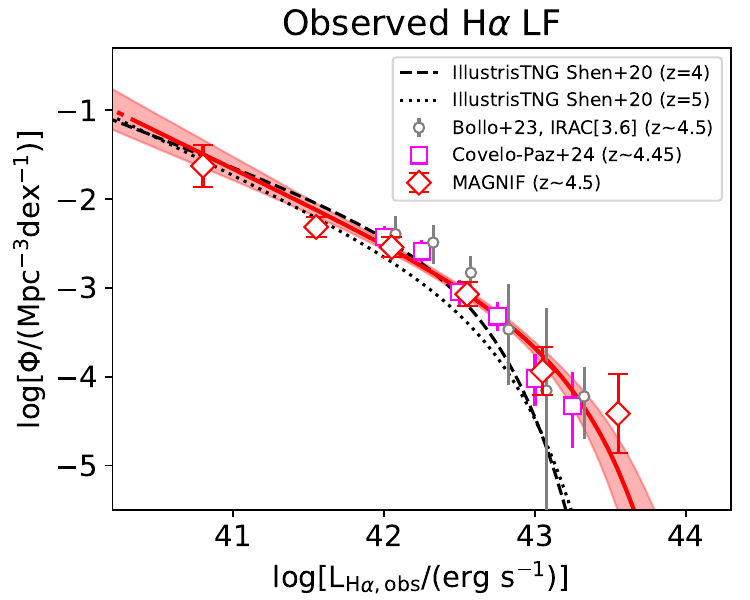}{0.49\textwidth}{}
          \fig{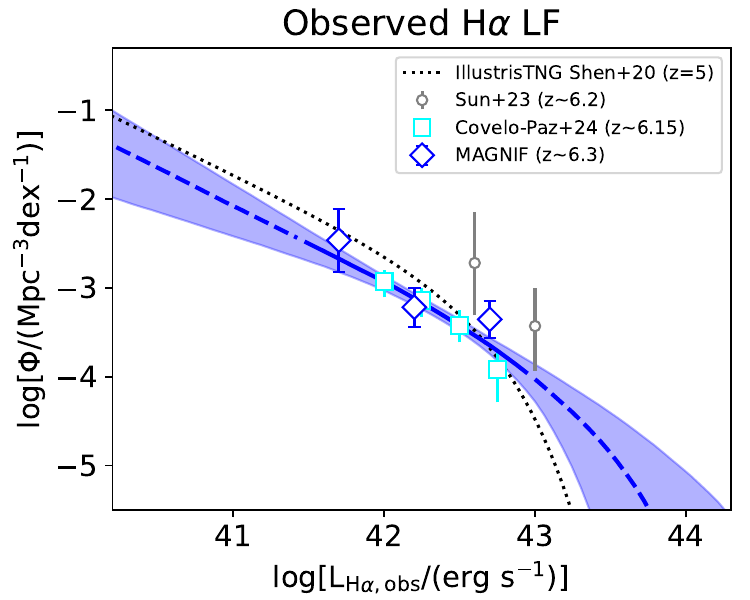}{0.49\textwidth}{}}
\caption{Observed \ha\ LFs at $z\sim 4.5$ (left) and $z\sim 6.3$ (right). The red (blue) line with the corresponding shaded region shows the best-fit Schechter function and the $1\sigma$ confidence interval. The dashed lines indicate the regions extrapolated from the best-fit functions. The fitting incorporates the \citet{Covelo-Paz_2024} \ha\ LF data at $z\sim4.45$ and 6.2 (magenta and cyan squares), which are also derived from NIRCam WFSS observations (FRESCO and CONGRESS). Also shown are the \ha\ LFs at $z\sim4$ and $z\sim5$ from the IllustrisTNG simulation \citep{Shen_2020} and the \ha\ LFs at $z \sim 4.5$ based on Spitzer IRAC photometry \citep{Bollo_2023}.   \label{fig:HaLF_obs}}
\end{figure*}

\begin{figure*}[!t]
\gridline{\fig{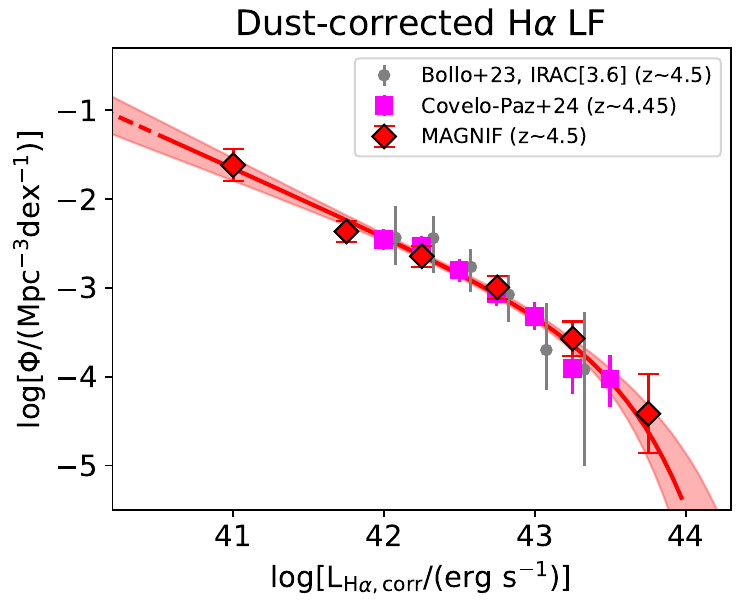}{0.49\textwidth}{}
          \fig{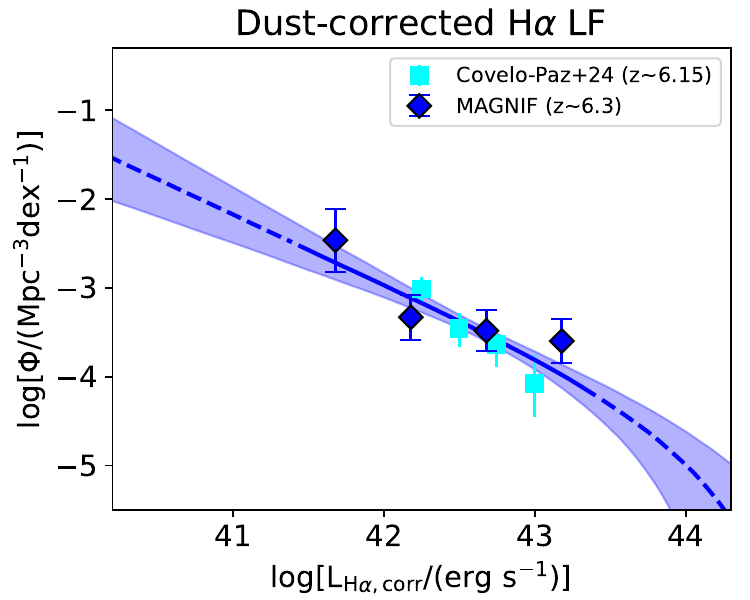}{0.49\textwidth}{}}
\caption{Dust-corrected \ha\ LFs at $z\sim 4.5$ (left) and $z\sim 6.3$ (right). The red (blue) lines with the corresponding shaded region show the best-fit Schechter function and the $1\sigma$ confidence interval. The dashed lines indicate the region extrapolated from the best-fit functions. The fitting incorporates the \citet{Covelo-Paz_2024} $z\sim4.45$ and 6.2 dust-corrected \ha\ LF data (magenta and cyan squares). The dust-corrected \ha\ LFs from \citet{Bollo_2023} at $z \sim 4.5$ is also shown for comparison. \label{fig:HaLF_corr}}
\end{figure*}

\subsection{\ha\ Luminosity Functions} \label{subsec:HaLF}

The \ha\ LFs are calculated using the direct $1/V_\mathrm{max}$ method \citep{Schmidt_1968} as 
\begin{equation}
\Phi(L) = \frac{1}{d\log{L}}\sum_i\frac{1}{C_iV_{\mathrm{max},i}},
\end{equation}
where $C_i$ is the completeness of the $i$-th source in the luminosity bin (see Section~\ref{subsec:complt}), and $V_{\mathrm{max},i}$ is the maximum observable volume of the $i$th source (see Section~\ref{subsec:vmax}).

We calculated observed \ha\ luminosity functions and the dust-corrected luminosity functions at $z \sim 4.5$ and $z \sim 6.3$. Both `good' samples and `possible' samples are used in the following calculation of the luminosity functions; removing `possible' objects from the analyses only causes a $\lesssim 0.1$~dex decrease in each bin. 
We identify 11 groups of possible mergers or interacting galaxies. Each group contains two or three \ha\ emitters located at the same redshift with a separation less than $1\arcsec$. When calculating the luminosity functions, we treat each group as a single source. The final sample size for our LF measurement is 190 ($z\sim4.5$) $+$ 15 ($z\sim6.3$) $=$205.

The resultant LFs are reported in Table~\ref{tab:HaLF} and shown in Figure~\ref{fig:HaLF_obs} (observed \ha\ LFs) and Figure~\ref{fig:HaLF_corr} (dust-corrected LFs).
When calculating errors, we consider Poisson noise for the small number statistics and also include the uncertainty caused by cosmic variance following the prescription in \citet{Moster_2011}, which estimates cosmic variance as the product of the galaxy bias \citep{Qiu_2018,Dalmasso_2024} and dark matter cosmic variance for a given survey geometry. The resultant cosmic variances are divided by the square root of the number of sightlines, which are in consistency with those derived from the simple formula by \citet{Driver_2010}.
We further assess cosmic variance using the field-to-field variation of our sample in Section~\ref{subsec:cos_var}.

We model the \ha\ LFs using a Schechter function defined as:
\begin{equation}
\Phi(L)=\ln(10)\phi^* \left( \frac{L}{L^*} \right)^\alpha e^{-L/L^*}\left(\frac{L}{L^*} \right)
\end{equation}
where $\phi^*$ represents the number density normalization, $L^*$ is the characteristic (break) luminosity, and $\alpha$ is the faint-end slope. 

We obtain Monte Carlo Markov Chain (MCMC) fitting of the function above using \texttt{EMCEE} package.
During the fitting process, we find that the break luminosity $L^*$ is poorly constrained due to the limited survey volume and thus a small sample size at the bright end. We obtain unrealistically large break luminosities if we allow all three Schechter parameters to vary.
To address this, we incorporate data from \citet{Covelo-Paz_2024}, who obtain \ha\ LF measurements at $z=4-6.5$ from NIRCam WFSS surveys (FRESCO and CONGRESS) over blank fields and larger area.
This allows us to better constrain the bright end of the LFs. 
\ha\ emitters with broad components, which likely host AGNs, were removed from their sample.
We apply the same cosmic variance estimation method as above to the LF data from \citet{Covelo-Paz_2024}, which results in an error floor of approximately $\sim 0.1$\,dex.
The allowed ranges for the Schechter parameters are $40<\log{L^*}<45$, $-3<\alpha<-1$, and $-6<\log{\phi^*}<-2$. All prior distributions are flat.

The Schechter functions and the $1\sigma$ confidence intervals are shown in Figure~\ref{fig:HaLF_obs} and Figure~\ref{fig:HaLF_corr}. 
The best-fit results are summarized in Table~\ref{tab:HaLF_para}. 

\begin{figure*}[!t]
\gridline{\fig{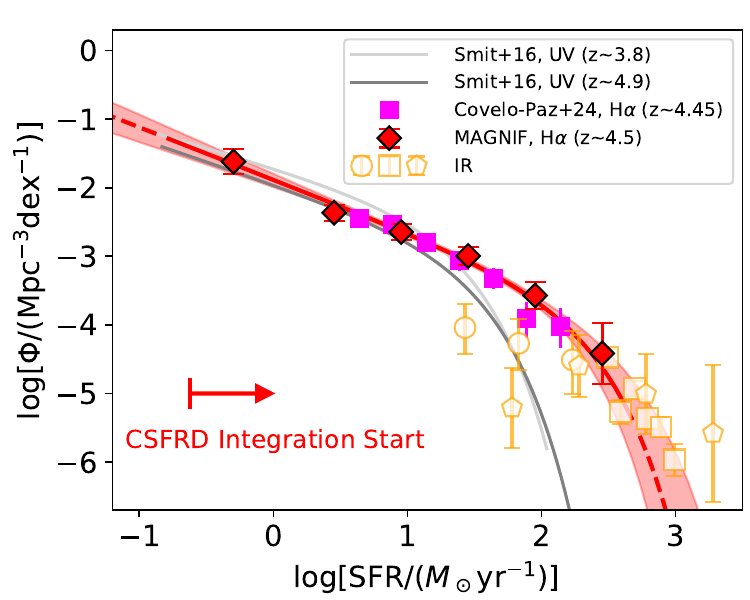}{0.49\textwidth}{}
          \fig{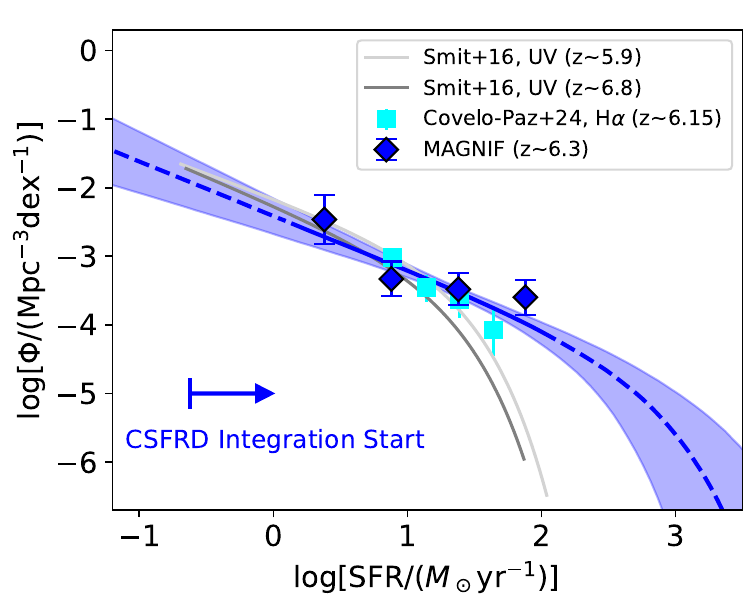}{0.49\textwidth}{}}
\caption{Star formation rate functions at $z\sim 4.5$ (left) and $z\sim 6.3$ (right) calculated based on the dust-corrected \ha\ LFs. The color scheme is the same as in Figure~\ref{fig:HaLF_corr}. The lower integration limit for the CSFRD calculation is $0.24\ M_\odot\ \mathrm{yr}^{-1}$ as indicated by the arrow. Results at similar redshifts from the literature are shown for comparison, including SFRFs based on \ha\  \citep{Covelo-Paz_2024}, UV \citep{Smit_2016} and infrared LFs \citep{Koprowski_2017,Traina_2024,sun_2025}. All results are dust-corrected and scaled to a \citet{Chabrier_2003} IMF.  \label{fig:SFRF}}
\end{figure*}

\subsection{SFRF and CSFRD} \label{subsec:sfrf}

The SFRF can be directly converted from the dust-corrected \ha\ luminosity functions assuming the \citet{Kennicutt_2012} relation $\log[\mathrm{SFR} / (M_\odot\ \mathrm{yr}^{-1})]=\log[L_{\mathrm{H}\alpha}/(\mathrm{erg\ s}^{-1})]-41.27$, multiplied by a constant factor 0.94 to consider a \citet{Chabrier_2003} IMF \citep{Madau_2014}.
The results are shown in Figure~\ref{fig:SFRF}. The SFRFs from the literature at similar redshift are also shown for comparison.
We apply scaling factors of 0.63 and 0.94 to the literature SFRs to convert the IMF from Salpeter \citep{Salpeter_1955} or Kroupa \citep{Kroupa_2001} to \citet{Chabrier_2003}, respectively.
Our estimated SFRFs are in good agreement with that from \citet{Bollo_2023} based on a \ha\ LF inferred from Spitzer/IRAC broad-band photometry, as well as those from \citet{Covelo-Paz_2024}. We also show the SFRFs calculated from the UV \citep[][extinction corrected]{Smit_2016} and infrared LFs \citep{Koprowski_2017,Traina_2024,sun_2025} at similar redshift for comparison. 

We find that in general the \ha-based SFRFs agree with UV-based SFRFs in the low-SFR regime ($\mathrm{SFR} \simeq 0.1 - 30\ M_\odot\ \mathrm{yr}^{-1}$; corresponding to absolute UV magnitude $-22 \lesssim M_\mathrm{UV} \lesssim -16$\,mag) and with IR-based SFRFs in the high-SFR regime ($\mathrm{SFR} \gtrsim 100\ M_\odot\ \mathrm{yr}^{-1}$). 
This highlights the ability of \ha\ emission in tracing the star formation across a remarkably wide range of galaxy populations over a $\gtrsim$3-dex span of SFR, including dwarf star-forming galaxies that often remain undetected in infrared surveys and dusty star-forming galaxies that are frequently missed in UV surveys. 
We discuss in more details the small difference in the faint-end slope between \ha-based SFRF and UV-based SFRF in Section~\ref{subsec:evo}.

\begin{figure*}[!ht] 
    \centering
    \includegraphics[width=\textwidth]{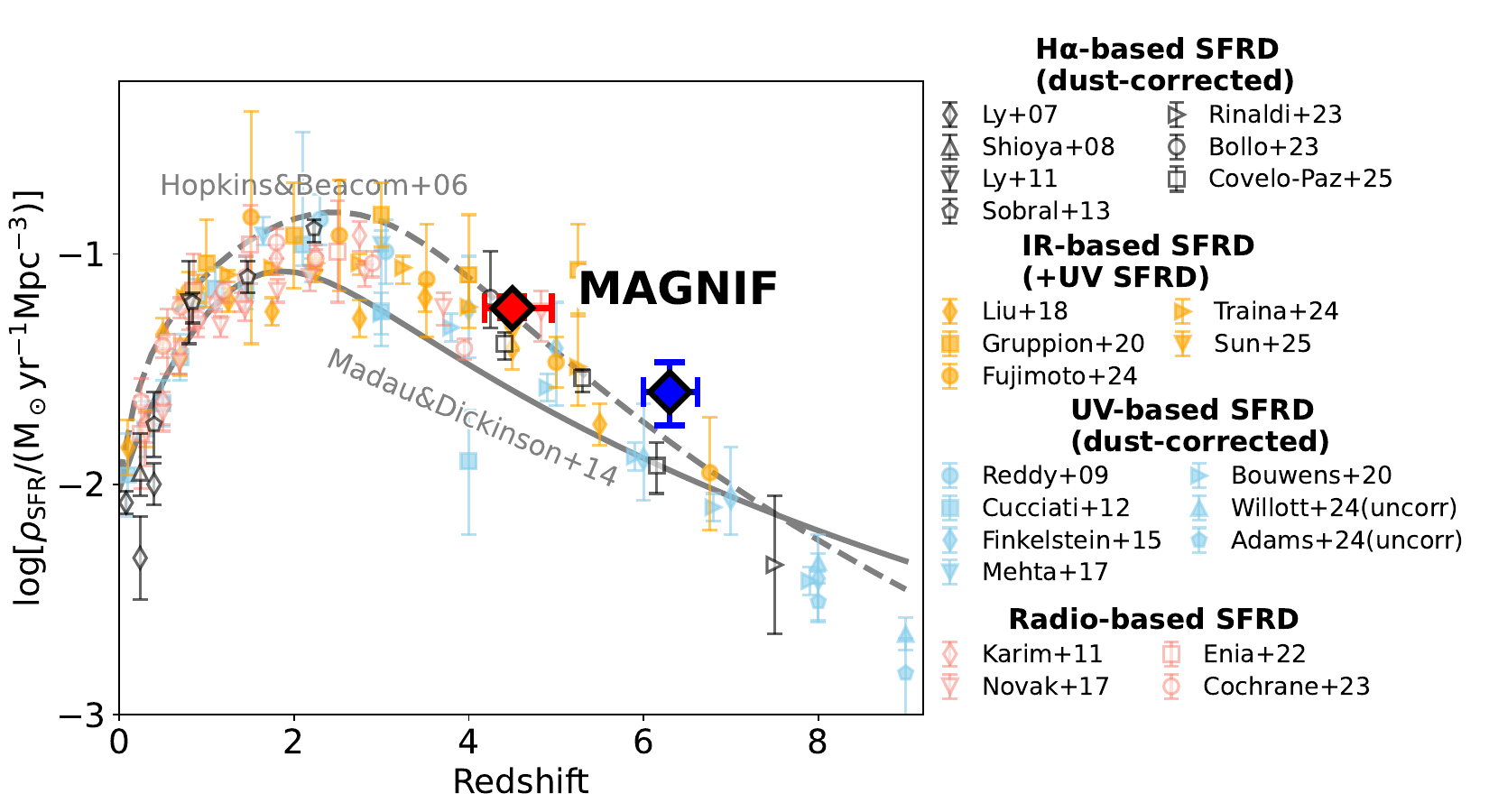} 
    \caption{Cosmic evolution of the CSFRD. Our results are compared to previous measurements based on: \ha\ surveys \citep{Ly_2007, Shioya_2008,Ly_2011,Sobral_2013, Bollo_2023, Rinaldi_2023, Covelo-Paz_2024}; infrared surveys \citep{Liu_2018,gruppioni_alpine-alma_2020,fujimoto_alma_2024,Traina_2024,sun_2025}; dust-corrected UV LFs \citep{Reddy_2009,Cucciati_2012,Finkelstein_2015,Mehta_2017,Bouwens_2020}; radio surveys \citep{Karim_2011,novak_vla-cosmos_2017,Enia_2022,Cochrane_2023}.
    The solid and dashed gray lines are the fits by \citet{Madau_2014} and \citet{Hopkins_2006}, respectively. All literature results were corrected for a \citet{Chabrier_2003} IMF.}
    \label{fig:CSFRD}
\end{figure*}

We calculate the CSFRD ($\rho_\mathrm{SFR}$) by integrating the best-fit Schechter function of SFRFs. 
For consistency with previous works, we set the lower integration limit to $0.24\ M_\odot\ \mathrm{yr}^{-1}$, labeled in Figure~\ref{fig:SFRF}, corresponding to an absolute UV magnitude limit of $M_{\mathrm{UV}}=-17$ \citep[e.g.,][]{Bouwens_2015,Adams_2024,Willott_2024}. This limit was set because galaxies are commonly observed at $z \geq 6$ down to magnitudes of $M_{\mathrm{UV}}=-17$. 
Note that we now find marginal spectroscopic evidence of star formation below $0.24\ M_\odot\ \mathrm{yr}^{-1}$, consistent with previous photometric UV observations of Hubble Frontier Fields, which rule out the presence of a turn-over brighter than $M_{\mathrm{UV}} = -13.1$ at $z \sim 3$, $M_{\mathrm{UV}} = -14.3$ at $z \sim 6$, and $M_{\mathrm{UV}} = -15.5$ at all other redshifts between $z \sim 9$ and $z \sim 2$ \citep{Livermore_2017,Bouwens_2022c}.
Here we use the same lower SFR limit $0.24\ M_\odot\ \mathrm{yr}^{-1}$ to fairly compare the UV- and \ha-based CSFRD estimates, and we discuss the difference in CSFRD by adopting different limits in Section~\ref{subsec:evo}. 

For each redshift bin, we run MC sampling of the posterior distribution of SFRF parameters and integrate the SFRF to derive the 16-50-84th confidence intervals of CSFRD.
We find that the CSFRD declines from $\rho_\mathrm{SFR}=0.058^{+0.008}_{-0.006}\ \ M_\odot\ \mathrm{yr}^{-1}\ \mathrm{Mpc}^{-3}$ at $z \sim 4.5$ to $\rho_\mathrm{SFR}=0.025^{+0.009}_{-0.007}\ \ M_\odot\ \mathrm{yr}^{-1}\ \mathrm{Mpc}^{-3}$ at $z\sim 6.3$. 
As shown in Figure~\ref{fig:CSFRD}, the CSFRD values we derived at $z\sim 4.5$ and $z\sim 6.3$ are both approximately 2.2 times higher than the values predicted based on the \citet{Madau_2014} relation, which is also suggested recently by \citet{sun_2025} and \citet{fujimoto_alma_2024} based on infrared surveys. 
Note that the higher IR selected or \ha-based CSFRD was first alluded to by \citet{Hopkins_2006}. 
The radio-based CSFRD is also higher than the \citet{Madau_2014} relation and is similar to the IR and \ha\ -based CSFRD \citep[e.g.,][]{Haarsma_2000, novak_vla-cosmos_2017}. This is expected as the radio synchrotron emission is not affected by dust extinction, and the production timescale of \ha\ emission generated by massive stars is similar to the production timescale of the radio synchrotron emission generated by SN remnants.

\section{Discussion} \label{sec:discuss}

In this section, we first discuss the potential systematic uncertainties of our LF and CSFRD determinations.
We then discuss the potential redshift evolution of faint-end slope $\alpha_\mathrm{H\alpha}$ and the physics behind.

\subsection{Different lens models} \label{subsec:difflensmod}

\begin{figure*}[!t]
\centering
\includegraphics[width=0.8\linewidth]{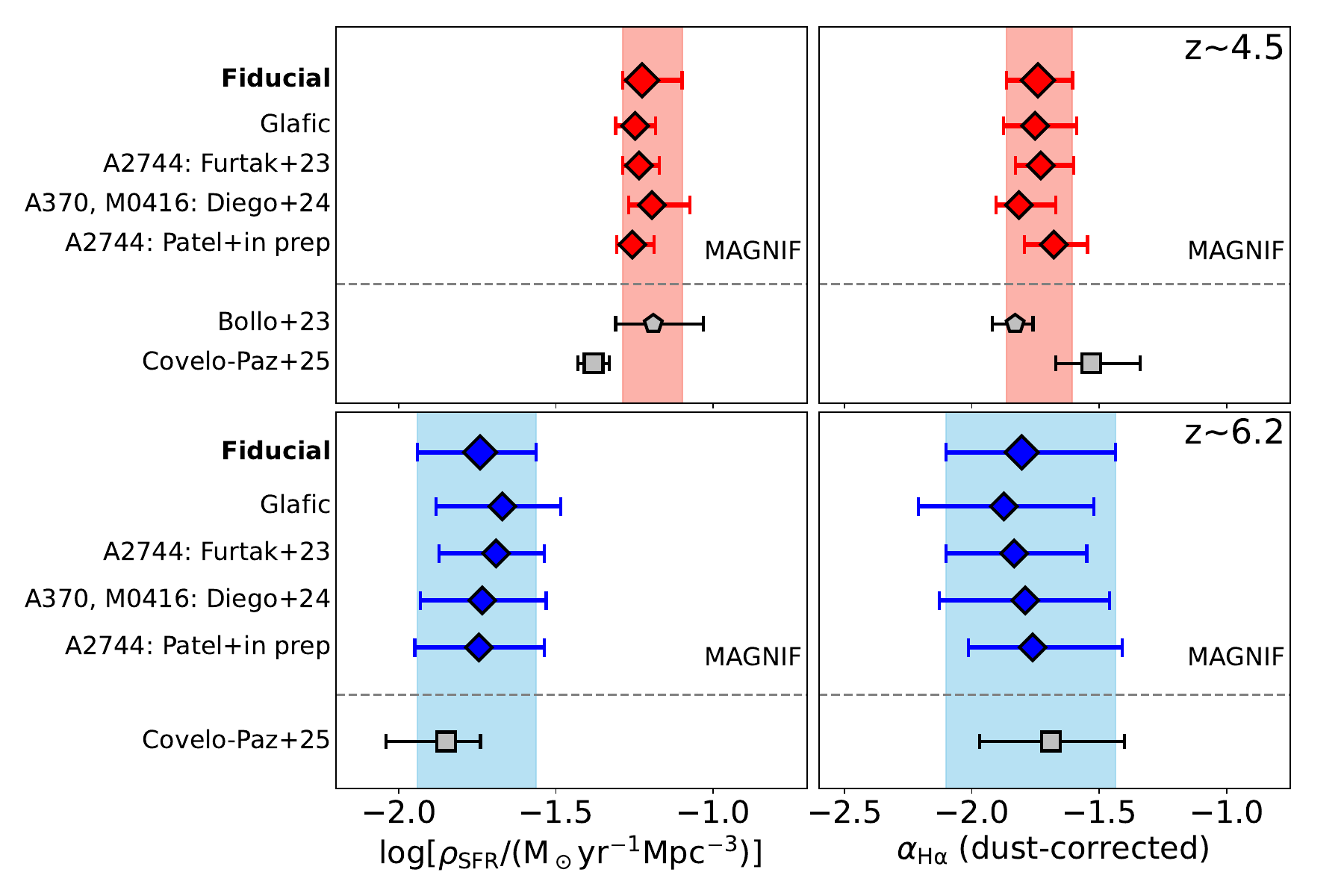}
\caption{The 16--50--84\%\ credible-interval constraints on CSFRD (left) and $\alpha_\mathrm{H\alpha}$ (right, dust-corrected).
Results from MAGNIF with assumptions of different sets of lens models (see Section~\ref{subsec:difflensmod}) are shown  as the solid diamonds (upper panels: $z\sim4.5$; lower panels: $z\sim6.3$).
Literature results based on \ha\ LF measurements are shown in the open symbols \citep{Bollo_2023,Covelo-Paz_2024}.
}
\label{fig:diff_lens}
\end{figure*}

The faint-end slope of \ha\ LF and resultant integral CSFRD are sensitive to the lens models.
To assess the uncertainty of lens models and the propagated systematic biases in our LF and CSFRD measurements, we repeat our measurements using different sets of lens models. 
In addition to the fiducial lens models (\romannumeral1) discussed in Section~\ref{subsec:lensmod}, we also experiment with
(\romannumeral2) \textsc{glafic} lens models \citep[][]{Oguri_2010,Kawamata_2016,Kawamata_2018,Okabe_2020} with updates for M0416 to include new multiple images and spectroscopic redshifts from MUSE \citep{Richard_2021} and for A2744 to include the extended region observed with JWST based on multiple image identifications of \citet{Furtak_2023} and \citet{Price_2024}; 
(\romannumeral3) same as the fiducial models but replacing A2744 with the \citet{Furtak_2023} model produced by the UNCOVER team, which includes the new UNCOVER \citep{Price_2024} and ALT \citep{Naidu_2024} spectroscopic redshifts, especially relevant for the two sub-clusters in the north and north-west; 
(\romannumeral4) same as the fiducial models but replacing A370 and M0416 with the models produced by \citet[][]{Diego_2024}, with updates from the JWST observations of A370 and M0416; 
(\romannumeral5) same as the fiducial models but replacing A2744 with the Patel et al. (2025, in preparation) modeled by the BUFFALO team, in which the latest multiple images from the UNCOVER team are also used to model the core, the north and north-west substructures of the cluster simultaneously. 

Figure~\ref{fig:diff_lens} shows the comparison of CSFRD and faint-end slope $\alpha_\mathrm{H\alpha}$ (dust-corrected) measured with different sets of lens models.
Results computed based on different lens model sets are consistent within $1\sigma$ uncertainty.
Therefore, we conclude that our fiducial measurements are relatively robust against the uncertainties of lens models.

\subsection{Cosmic variance}  \label{subsec:cos_var}

\begin{figure}[!t] 
    \centering
    \includegraphics[width=\linewidth]{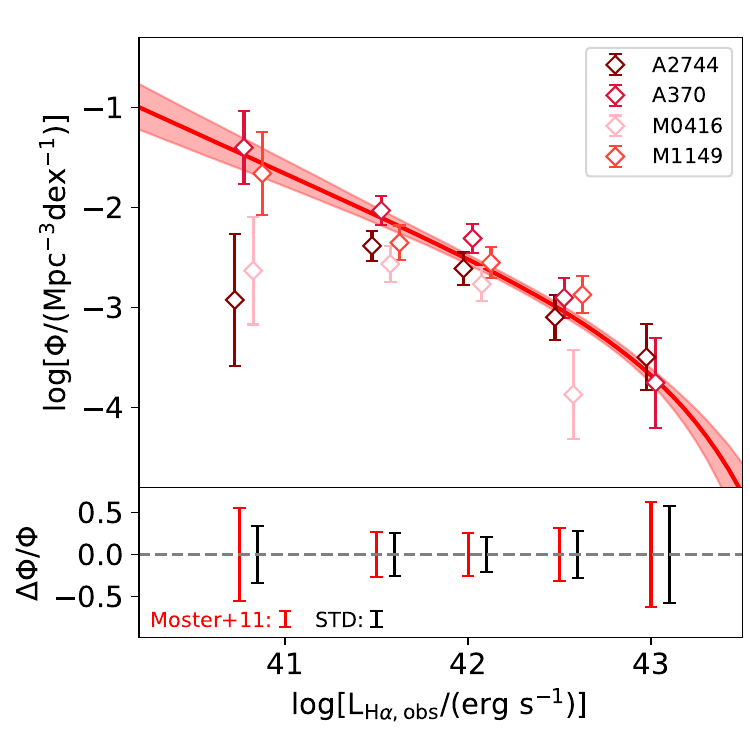} 
    \caption{Upper: \ha\ LFs at $z \sim 4.5$ calculated from four individual cluster fields. The LF derived from our entire sample with all parameters free, along with its $1\sigma$ confidence interval, is represented by the red solid line and the red shaded region, respectively. Bottom: The relative uncertainties of the $z \sim 4.5$ \ha\ LF derived from Poisson noise and the simple calculation of cosmic variance (red), as described in Section~\ref{subsec:HaLF}, and those derived from the standard deviation of the four LFs in the four fields (black). For clarity in the presentation, we slightly offset the markers along the $x$-axis.
\label{fig:LF_four_field}}
\end{figure}

Pencil-beam surveys of galaxies are subject to the cosmic variance arising from the large-scale structures of the Universe. 
This has been demonstrated in Table~\ref{tab:sample} as the strong field-to-field variation of numbers of \ha\ emitters.
Indeed, we designed the MAGNIF survey to observe four independent sightlines to quantify and mitigate the impact from cosmic variance on our LF measurements. 

In Figure~\ref{fig:LF_four_field} we show the \ha\ LFs derived using each of the four cluster fields at $z\sim4.5$. 
The LFs exhibit significant variation across the four cluster fields, particularly at the faint end, where the cosmic volume of a single field is limited.
We compare the relative uncertainties of the $z \sim 4.5$ \ha\ LF $\Delta\Phi/\Phi$ derived from Poisson noise and the simple calculation of cosmic variance, as described in Section~\ref{subsec:HaLF}, with those derived from the standard deviation of the four LFs in the four fields, as shown in the lower panel of Figure~\ref{fig:LF_four_field}. 
We find that the calculation of cosmic variance based on the method of \citet{Moster_2011} gives an excellent estimation for the cosmic variance, ensuring that we do not underestimate the uncertainty of our LF or CSFRD measurements.

\subsection{AGN contamination} \label{subsec:broad_Ha}

\begin{figure}[!t] 
    \centering
    \includegraphics[width=\linewidth]{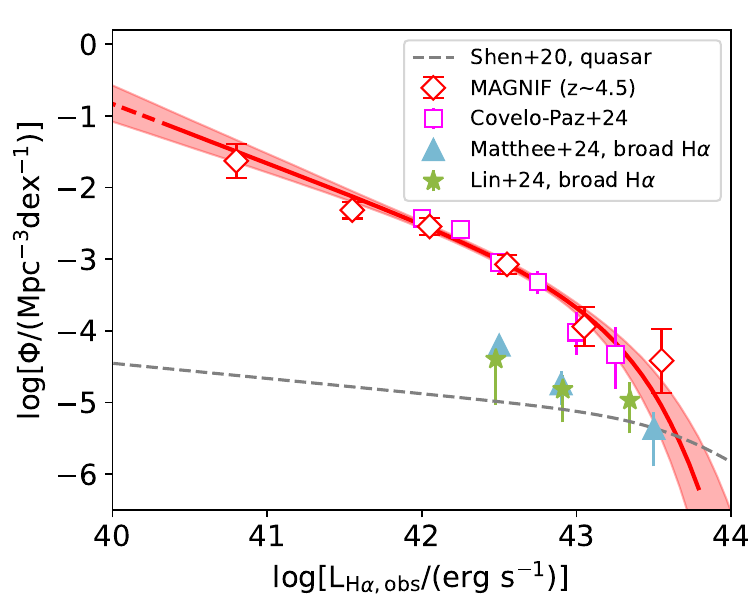} 
    \caption{\ha\ LF at $z\sim 4.5$  compared with the broad \ha\ line LFs at $z=4.2-5.5$ derived from EIGER and FRESCO surveys \citep{Matthee_2024} and at $z=4-5$ derived from ASPIRE survey \citep{LinX_2024}.  
    The \ha\ LF of quasars converted from the quasar bolometric LF at $z \sim 4$ (\citealt{Shen_2020_quasarLF}, extended to low-luminosity end) is also shown as the gray dashed line. 
    The AGN contribution to \ha\ LF at $L<10^{43}\ \mathrm{erg}\ \mathrm{s}^{-1}$ is expected to be minor.
    \label{fig:Ha_vs_AGN}}
\end{figure}

The broad-line region of AGN can emit luminous \ha\ line and contribute to the \ha\ LF.
Therefore, it is necessary to assess the AGN contamination in our \ha\ emitter sample and resultant CSFRD measurements.

We do not find broad line components among \ha\ emitters in our sample, indicating a lack of significant unobscured AGN contribution.
This may be a combined result of (\romannumeral1) a relatively low volume density of broad-line AGN at relatively low $L_\mathrm{H\alpha}$ ($\lesssim10^{42}\ \mathrm{erg}\ \mathrm{s}^{-1}$; in other words, the AGN luminosity function is flatter than that of galaxies at the faint end), and (\romannumeral2) a selection bias against broad-line AGN with our survey design and selection methodology.
We note that one \ha\ emitter at $ z = 6.25 $ in our sample was previously reported by \citet{Greene_2024} as a so-called ``little red dot'' (MSAID 35488), and the authors measured a broad \ha\ component ($\mathrm{FWHM} = 1900 \pm 210\ \mathrm{km\ s^{-1}}$) using the low-resolution ($R \sim 100$) NIRSpec/PRISM.
However, in our $R \sim 1600$ NIRCam slitless spectra, we do not detect the broad component, and a single Gaussian fitting reveals a $\mathrm{FWHM} = 251 \pm 39\ \mathrm{km\ s^{-1}}$ after correcting for instrumental broadening.

To further explore this non-detection of broad lines, we compare our \ha\ LFs with the broad \ha\ line LFs derived from other NIRCam WFSS surveys at similar redshifts \citep{Matthee_2024, LinX_2024}, as shown in Figure~\ref{fig:Ha_vs_AGN}. 
The \ha\ LF of quasars converted from the quasar bolometric LF at $z \sim 4$ \citep{Shen_2020_quasarLF} assuming $L_\mathrm{5100}=10^{44}[L_\mathrm{H\alpha}/(5.25\times 10^{42}\ \mathrm{erg}\ \mathrm{s}^{-1})]^{1/1.157}\ \mathrm{erg}\ \mathrm{s}^{-1}$ \citep{Greene_2005} and $L_\mathrm{bol}=10.33\times L_\mathrm{5100}$ \citep{Richards_2006} is also shown for comparison.
Our sample occupies a much fainter \ha\ luminosity range compared to the broad-line \ha\ emitter samples. 
Only three galaxies in our most luminous bin ($42.5 < \log[L_\mathrm{H\alpha}/(\mathrm{erg}\ \mathrm{s}^{-1})] < 43.0$) have \ha\ luminosities comparable to those of the broad-line AGNs. 
The number density of broad-line \ha\ emitters in this bin is approximately 1/8 of that of the \ha\ emitters without broad lines.
Therefore, we conclude that non-detection of broad \ha\ line in our sample is somewhat expected, and the AGN contamination to our CSFRD determination is negligible.

We also consider the contribution from obscured Type-2 AGN. 
\citet{Scholtz_2023} reports that $\sim20$\%\ of galaxies surveyed with JADES-NIRSpec host Type-2 AGNs (see also \citealt{Lyu_2024}).
However, most of these AGNs are identified through high-ionization lines (mostly in the rest-frame UV) or mid-IR SED modeling, and conventional optical diagnostics involving \ha\ emission (e.g., N2-BPT diagram; \citealt{Baldwin_1981}) are found less effective for high-redshift galaxies because of intrinsically lower metallicity.
Recent spatially resolved studies at low-redshift also suggest that about half of \ha\ emissions in AGN host galaxies still originate from star formation instead of AGN ionization \citep[e.g.,][]{deMellos_2024}.
Therefore, we conclude a minor ($\lesssim10$\%) contribution of Type-2 AGN to the observed \ha\ luminosity density, which is less a concern for CSFRD measurements compared with other aforementioned uncertainties.
Further JWST rest-frame UV spectroscopy and mid-IR imaging will be essential to fully assess the fraction of Type-2 AGN among our sample.

\subsection{Faint end of the \ha\ luminosity function} \label{subsec:evo}

\begin{figure}[!t] 
    \centering
    \includegraphics[width=\linewidth]{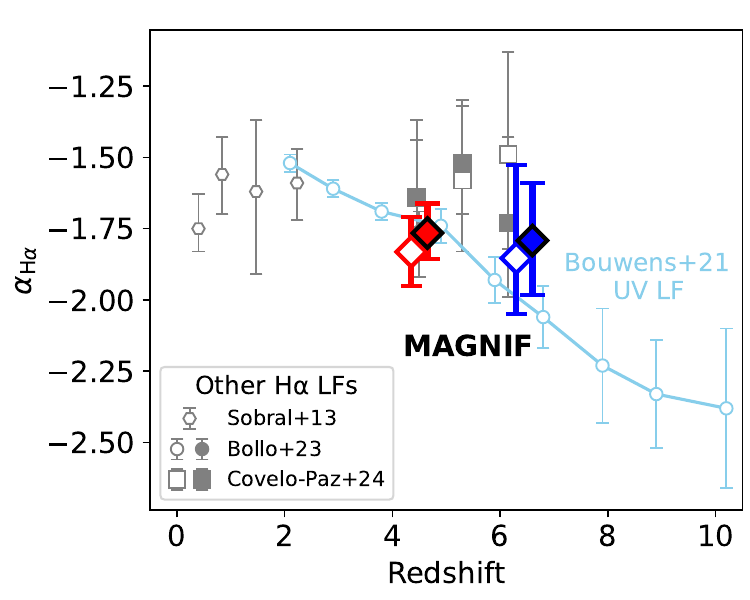} 
    \caption{No or weak redshift evolution of the faint-end slope of \ha\ LF across $z\simeq 0.4 - 6.3$. Our best-fit values (incorporating the \ha\ LF data from \citealt{Covelo-Paz_2024}) are compared with previous results: \citet{Sobral_2013} at $z \sim 0.4,\ 0.84,\ 1.47$, and $2.23$, \citet{Bollo_2023} at $z \sim 4.5$, and \citet{Covelo-Paz_2024} at $z \sim 4.45,\ 5.30$ and $6.15$. The open circles represent the parameters of the observed \ha\ LFs, while the filled circles are for the dust-corrected fits. 
    For clarity in the presentation, we slightly offset the markers along the $x$-axis.
    We also plot the faint-end slope of UV LFs \citep[][open skyblue circles]{Bouwens_2021} for comparison.}\label{fig:schechter_parameter_alpha}
\end{figure}

Figure~\ref{fig:schechter_parameter_alpha} shows the redshift evolution of the faint-end slope $\alpha$ of both observed and dust-corrected \ha\ LFs.
After dust correction, the faint-end slopes remain almost unchanged, while $L^*$ become larger. The results suggest that the majority of the dust-obscured star formation occurs in the most massive star-forming galaxies, rather than galaxies at the low-mass end.

We find that the faint-end slope of our dust-corrected \ha\ LFs at $z \sim 4.5$ is consistent with the faint-end slope of the UV LF \citep{Bouwens_2021}, as also shown in Figure~\ref{fig:SFRF}. 
Because of heavier dust attenuation at the rest-frame UV compared to that at the \ha\ wavelength, the observed UV LF is expected to be steeper than that of dust-corrected \ha\ LF \citep[e.g.,][]{Finkelstein_2012}.
However, galaxies with lower UV luminosities tend to have higher ratios of \ha\ luminosities to the luminosities of non-ionizing (UV) photons \citep[$L_\mathrm{H\alpha} / L_\mathrm{UV}$; e.g.,][]{Matthee_2016,Maseda_2020,Atek_star_2022}, leading to a steeper faint-end slope of the \ha\ LF.
The high $L_\mathrm{H\alpha} / L_\mathrm{UV}$ is a combined result of (\romannumeral1) higher ionizing photon production efficiency $\xi_\mathrm{ion}$ due to lower metallicity \citep[e.g.,][]{Kimm_2019} and (\romannumeral2) more bursty star formation in low-mass galaxies \citep[e.g.,][]{Shen_baryon_2014,Emami_closer_2019}, especially the prevalence of up-turns in the recent SFHs ($\lesssim10$\,Myr) among \ha-selected galaxies. 
Therefore, the SFR of these low-mass galaxies estimated through $L_\mathrm{H\alpha} - SFR$ conversion \citep{Kennicutt_2012} could be somewhat overestimated, which may lead to a potentially overestimated CSFRD.
We note that the metallicity-dependent $L_\mathrm{H\alpha} - {\rm SFR}$ conversion is less a concern at the bright end, where the \ha-based SFRF is consistent with the SFRF measured from dust-continuum-selected (and thus metal-rich) galaxies (Figure~\ref{fig:SFRF}).

At $z\sim6.3$, the faint-end slope measured from our dust-corrected \ha\ LF is consistent with the slope at $z\sim4.5$. 
Combining the measurements at $z\simeq0-2.3$ \citep[$\alpha_\mathrm{H\alpha}\sim-1.6$; e.g.,][]{Sobral_2013}, we conclude no or weak redshift evolution of $\alpha_\mathrm{H\alpha}$ out to the epoch of reionization (Figure~\ref{fig:schechter_parameter_alpha}).

The faint-end slope of \ha\ LF is somewhat flatter than that of UV LF at $z\sim6.3$, although caution shall be taken given the relatively large uncertainty of $\alpha_\mathrm{H\alpha}$.
We notice that at $z\sim6$, the so-called ``mini-quenched'' galaxies \citep[e.g.,][]{Looser_24a,Dome_2024} may constitute a substantial fraction (10--35\%) among dwarf galaxies with stellar mass $\simeq 10^8-10^9\ M_\odot$ \citep{Trussler_2025}.
These galaxies will contribute to the faint end of UV LF with a volume density of $\gtrsim 10^{-3}\ \mathrm{Mpc}^{-3}\ \mathrm{dex}^{-1}$ (from FLARES simulation; \citealt{Lovell_2021,Vijayan_2021}), but are low in \ha\ equivalent widths (EWs) and luminosities and thus have negligible contribution to the \ha\ LF.
The short-timescale dips in SFH (e.g., as observed through \ha) are known to be preferentially averaged out through SFR indicators like UV luminosities tracing longer timescale \citep[e.g.,][]{caplar_2019,Tacchella_2020a}.
The substantial presence of ``mini-quenched'' galaxies (or in other words, a large scattering in \ha\ EWs in particularly for dwarf galaxies; e.g., \citealt{Endsley_2024}) can be responsible for a flattened faint end of \ha\ LF.
This scenario can be tested through the measurements of spectroscopically complete UV LFs using the same data, which will be a subject of future work from the collaboration.

Finally, as mentioned in Section~\ref{subsec:sfrf}, we default to set the lower integration limit of CSFRD to $0.24$\,\smpy\ (corresponding to $M_{\mathrm{UV}}=-17$ or $L_\mathrm{H\alpha}=10^{40.65}\ \mathrm{erg}\ \mathrm{s}^{-1}$) for fair comparison with previous UV-based CSFRD measurements. 
\citet{Bouwens_2022c} derived an upper limit of the turn-over luminosity of the UV LFs over $z=2-9$ as $M_{\mathrm{UV}} > -15.5$.
If we integrate the SFRF down to such a lower limit (corresponding to $\mathrm{SFR}=0.06$\,\smpy) the resulting CSFRD increases to $0.063^{+0.010}_{-0.007}$ at $z \sim 4.5$ and $0.028^{+0.010}_{-0.008}$ at $z \sim 6.3$, which are both $\sim1.1$ times higher than our initial estimates.

\section{Summary} \label{sec:summary}
In this work, we search for high-redshift \ha\ emitters in four Hubble Frontier Fields using the NIRCam grism slitless  spectra obtained by the JWST MAGNIF program. 
Our results and conclusions are summarized as follows:

\begin{enumerate} 
\item We identify 222 \ha\ emitters at $z=4.18-4.95$ and 26 \ha\ emitters at $z=6.00-6.62$. Using \textsc{Lenstool} we identify 14 multiply imaged systems (33 images in total) of \ha\ emitters.
We find that the number of galaxies in the four fields differs significantly, highlighting the importance of observing multiple independent sightlines to mitigate the strong cosmic variance with pencil-beam surveys.

\item We perform SED modeling and validate \ha\ as a reliable tracer of SFR within the last 10\,Myr. Additionally, we derive dust attenuation of the \ha\ lines, which show a clear positive correlation with \ha\ luminosity. We conclude that more luminous \ha\ emitters generally have heavier dust attenuation, likely due to the greater amount of gas and dust in their environments.

\item We measure \ha\ LFs at $z\sim4.5$ and 6.3, both with and without dust corrections, down to $L_{\mathrm{H}\alpha} \sim 10^{40.3}\,\mathrm{erg}\,\mathrm{s}^{-1}$ at $z \sim 4.5$ and $10^{41.5}\,\mathrm{erg}\,\mathrm{s}^{-1}$ at $z \sim 6.3$.
These luminosities correspond to SFRs of $\sim$\,0.1 and 1.7\,\smpy, respectively. 
Through a remarkable 3-dex span in SFR, our \ha-based SFRFs bridge the gap between UV-based SFRFs (that may miss dust-obscured massive galaxy population) and IR-based SFRFs (that may miss dust-poor dwarf galaxies) around $\mathrm{SFR}\simeq 30 - 100$\,\smpy.

\item  We measure the faint-end slopes of observed \ha\ LFs $\alpha=-1.83^{+0.13}_{-0.13}$ at $z\sim4.5$ and $\alpha=-1.85^{+0.33}_{-0.19}$ at $z\sim 6.3$, and after dust-correction the slopes become $\alpha=-1.76^{+0.11}_{-0.10}$ at $z\sim4.5$ and $\alpha=-1.79^{+0.20}_{-0.19}$ at $z\sim 6.3$. 
The $\alpha_\mathrm{H\alpha}$ shows no or weak redshift evolution across the low-redshift Universe to the Epoch of Reionization.

\item The faint-end slope of \ha\ LF is consistent than that of UV LF at $z\sim4.5$, which is likely a combined results of lower dust attenuation with \ha\ and higher $L_\mathrm{H\alpha} / L_\mathrm{UV}$ among low-mass \ha\ emitters (because of high ionizing photon production efficiency and prevalence of upturns in their recent bursty star-formation history).
The faint-end slope of \ha\ LF is somewhat flatter than that of UV LF at $z\sim6.3$, which may be caused by the substantial existence of ``mini-quenched'' galaxies at this redshift.

\item We calculate the cosmic star formation rate densities, finding $\rho_\mathrm{SFR} = 0.058^{+0.008}_{-0.006}\ M_\odot\, \mathrm{yr}^{-1}\, \mathrm{Mpc}^{-3}$ at $z \sim 4.5$ and $\rho_\mathrm{SFR} = 0.025^{+0.009}_{-0.007}\ M_\odot\, \mathrm{yr}^{-1}\, \mathrm{Mpc}^{-3}$ at $z \sim 6.3$. These values are significantly higher than previous estimates based on dust-corrected UV LFs but are in good agreement with recent measurements obtained from infrared surveys.

\item We conduct a series of tests and confirm that our results remain robust against potential uncertainties arising from lens models, cosmic variance, and the contribution of both obscured and unobscured AGNs.

\end{enumerate}

Our work demonstrates the efficiency and effectiveness of JWST NIRCam/grism, combined with gravitational lensing, in constructing highly complete emission-line galaxy samples. This spectroscopically confirmed \ha\ emitter sample, along with extensive photometric data in the cluster fields, enables in-depth studies of high-redshift galaxy properties, including star formation rate tracers, ionizing photon production efficiency, the star-forming main sequence, and star formation burstiness. Future JWST slitless spectroscopic programs, such as COSMOS-3D (PID 5893, \citealt{Kakiichi_2024jwstprop}), SAPPHIRES (PID 6434, \citealt{Egami_2024jwstprop}), and NEXUS (PID 5105, \citealt{Shen_2024jwstprop}), will further probe both the bright and faint end of the \ha\ luminosity function in blank fields, enhancing our understanding of galaxy formation and evolution at high redshifts.

\section*{Acknowledgment}

We acknowledge support from the National Key R\&D Program of China (2022YFF0503401) and the National Science Foundation of China (12225301). FS, EE, DJE and YZ acknowledge funding from JWST/NIRCam contract to the University of Arizona, NAS5-02105. FEB acknowledges support from ANID-Chile BASAL CATA FB210003, FONDECYT Regular 1241005, and Millennium Science Initiative, AIM23-0001.
MO acknowledges support from JSPS KAKENHI Grant Numbers JP23K22531, JP22K21349, JP20H05856.
RAW acknowledges support from NASA JWST Interdisciplinary Scientist grants
NAG5-12460, NNX14AN10G and 80NSSC18K0200 from GSFC.
FW acknowledges support from NSF award AST-2513040.
AZ acknowledges support by Grant No. 2020750 from the United States-Israel Binational Science Foundation (BSF) and Grant No. 2109066 from the United States National Science Foundation (NSF); and by the Israel Science Foundation Grant No. 864/23.
C.-C.C. acknowledges support from the National Science and Technology Council of Taiwan (111-2112-M-001-045-MY3), as well as Academia Sinica through the Career Development Award (AS-CDA-112-M02).
MJ and NP are supported by the United Kingdom Research and Innovation (UKRI) Future Leaders Fellowship `Using Cosmic Beasts to uncover the Nature of Dark Matter' (grant number MR/X006069/1).
KK acknowledges support from JSPS KAKENHI Grant Numbers JP22H04939, JP23K20035, and JP24H00004. 
We thank Weston Eck, our JWST-GO-2883 program coordinator, for the valuable support to our observations.
FS acknowledges helpful discussions with James Trussler.

This work is based on observations made with the NASA/ESA/CSA James Webb Space Telescope. The data were obtained from the Mikulski Archive for Space Telescopes at the Space Telescope Science Institute, which is operated by the Association of Universities for Research in Astronomy, Inc., under NASA contract NAS 5-03127 for JWST. These observations are associated with program \#1176, 1199, 1208, 1324, 2514, 2561, 2756, 2883, 3362, 3516, 3538, 4111.
The authors acknowledge the 1324 (GLASS), 2514 (PANORAMIC), 2561 (UNCOVER), 3362 (Technicolor), 3516 (ALT), 3538, 4111 (MegaScience) for developing their observing program with a zero-exclusive-access period.
Support for program \#2883 was provided by NASA through a grant from the Space Telescope Science Institute, which is operated by the Association of Universities for Research in Astronomy, Inc., under NASA contract NAS 5-03127.

\vspace{5mm}
\facilities{JWST (NIRCam and NIRISS), HST (ACS)}

\software{astropy \citep{astropy_2013},  
          photutils \citep{photutils},
          cigale \citep{Boquien_2019},
          emcee \citep{Foreman-Mackey_2013},
          JWST \citep{jwst_bushouse_2023},
          Lenstool \citep{Jullo_2007},
          Webbpsf \citep{webbpsf}
          }

\bibliography{sample631}{}

\begin{thebibliography}{}
\expandafter\ifx\csname natexlab\endcsname\relax\def\natexlab#1{#1}\fi
\providecommand{\url}[1]{\href{#1}{#1}}
\providecommand{\dodoi}[1]{doi:~\href{http://doi.org/#1}{\nolinkurl{#1}}}
\providecommand{\doeprint}[1]{\href{http://ascl.net/#1}{\nolinkurl{http://ascl.net/#1}}}
\providecommand{\doarXiv}[1]{\href{https://arxiv.org/abs/#1}{\nolinkurl{https://arxiv.org/abs/#1}}}

\bibitem[{{Adams} {et~al.}(2024){Adams}, {Conselice}, {Austin}, {Harvey},
  {Ferreira}, {Trussler}, {Juod{\v{z}}balis}, {Li}, {Windhorst}, {Cohen},
  {Jansen}, {Summers}, {Tompkins}, {Driver}, {Robotham}, {D'Silva}, {Yan},
  {Coe}, {Frye}, {Grogin}, {Koekemoer}, {Marshall}, {Pirzkal}, {Ryan},
  {Maksym}, {Rutkowski}, {Willmer}, {Hammel}, {Nonino}, {Bhatawdekar},
  {Wilkins}, {Bradley}, {Broadhurst}, {Cheng}, {Dole}, {Hathi}, \&
  {Zitrin}}]{Adams_2024}
{Adams}, N.~J., {Conselice}, C.~J., {Austin}, D., {et~al.} 2024, \apj, 965,
  169, \dodoi{10.3847/1538-4357/ad2a7b}

\bibitem[{Atek {et~al.}(2022)Atek, Furtak, Oesch, van Dokkum, Reddy, Contini,
  Illingworth, \& Wilkins}]{Atek_star_2022}
Atek, H., Furtak, L.~J., Oesch, P., {et~al.} 2022, Monthly Notices of the Royal
  Astronomical Society, 511, 4464, \dodoi{10.1093/mnras/stac360}

\bibitem[{{Atek} {et~al.}(2015){Atek}, {Richard}, {Kneib}, {Jauzac},
  {Schaerer}, {Clement}, {Limousin}, {Jullo}, {Natarajan}, {Egami}, \&
  {Ebeling}}]{Atek_2015}
{Atek}, H., {Richard}, J., {Kneib}, J.-P., {et~al.} 2015, \apj, 800, 18,
  \dodoi{10.1088/0004-637X/800/1/18}

\bibitem[{Backhaus {et~al.}(2024)Backhaus, Trump, Pirzkal, Barro, Finkelstein,
  Haro, Simons, Wessner, Cleri, Bagley, Hirschmann, Nicholls, Dickinson,
  Kartaltepe, Papovich, Kocevski, Koekemoer, Bisigello, Jaskot, Lucas, Jung,
  Wilkins, Yung, Ferguson, Fontana, Grazian, Grogin, Kewley, Kirkpatrick, Lotz,
  Pentericci, P{\'e}rez-Gonz{\'a}lez, Ravindranath, Somerville, Yang, Holwerda,
  Kurczynski, Hathi, Rose, \& Davis}]{Backhaus_2024}
Backhaus, B.~E., Trump, J.~R., Pirzkal, N., {et~al.} 2024, The Astrophysical
  Journal, 962, 195, \dodoi{10.3847/1538-4357/ad1520}

\bibitem[{{Baldwin} {et~al.}(1981){Baldwin}, {Phillips}, \&
  {Terlevich}}]{Baldwin_1981}
{Baldwin}, J.~A., {Phillips}, M.~M., \& {Terlevich}, R. 1981, \pasp, 93, 5,
  \dodoi{10.1086/130766}

\bibitem[{{Bergamini} {et~al.}(2023{\natexlab{a}}){Bergamini}, {Acebron},
  {Grillo}, {Rosati}, {Caminha}, {Mercurio}, {Vanzella}, {Mason}, {Treu},
  {Angora}, {Brammer}, {Meneghetti}, {Nonino}, {Boyett}, {Brada{\v{c}}},
  {Castellano}, {Fontana}, {Morishita}, {Paris}, {Prieto-Lyon},
  {Roberts-Borsani}, {Roy}, {Santini}, {Vulcani}, {Wang}, \&
  {Yang}}]{Bergamini_2023}
{Bergamini}, P., {Acebron}, A., {Grillo}, C., {et~al.} 2023{\natexlab{a}},
  \apj, 952, 84, \dodoi{10.3847/1538-4357/acd643}

\bibitem[{{Bergamini} {et~al.}(2023{\natexlab{b}}){Bergamini}, {Acebron},
  {Grillo}, {Rosati}, {Caminha}, {Mercurio}, {Vanzella}, {Angora}, {Brammer},
  {Meneghetti}, \& {Nonino}}]{Bergamini_2023A}
---. 2023{\natexlab{b}}, \aap, 670, A60, \dodoi{10.1051/0004-6361/202244575}

\bibitem[{{Bertin} \& {Arnouts}(1996)}]{Bertin_1996}
{Bertin}, E., \& {Arnouts}, S. 1996, \aaps, 117, 393,
  \dodoi{10.1051/aas:1996164}

\bibitem[{Bezanson {et~al.}(2024)Bezanson, Labbe, Whitaker, Leja, Price, Franx,
  Brammer, Marchesini, Zitrin, Wang, Weaver, Furtak, Atek, Coe, Cutler, Dayal,
  van Dokkum, Feldmann, Schreiber, Fujimoto, Geha, Glazebrook, de~Graaff,
  Greene, Juneau, Kassin, Kriek, Khullar, Maseda, Mowla, Muzzin, Nanayakkara,
  Nelson, Oesch, Pacifici, Pan, Papovich, Setton, Shapley, Smit, Stefanon,
  Taylor, \& Williams}]{Bezanson_2024}
Bezanson, R., Labbe, I., Whitaker, K.~E., {et~al.} 2024, The JWST UNCOVER
  Treasury survey: Ultradeep NIRSpec and NIRCam ObserVations before the Epoch
  of Reionization.
\newblock \doarXiv{2212.04026}

\bibitem[{{Bollo} {et~al.}(2023){Bollo}, {Gonz{\'a}lez}, {Stefanon}, {Oesch},
  {Bouwens}, {Smit}, {Illingworth}, \& {Labb{\'e}}}]{Bollo_2023}
{Bollo}, V., {Gonz{\'a}lez}, V., {Stefanon}, M., {et~al.} 2023, \apj, 946, 117,
  \dodoi{10.3847/1538-4357/acbc79}

\bibitem[{{Boquien} {et~al.}(2019){Boquien}, {Burgarella}, {Roehlly}, {Buat},
  {Ciesla}, {Corre}, {Inoue}, \& {Salas}}]{Boquien_2019}
{Boquien}, M., {Burgarella}, D., {Roehlly}, Y., {et~al.} 2019, \aap, 622, A103,
  \dodoi{10.1051/0004-6361/201834156}

\bibitem[{{Bouwens} {et~al.}(2020){Bouwens}, {Gonz{\'a}lez-L{\'o}pez},
  {Aravena}, {Decarli}, {Novak}, {Stefanon}, {Walter}, {Boogaard}, {Carilli},
  {Dudzevi{\v{c}}i{\={u}}t{\.{e}}}, {Smail}, {Daddi}, {da Cunha}, {Ivison},
  {Nanayakkara}, {Cortes}, {Cox}, {Inami}, {Oesch}, {Popping}, {Riechers}, {van
  der Werf}, {Weiss}, {Fudamoto}, \& {Wagg}}]{Bouwens_2020}
{Bouwens}, R., {Gonz{\'a}lez-L{\'o}pez}, J., {Aravena}, M., {et~al.} 2020,
  \apj, 902, 112, \dodoi{10.3847/1538-4357/abb830}

\bibitem[{{Bouwens} {et~al.}(2022){Bouwens}, {Illingworth}, {Ellis}, {Oesch},
  \& {Stefanon}}]{Bouwens_2022c}
{Bouwens}, R.~J., {Illingworth}, G., {Ellis}, R.~S., {Oesch}, P., \&
  {Stefanon}, M. 2022, \apj, 940, 55, \dodoi{10.3847/1538-4357/ac86d1}

\bibitem[{{Bouwens} {et~al.}(2015){Bouwens}, {Illingworth}, {Oesch}, {Trenti},
  {Labb{\'e}}, {Bradley}, {Carollo}, {van Dokkum}, {Gonzalez}, {Holwerda},
  {Franx}, {Spitler}, {Smit}, \& {Magee}}]{Bouwens_2015}
{Bouwens}, R.~J., {Illingworth}, G.~D., {Oesch}, P.~A., {et~al.} 2015, \apj,
  803, 34, \dodoi{10.1088/0004-637X/803/1/34}

\bibitem[{{Bouwens} {et~al.}(2021){Bouwens}, {Oesch}, {Stefanon},
  {Illingworth}, {Labb{\'e}}, {Reddy}, {Atek}, {Montes}, {Naidu},
  {Nanayakkara}, {Nelson}, \& {Wilkins}}]{Bouwens_2021}
{Bouwens}, R.~J., {Oesch}, P.~A., {Stefanon}, M., {et~al.} 2021, \aj, 162, 47,
  \dodoi{10.3847/1538-3881/abf83e}

\bibitem[{Bradley {et~al.}(2024)Bradley, Sip{\H o}cz, Robitaille, Tollerud,
  Vin{\'{\i}}cius, Deil, Barbary, Wilson, Busko, Donath, G{\"u}nther, Cara,
  Lim, Me{\ss}linger, Burnett, Conseil, Droettboom, Bostroem, Bray, Bratholm,
  Jamieson, Ginsburg, Barentsen, Craig, Pascual, Rathi, Perrin, Morris, \&
  Perren}]{photutils}
Bradley, L., Sip{\H o}cz, B., Robitaille, T., {et~al.} 2024, astropy/photutils:
  1.12.0, 1.12.0,  Zenodo, \dodoi{10.5281/zenodo.10967176}

\bibitem[{{Brammer} {et~al.}(2008){Brammer}, {van Dokkum}, \&
  {Coppi}}]{Brammer_2008}
{Brammer}, G.~B., {van Dokkum}, P.~G., \& {Coppi}, P. 2008, \apj, 686, 1503,
  \dodoi{10.1086/591786}

\bibitem[{{Brinchmann} {et~al.}(2004){Brinchmann}, {Charlot}, {White},
  {Tremonti}, {Kauffmann}, {Heckman}, \& {Brinkmann}}]{Brinchmann_2004}
{Brinchmann}, J., {Charlot}, S., {White}, S.~D.~M., {et~al.} 2004, \mnras, 351,
  1151, \dodoi{10.1111/j.1365-2966.2004.07881.x}

\bibitem[{{Bruzual} \& {Charlot}(2003)}]{Bruzual_2003}
{Bruzual}, G., \& {Charlot}, S. 2003, \mnras, 344, 1000,
  \dodoi{10.1046/j.1365-8711.2003.06897.x}

\bibitem[{{Buat} {et~al.}(2018){Buat}, {Boquien}, {Ma{\l}ek}, {Corre}, {Salas},
  {Roehlly}, {Shirley}, \& {Efstathiou}}]{Buat_2018}
{Buat}, V., {Boquien}, M., {Ma{\l}ek}, K., {et~al.} 2018, \aap, 619, A135,
  \dodoi{10.1051/0004-6361/201833841}

\bibitem[{Bushouse {et~al.}(2023)Bushouse, Eisenhamer, Dencheva, Davies,
  Greenfield, Morrison, Hodge, Simon, Grumm, Droettboom, Slavich, Sosey, Pauly,
  Miller, Jedrzejewski, Hack, Davis, Crawford, Law, Gordon, Regan, Cara,
  MacDonald, Bradley, Shanahan, Jamieson, Teodoro, \&
  Williams}]{jwst_bushouse_2023}
Bushouse, H., Eisenhamer, J., Dencheva, N., {et~al.} 2023, JWST Calibration
  Pipeline, 1.10.2,  Zenodo, \dodoi{10.5281/zenodo.7829329}

\bibitem[{{Calzetti} {et~al.}(2000){Calzetti}, {Armus}, {Bohlin}, {Kinney},
  {Koornneef}, \& {Storchi-Bergmann}}]{Calzetti_2000}
{Calzetti}, D., {Armus}, L., {Bohlin}, R.~C., {et~al.} 2000, \apj, 533, 682,
  \dodoi{10.1086/308692}

\bibitem[{{Caplar} \& {Tacchella}(2019)}]{caplar_2019}
{Caplar}, N., \& {Tacchella}, S. 2019, \mnras, 487, 3845,
  \dodoi{10.1093/mnras/stz1449}

\bibitem[{{Chabrier}(2003)}]{Chabrier_2003}
{Chabrier}, G. 2003, \pasp, 115, 763, \dodoi{10.1086/376392}

\bibitem[{{Chen} {et~al.}(2022){Chen}, {Kelly}, {Treu}, {Williams},
  {Broadhurst}, {Castellano}, {Diego}, {Filippenko}, {Glazebrook}, {Koekemoer},
  {Morishita}, {Nanayakkara}, {Nonino}, {Pierel}, {Rieck}, {Strausbaugh},
  {Trenti}, {Wang}, {Wang}, {Windhorst}, \&
  {Zitrin}}]{ChenW_2022jwst.prop.2756C}
{Chen}, W., {Kelly}, P., {Treu}, T.~L., {et~al.} 2022, {Imaging and
  Spectroscopic Follow-up of a Supernova at Redshift z=3.47}, JWST Proposal.
  Cycle 1, ID. \#2756

\bibitem[{{Cochrane} {et~al.}(2023){Cochrane}, {Kondapally}, {Best}, {Sabater},
  {Duncan}, {Smith}, {Hardcastle}, {R{\"o}ttgering}, {Prandoni}, {Haskell},
  {G{\"u}rkan}, \& {Miley}}]{Cochrane_2023}
{Cochrane}, R.~K., {Kondapally}, R., {Best}, P.~N., {et~al.} 2023, \mnras, 523,
  6082, \dodoi{10.1093/mnras/stad1602}

\bibitem[{{Covelo-Paz} {et~al.}(2025){Covelo-Paz}, {Giovinazzo}, {Oesch},
  {Meyer}, {Weibel}, {Brammer}, {Fudamoto}, {Kerutt}, {Lin}, {Matharu},
  {Naidu}, {Velichko}, {Bollo}, {Bouwens}, {Chisholm}, {Illingworth},
  {Kramarenko}, {Magee}, {Maseda}, {Matthee}, {Nelson}, {Reddy}, {Schaerer},
  {Stefanon}, \& {Xiao}}]{Covelo-Paz_2024}
{Covelo-Paz}, A., {Giovinazzo}, E., {Oesch}, P.~A., {et~al.} 2025, \aap, 694,
  A178, \dodoi{10.1051/0004-6361/202452363}

\bibitem[{{Cucciati} {et~al.}(2012){Cucciati}, {Tresse}, {Ilbert}, {Le
  F{\`e}vre}, {Garilli}, {Le Brun}, {Cassata}, {Franzetti}, {Maccagni},
  {Scodeggio}, {Zucca}, {Zamorani}, {Bardelli}, {Bolzonella}, {Bielby},
  {McCracken}, {Zanichelli}, \& {Vergani}}]{Cucciati_2012}
{Cucciati}, O., {Tresse}, L., {Ilbert}, O., {et~al.} 2012, \aap, 539, A31,
  \dodoi{10.1051/0004-6361/201118010}

\bibitem[{{Dalmasso} {et~al.}(2024){Dalmasso}, {Leethochawalit}, {Trenti}, \&
  {Boyett}}]{Dalmasso_2024}
{Dalmasso}, N., {Leethochawalit}, N., {Trenti}, M., \& {Boyett}, K. 2024,
  \mnras, 533, 2391, \dodoi{10.1093/mnras/stae2006}

\bibitem[{{de Mellos} {et~al.}(2024){de Mellos}, {Riffel}, {Schimoia},
  {Rembold}, {Riffel}, {Storchi-Bergmann}, {Wylezalek}, {Ilha}, {Alb{\'a}n},
  {Dors}, {Gatto}, {Krabbe}, {Mallmann}, \& {Trevisan}}]{deMellos_2024}
{de Mellos}, M. S.~Z., {Riffel}, R.~A., {Schimoia}, J.~S., {et~al.} 2024,
  \mnras, 535, 123, \dodoi{10.1093/mnras/stae2352}

\bibitem[{Dey {et~al.}(2019)Dey, Schlegel, Lang, Blum, Burleigh, Fan, Findlay,
  Finkbeiner, Herrera, Juneau, Landriau, Levi, McGreer, Meisner, Myers,
  Moustakas, Nugent, Patej, Schlafly, Walker, Valdes, Weaver, Y{\`e}che, Zou,
  Zhou, Abareshi, Abbott, Abolfathi, Aguilera, Alam, Allen, Alvarez, Annis,
  Ansarinejad, Aubert, Beechert, Bell, BenZvi, Beutler, Bielby, Bolton,
  Brice{\~n}o, Buckley-Geer, Butler, Calamida, Carlberg, Carter, Casas,
  Castander, Choi, Comparat, Cukanovaite, Delubac, DeVries, Dey, Dhungana,
  Dickinson, Ding, Donaldson, Duan, Duckworth, Eftekharzadeh, Eisenstein,
  Etourneau, Fagrelius, Farihi, Fitzpatrick, Font-Ribera, Fulmer, G{\"a}nsicke,
  Gaztanaga, George, Gerdes, Gontcho, Gorgoni, Green, Guy, Harmer, Hernandez,
  Honscheid, Huang, James, Jannuzi, Jiang, Joyce, Karcher, Karkar, Kehoe,
  Jean-Paul, Kueter-Young, Lan, Lauer, Guillou, Suu, Lee, Lesser, Levasseur,
  Li, Mann, Marshall, Mart{\'\i}nez-V{\'a}zquez, Martini, du~Mas~des Bourboux,
  McManus, Meier, M{\'e}nard, Metcalfe, Mu{\~n}oz-Guti{\'e}rrez, Najita,
  Napier, Narayan, Newman, Nie, Nord, Norman, Olsen, Paat,
  Palanque-Delabrouille, Peng, Poppett, Poremba, Prakash, Rabinowitz, Raichoor,
  Rezaie, Robertson, Roe, Ross, Ross, Rudnick, Gaines, Saha, S{\'a}nchez,
  Savary, Schweiker, Scott, Seo, Shan, Silva, Slepian, Soto, Sprayberry,
  Staten, Stillman, Stupak, Summers, Tie, Tirado, Vargas-Maga{\~n}a, Vivas,
  Wechsler, Williams, Yang, Yang, Yapici, Zaritsky, Zenteno, Zhang, Zhang,
  Zhou, \& Zhou}]{Dey_2019}
Dey, A., Schlegel, D.~J., Lang, D., {et~al.} 2019, The Astronomical Journal,
  157, 168, \dodoi{10.3847/1538-3881/ab089d}

\bibitem[{{Diego} {et~al.}(2024){Diego}, {Adams}, {Willner}, {Harvey},
  {Broadhurst}, {Cohen}, {Jansen}, {Summers}, {Windhorst}, {D'Silva},
  {Koekemoer}, {Coe}, {Conselice}, {Driver}, {Frye}, {Grogin}, {Marshall},
  {Nonino}, {Ortiz}, {Pirzkal}, {Robotham}, {Ryan}, {Willmer}, {Yan}, {Sun},
  {Hainline}, {Berkheimer}, {Polletta}, \& {Zitrin}}]{Diego_2024}
{Diego}, J.~M., {Adams}, N.~J., {Willner}, S.~P., {et~al.} 2024, \aap, 690,
  A114, \dodoi{10.1051/0004-6361/202349119}

\bibitem[{{Dome} {et~al.}(2024){Dome}, {Tacchella}, {Fialkov}, {Ceverino},
  {Dekel}, {Ginzburg}, {Lapiner}, \& {Looser}}]{Dome_2024}
{Dome}, T., {Tacchella}, S., {Fialkov}, A., {et~al.} 2024, \mnras, 527, 2139,
  \dodoi{10.1093/mnras/stad3239}

\bibitem[{{Driver} \& {Robotham}(2010)}]{Driver_2010}
{Driver}, S.~P., \& {Robotham}, A. S.~G. 2010, \mnras, 407, 2131,
  \dodoi{10.1111/j.1365-2966.2010.17028.x}

\bibitem[{{Eddington}(1913)}]{eddington1913}
{Eddington}, A.~S. 1913, \mnras, 73, 359, \dodoi{10.1093/mnras/73.5.359}

\bibitem[{{Egami} {et~al.}(2023){Egami}, {Sun}, {Alberts}, {Baum}, {Boyett},
  {Bunker}, {Cameron}, {Carniani}, {Charlot}, {Chen}, {Chevallard}, {Curti},
  {D'Eugenio}, {Danhaive}, {DeCoursey}, {Dudzeviciute}, {Eisenstein},
  {Hainline}, {Helton}, {Ji}, {Johnson}, {Kumari}, {Looser}, {Lyu}, {Ma},
  {Maiolino}, {Maseda}, {Nelson}, {Rawle}, {Rieke}, {Robertson}, {Sandles},
  {Shivaei}, {Smit}, {Suess}, {Tacchella}, {Uebler}, {Whitler}, {Williams},
  {Willmer}, {Willott}, {Witstok}, \& {de Graaff}}]{Egami_2023}
{Egami}, E., {Sun}, F., {Alberts}, S., {et~al.} 2023, {Complete NIRCam Grism
  Redshift Survey (CONGRESS)}, JWST Proposal. Cycle 2, ID. \#3577

\bibitem[{{Egami} {et~al.}(2024){Egami}, {Fan}, {Sun}, {Wang}, {Yang},
  {Bunker}, {Cai}, {Charlot}, {DeCoursey}, {Eisenstein}, {Hainline},
  {Harikane}, {Ji}, {Kakiichi}, {Ma}, {Maiolino}, {Rieke}, \&
  {Willmer}}]{Egami_2024jwstprop}
{Egami}, E., {Fan}, X., {Sun}, F., {et~al.} 2024, {SAPPHIRES: Slitless Areal
  Pure-Parallel High-Redshift Emission Survey}, JWST Proposal. Cycle 3, ID.
  \#6434

\bibitem[{Emami {et~al.}(2019)Emami, Siana, Weisz, Johnson, Ma, \&
  El-Badry}]{Emami_closer_2019}
Emami, N., Siana, B., Weisz, D.~R., {et~al.} 2019, The Astrophysical Journal,
  881, 71, \dodoi{10.3847/1538-4357/ab211a}

\bibitem[{{Endsley} {et~al.}(2024){Endsley}, {Stark}, {Whitler}, {Topping},
  {Johnson}, {Robertson}, {Tacchella}, {Alberts}, {Baker}, {Bhatawdekar},
  {Boyett}, {Bunker}, {Cameron}, {Carniani}, {Charlot}, {Chen}, {Chevallard},
  {Curtis-Lake}, {Danhaive}, {Egami}, {Eisenstein}, {Hainline}, {Helton}, {Ji},
  {Looser}, {Maiolino}, {Nelson}, {Pusk{\'a}s}, {Rieke}, {Rieke}, {Rix},
  {Sandles}, {Saxena}, {Simmonds}, {Smit}, {Sun}, {Williams}, {Willmer},
  {Willott}, \& {Witstok}}]{Endsley_2024}
{Endsley}, R., {Stark}, D.~P., {Whitler}, L., {et~al.} 2024, \mnras, 533, 1111,
  \dodoi{10.1093/mnras/stae1857}

\bibitem[{Enia {et~al.}(2022)Enia, Talia, Pozzi, Cimatti, Delvecchio, Zamorani,
  D'Amato, Bisigello, Gruppioni, Rodighiero, Calura, Dallacasa, Giulietti,
  Barchiesi, Behiri, \& Romano}]{Enia_2022}
Enia, A., Talia, M., Pozzi, F., {et~al.} 2022, The Astrophysical Journal, 927,
  204, \dodoi{10.3847/1538-4357/ac51ca}

\bibitem[{{Faisst} {et~al.}(2019){Faisst}, {Capak}, {Emami}, {Tacchella}, \&
  {Larson}}]{Faisst_2019}
{Faisst}, A.~L., {Capak}, P.~L., {Emami}, N., {Tacchella}, S., \& {Larson},
  K.~L. 2019, \apj, 884, 133, \dodoi{10.3847/1538-4357/ab425b}

\bibitem[{{Fazio} {et~al.}(2004){Fazio}, {Hora}, {Allen}, {Ashby}, {Barmby},
  {Deutsch}, {Huang}, {Kleiner}, {Marengo}, {Megeath}, {Melnick}, {Pahre},
  {Patten}, {Polizotti}, {Smith}, {Taylor}, {Wang}, {Willner}, {Hoffmann},
  {Pipher}, {Forrest}, {McMurty}, {McCreight}, {McKelvey}, {McMurray}, {Koch},
  {Moseley}, {Arendt}, {Mentzell}, {Marx}, {Losch}, {Mayman}, {Eichhorn},
  {Krebs}, {Jhabvala}, {Gezari}, {Fixsen}, {Flores}, {Shakoorzadeh}, {Jungo},
  {Hakun}, {Workman}, {Karpati}, {Kichak}, {Whitley}, {Mann}, {Tollestrup},
  {Eisenhardt}, {Stern}, {Gorjian}, {Bhattacharya}, {Carey}, {Nelson},
  {Glaccum}, {Lacy}, {Lowrance}, {Laine}, {Reach}, {Stauffer}, {Surace},
  {Wilson}, {Wright}, {Hoffman}, {Domingo}, \& {Cohen}}]{Fazio_2004}
{Fazio}, G.~G., {Hora}, J.~L., {Allen}, L.~E., {et~al.} 2004, \apjs, 154, 10,
  \dodoi{10.1086/422843}

\bibitem[{{Finkelstein} {et~al.}(2012){Finkelstein}, {Papovich}, {Ryan},
  {Pawlik}, {Dickinson}, {Ferguson}, {Finlator}, {Koekemoer}, {Giavalisco},
  {Cooray}, {Dunlop}, {Faber}, {Grogin}, {Kocevski}, \&
  {Newman}}]{Finkelstein_2012}
{Finkelstein}, S.~L., {Papovich}, C., {Ryan}, R.~E., {et~al.} 2012, \apj, 758,
  93, \dodoi{10.1088/0004-637X/758/2/93}

\bibitem[{{Finkelstein} {et~al.}(2015){Finkelstein}, {Ryan}, {Papovich},
  {Dickinson}, {Song}, {Somerville}, {Ferguson}, {Salmon}, {Giavalisco},
  {Koekemoer}, {Ashby}, {Behroozi}, {Castellano}, {Dunlop}, {Faber}, {Fazio},
  {Fontana}, {Grogin}, {Hathi}, {Jaacks}, {Kocevski}, {Livermore}, {McLure},
  {Merlin}, {Mobasher}, {Newman}, {Rafelski}, {Tilvi}, \&
  {Willner}}]{Finkelstein_2015}
{Finkelstein}, S.~L., {Ryan}, Russell~E., J., {Papovich}, C., {et~al.} 2015,
  \apj, 810, 71, \dodoi{10.1088/0004-637X/810/1/71}

\bibitem[{Foreman-Mackey {et~al.}(2013)Foreman-Mackey, Hogg, Lang, \&
  Goodman}]{Foreman-Mackey_2013}
Foreman-Mackey, D., Hogg, D.~W., Lang, D., \& Goodman, J. 2013, Publications of
  the Astronomical Society of the Pacific, 125, 306, \dodoi{10.1086/670067}

\bibitem[{{Fujimoto} {et~al.}(2023){Fujimoto}, {Bezanson}, {Labbe}, {Brammer},
  {Price}, {Wang}, {Weaver}, {Fudamoto}, {Oesch}, {Williams}, {Dayal},
  {Feldmann}, {Greene}, {Leja}, {Whitaker}, {Zitrin}, {Cutler}, {Furtak},
  {Pan}, {Chemerynska}, {Kokorev}, {Miller}, {Atek}, {van Dokkum}, {Juneau},
  {Kassin}, {Khullar}, {Marchesini}, {Maseda}, {Nelson}, {Setton}, \&
  {Smit}}]{Fujimoto_2023}
{Fujimoto}, S., {Bezanson}, R., {Labbe}, I., {et~al.} 2023, arXiv e-prints,
  arXiv:2309.07834, \dodoi{10.48550/arXiv.2309.07834}

\bibitem[{Fujimoto {et~al.}(2024)Fujimoto, Kohno, Ouchi, Oguri, Kokorev,
  Brammer, Sun, Gonzalez-Lopez, Bauer, Caminha, Hatsukade, Richard, Smail,
  Tsujita, Ueda, Uematsu, Zitrin, Coe, Kneib, Postman, Umetsu, Lagos, Popping,
  Ao, Bradley, Caputi, Dessauges-Zavadsky, Egami, Espada, Ivison, Jauzac,
  Knudsen, Koekemoer, Magdis, Mahler, Arancibia, Rawle, Shimasaku, Toft,
  Umehata, Valentino, Wang, \& Wang}]{fujimoto_alma_2024}
Fujimoto, S., Kohno, K., Ouchi, M., {et~al.} 2024, {ALMA} {Lensing} {Cluster}
  {Survey}: {Deep} 1.2 mm {Number} {Counts} and {Infrared} {Luminosity}
  {Functions} at \$z{\textbackslash}simeq1-8\$,  arXiv,
  \dodoi{10.48550/arXiv.2303.01658}

\bibitem[{{Furtak} {et~al.}(2023){Furtak}, {Zitrin}, {Weaver}, {Atek},
  {Bezanson}, {Labb{\'e}}, {Whitaker}, {Leja}, {Price}, {Brammer}, {Wang},
  {Marchesini}, {Pan}, {Dayal}, {van Dokkum}, {Feldmann}, {Fujimoto}, {Franx},
  {Khullar}, {Nelson}, \& {Mowla}}]{Furtak_2023}
{Furtak}, L.~J., {Zitrin}, A., {Weaver}, J.~R., {et~al.} 2023, \mnras, 523,
  4568, \dodoi{10.1093/mnras/stad1627}

\bibitem[{{Greene} \& {Ho}(2005)}]{Greene_2005}
{Greene}, J.~E., \& {Ho}, L.~C. 2005, \apj, 630, 122, \dodoi{10.1086/431897}

\bibitem[{{Greene} {et~al.}(2024){Greene}, {Labbe}, {Goulding}, {Furtak},
  {Chemerynska}, {Kokorev}, {Dayal}, {Volonteri}, {Williams}, {Wang}, {Setton},
  {Burgasser}, {Bezanson}, {Atek}, {Brammer}, {Cutler}, {Feldmann}, {Fujimoto},
  {Glazebrook}, {de Graaff}, {Khullar}, {Leja}, {Marchesini}, {Maseda},
  {Matthee}, {Miller}, {Naidu}, {Nanayakkara}, {Oesch}, {Pan}, {Papovich},
  {Price}, {van Dokkum}, {Weaver}, {Whitaker}, \& {Zitrin}}]{Greene_2024}
{Greene}, J.~E., {Labbe}, I., {Goulding}, A.~D., {et~al.} 2024, \apj, 964, 39,
  \dodoi{10.3847/1538-4357/ad1e5f}

\bibitem[{{Greene} {et~al.}(2017){Greene}, {Kelly}, {Stansberry}, {Leisenring},
  {Egami}, {Schlawin}, {Chu}, {Hodapp}, \& {Rieke}}]{Greene_2017}
{Greene}, T.~P., {Kelly}, D.~M., {Stansberry}, J., {et~al.} 2017, Journal of
  Astronomical Telescopes, Instruments, and Systems, 3, 035001,
  \dodoi{10.1117/1.JATIS.3.3.035001}

\bibitem[{Gruppioni {et~al.}(2020)Gruppioni, B{\'e}thermin, Loiacono,
  Le~F{\`e}vre, Capak, Cassata, Faisst, Schaerer, Silverman, Yan, Bardelli,
  Boquien, Carraro, Cimatti, Dessauges-Zavadsky, Ginolfi, Fujimoto, Hathi,
  Jones, Khusanova, Koekemoer, Lagache, Lemaux, Oesch, Pozzi, Riechers,
  Rodighiero, Romano, Talia, Vallini, Vergani, Zamorani, \&
  Zucca}]{gruppioni_alpine-alma_2020}
Gruppioni, C., B{\'e}thermin, M., Loiacono, F., {et~al.} 2020, Astronomy \&
  Astrophysics, 643, A8, \dodoi{10.1051/0004-6361/202038487}

\bibitem[{{Haarsma} {et~al.}(2000){Haarsma}, {Partridge}, {Windhorst}, \&
  {Richards}}]{Haarsma_2000}
{Haarsma}, D.~B., {Partridge}, R.~B., {Windhorst}, R.~A., \& {Richards}, E.~A.
  2000, \apj, 544, 641, \dodoi{10.1086/317225}

\bibitem[{{Hainline} {et~al.}(2023){Hainline}, {Robertson}, {Tacchella},
  {Rieke}, {Eisenstein}, {Helton}, {Whitler}, {Topping}, {Sun}, {Hviding}, \&
  {Jades Collaboration}}]{Hainline_2023}
{Hainline}, K., {Robertson}, B., {Tacchella}, S., {et~al.} 2023, in American
  Astronomical Society Meeting Abstracts, Vol. 242, American Astronomical
  Society Meeting Abstracts, 212.02

\bibitem[{{Hayashi} {et~al.}(2012){Hayashi}, {Kodama}, {Tadaki}, {Koyama}, \&
  {Tanaka}}]{Hayashi_2012}
{Hayashi}, M., {Kodama}, T., {Tadaki}, K.-i., {Koyama}, Y., \& {Tanaka}, I.
  2012, \apj, 757, 15, \dodoi{10.1088/0004-637X/757/1/15}

\bibitem[{{Hopkins} \& {Beacom}(2006)}]{Hopkins_2006}
{Hopkins}, A.~M., \& {Beacom}, J.~F. 2006, \apj, 651, 142,
  \dodoi{10.1086/506610}

\bibitem[{{Iani} {et~al.}(2023){Iani}, {Annunziatella}, {Bartosch Caminha},
  {Caputi}, {Costantin}, {Kerutt}, {Kokorev}, {Navarro}, {Perez-Gonzalez},
  {Rinaldi}, {Yang}, \& {van Mierlo}}]{Iani_2023}
{Iani}, E., {Annunziatella}, M., {Bartosch Caminha}, G., {et~al.} 2023,
  {Unveiling the properties of high-redshift low/intermediate-mass galaxies in
  Lensing fields with NIRCam Wide Field Slitless Spectroscopy}, JWST Proposal.
  Cycle 2, ID. \#3538

\bibitem[{{Ishigaki} {et~al.}(2018){Ishigaki}, {Kawamata}, {Ouchi}, {Oguri},
  {Shimasaku}, \& {Ono}}]{Ishigaki_2018}
{Ishigaki}, M., {Kawamata}, R., {Ouchi}, M., {et~al.} 2018, \apj, 854, 73,
  \dodoi{10.3847/1538-4357/aaa544}

\bibitem[{{Jauzac} {et~al.}(2016){Jauzac}, {Richard}, {Limousin}, {Knowles},
  {Mahler}, {Smith}, {Kneib}, {Jullo}, {Natarajan}, {Ebeling}, {Atek},
  {Cl{\'e}ment}, {Eckert}, {Egami}, {Massey}, \& {Rexroth}}]{Jauzac_2016}
{Jauzac}, M., {Richard}, J., {Limousin}, M., {et~al.} 2016, \mnras, 457, 2029,
  \dodoi{10.1093/mnras/stw069}

\bibitem[{{Jullo} {et~al.}(2007){Jullo}, {Kneib}, {Limousin},
  {El{\'\i}asd{\'o}ttir}, {Marshall}, \& {Verdugo}}]{Jullo_2007}
{Jullo}, E., {Kneib}, J.~P., {Limousin}, M., {et~al.} 2007, New Journal of
  Physics, 9, 447, \dodoi{10.1088/1367-2630/9/12/447}

\bibitem[{{Kakiichi} {et~al.}(2024){Kakiichi}, {Egami}, {Fan}, {Lyu}, {Wang},
  {Yang}, {Bechtel}, {Behroozi}, {Bosman}, {Cai}, {Champagne}, {Davies}, {De
  Rosa}, {Decarli}, {Eilers}, {Ellis}, {Endsley}, {Farina}, {Finkelstein},
  {Fujimoto}, {Hennawi}, {Inoue}, {Jiang}, {Jin}, {Khusanova}, {Kirkpatrick},
  {Kocevski}, {Kulkarni}, {Lee}, {Liu}, {Meyer}, {Ono}, {Onoue}, {Ouchi},
  {Papovich}, {Satyavolu}, {Schindler}, {Sun}, {Tee}, {Vestergaard}, {Zhang},
  \& {Zou}}]{Kakiichi_2024jwstprop}
{Kakiichi}, K., {Egami}, E., {Fan}, X., {et~al.} 2024, {COSMOS-3D: A Legacy
  Spectroscopic/Imaging Survey of the Early Universe}, JWST Proposal. Cycle 3,
  ID. \#5893

\bibitem[{{Karim} {et~al.}(2011){Karim}, {Schinnerer},
  {Mart{\'\i}nez-Sansigre}, {Sargent}, {van der Wel}, {Rix}, {Ilbert},
  {Smol{\v{c}}i{\'c}}, {Carilli}, {Pannella}, {Koekemoer}, {Bell}, \&
  {Salvato}}]{Karim_2011}
{Karim}, A., {Schinnerer}, E., {Mart{\'\i}nez-Sansigre}, A., {et~al.} 2011,
  \apj, 730, 61, \dodoi{10.1088/0004-637X/730/2/61}

\bibitem[{{Kashino} {et~al.}(2023){Kashino}, {Lilly}, {Matthee}, {Eilers},
  {Mackenzie}, {Bordoloi}, \& {Simcoe}}]{Kashino_2023}
{Kashino}, D., {Lilly}, S.~J., {Matthee}, J., {et~al.} 2023, \apj, 950, 66,
  \dodoi{10.3847/1538-4357/acc588}

\bibitem[{{Kashino} {et~al.}(2013){Kashino}, {Silverman}, {Rodighiero},
  {Renzini}, {Arimoto}, {Daddi}, {Lilly}, {Sanders}, {Kartaltepe}, {Zahid},
  {Nagao}, {Sugiyama}, {Capak}, {Carollo}, {Chu}, {Hasinger}, {Ilbert},
  {Kajisawa}, {Kewley}, {Koekemoer}, {Kova{\v{c}}}, {Le F{\`e}vre}, {Masters},
  {McCracken}, {Onodera}, {Scoville}, {Strazzullo}, {Symeonidis}, \&
  {Taniguchi}}]{Kashino_2013}
{Kashino}, D., {Silverman}, J.~D., {Rodighiero}, G., {et~al.} 2013, \apjl, 777,
  L8, \dodoi{10.1088/2041-8205/777/1/L8}

\bibitem[{{Kawamata} {et~al.}(2018){Kawamata}, {Ishigaki}, {Shimasaku},
  {Oguri}, {Ouchi}, \& {Tanigawa}}]{Kawamata_2018}
{Kawamata}, R., {Ishigaki}, M., {Shimasaku}, K., {et~al.} 2018, \apj, 855, 4,
  \dodoi{10.3847/1538-4357/aaa6cf}

\bibitem[{{Kawamata} {et~al.}(2016){Kawamata}, {Oguri}, {Ishigaki},
  {Shimasaku}, \& {Ouchi}}]{Kawamata_2016}
{Kawamata}, R., {Oguri}, M., {Ishigaki}, M., {Shimasaku}, K., \& {Ouchi}, M.
  2016, \apj, 819, 114, \dodoi{10.3847/0004-637X/819/2/114}

\bibitem[{{Kennicutt} \& {Evans}(2012)}]{Kennicutt_2012}
{Kennicutt}, R.~C., \& {Evans}, N.~J. 2012, \araa, 50, 531,
  \dodoi{10.1146/annurev-astro-081811-125610}

\bibitem[{Kimm {et~al.}(2019)Kimm, Blaizot, Garel, Michel-Dansac, Katz,
  Rosdahl, Verhamme, \& Haehnelt}]{Kimm_2019}
Kimm, T., Blaizot, J., Garel, T., {et~al.} 2019, Monthly Notices of the Royal
  Astronomical Society, 486, 2215, \dodoi{10.1093/mnras/stz989}

\bibitem[{{Koprowski} {et~al.}(2017){Koprowski}, {Dunlop}, {Micha{\l}owski},
  {Coppin}, {Geach}, {McLure}, {Scott}, \& {van der Werf}}]{Koprowski_2017}
{Koprowski}, M.~P., {Dunlop}, J.~S., {Micha{\l}owski}, M.~J., {et~al.} 2017,
  \mnras, 471, 4155, \dodoi{10.1093/mnras/stx1843}

\bibitem[{{Kriek} {et~al.}(2015){Kriek}, {Shapley}, {Reddy}, {Siana}, {Coil},
  {Mobasher}, {Freeman}, {de Groot}, {Price}, {Sanders}, {Shivaei}, {Brammer},
  {Momcheva}, {Skelton}, {van Dokkum}, {Whitaker}, {Aird}, {Azadi}, {Kassis},
  {Bullock}, {Conroy}, {Dav{\'e}}, {Kere{\v{s}}}, \& {Krumholz}}]{Kriek_2015}
{Kriek}, M., {Shapley}, A.~E., {Reddy}, N.~A., {et~al.} 2015, \apjs, 218, 15,
  \dodoi{10.1088/0067-0049/218/2/15}

\bibitem[{{Kron}(1980)}]{Kron_1980}
{Kron}, R.~G. 1980, \apjs, 43, 305, \dodoi{10.1086/190669}

\bibitem[{{Kroupa}(2001)}]{Kroupa_2001}
{Kroupa}, P. 2001, \mnras, 322, 231, \dodoi{10.1046/j.1365-8711.2001.04022.x}

\bibitem[{{Labb{\'e}} {et~al.}(2013){Labb{\'e}}, {Oesch}, {Bouwens},
  {Illingworth}, {Magee}, {Gonz{\'a}lez}, {Carollo}, {Franx}, {Trenti}, {van
  Dokkum}, \& {Stiavelli}}]{Labbe_2013}
{Labb{\'e}}, I., {Oesch}, P.~A., {Bouwens}, R.~J., {et~al.} 2013, \apjl, 777,
  L19, \dodoi{10.1088/2041-8205/777/2/L19}

\bibitem[{{Lagattuta} {et~al.}(2019){Lagattuta}, {Richard}, {Bauer},
  {Cl{\'e}ment}, {Mahler}, {Soucail}, {Carton}, {Kneib}, {Laporte}, {Martinez},
  {Patr{\'\i}cio}, {Payne}, {Pell{\'o}}, {Schmidt}, \& {de la
  Vieuville}}]{Lagattuta_2019}
{Lagattuta}, D.~J., {Richard}, J., {Bauer}, F.~E., {et~al.} 2019, \mnras, 485,
  3738, \dodoi{10.1093/mnras/stz620}

\bibitem[{{Lin} {et~al.}(2024){Lin}, {Wang}, {Fan}, {Cai}, {Champagne}, {Sun},
  {Volonteri}, {Yang}, {Hennawi}, {Ba{\~n}ados}, {Barth}, {Eilers}, {Farina},
  {Liu}, {Jin}, {Jun}, {Lupi}, {Kakiichi}, {Mazzucchelli}, {Onoue}, {Pan},
  {Pizzati}, {Rojas-Ruiz}, {Schindler}, {Trakhtenbrot}, {Shen}, {Trebitsch},
  {Zhuang}, {Endsley}, {Meyer}, {Li}, {Li}, {Pudoka}, {Tee}, {Wu}, \&
  {Zhang}}]{LinX_2024}
{Lin}, X., {Wang}, F., {Fan}, X., {et~al.} 2024, \apj, 974, 147,
  \dodoi{10.3847/1538-4357/ad6565}

\bibitem[{{Liu} {et~al.}(2018){Liu}, {Daddi}, {Dickinson}, {Owen}, {Pannella},
  {Sargent}, {B{\'e}thermin}, {Magdis}, {Gao}, {Shu}, {Wang}, {Jin}, \&
  {Inami}}]{Liu_2018}
{Liu}, D., {Daddi}, E., {Dickinson}, M., {et~al.} 2018, \apj, 853, 172,
  \dodoi{10.3847/1538-4357/aaa600}

\bibitem[{{Livermore} {et~al.}(2017){Livermore}, {Finkelstein}, \&
  {Lotz}}]{Livermore_2017}
{Livermore}, R.~C., {Finkelstein}, S.~L., \& {Lotz}, J.~M. 2017, \apj, 835,
  113, \dodoi{10.3847/1538-4357/835/2/113}

\bibitem[{{Looser} {et~al.}(2024){Looser}, {D'Eugenio}, {Maiolino}, {Witstok},
  {Sandles}, {Curtis-Lake}, {Chevallard}, {Tacchella}, {Johnson}, {Baker},
  {Suess}, {Carniani}, {Ferruit}, {Arribas}, {Bonaventura}, {Bunker},
  {Cameron}, {Charlot}, {Curti}, {de Graaff}, {Maseda}, {Rawle}, {Rix}, {Del
  Pino}, {Smit}, {{\"U}bler}, {Willott}, {Alberts}, {Egami}, {Eisenstein},
  {Endsley}, {Hausen}, {Rieke}, {Robertson}, {Shivaei}, {Williams}, {Boyett},
  {Chen}, {Ji}, {Jones}, {Kumari}, {Nelson}, {Perna}, {Saxena}, \&
  {Scholtz}}]{Looser_24a}
{Looser}, T.~J., {D'Eugenio}, F., {Maiolino}, R., {et~al.} 2024, \nat, 629, 53,
  \dodoi{10.1038/s41586-024-07227-0}

\bibitem[{{Lotz} {et~al.}(2017){Lotz}, {Koekemoer}, {Coe}, {Grogin}, {Capak},
  {Mack}, {Anderson}, {Avila}, {Barker}, {Borncamp}, {Brammer}, {Durbin},
  {Gunning}, {Hilbert}, {Jenkner}, {Khandrika}, {Levay}, {Lucas}, {MacKenty},
  {Ogaz}, {Porterfield}, {Reid}, {Robberto}, {Royle}, {Smith},
  {Storrie-Lombardi}, {Sunnquist}, {Surace}, {Taylor}, {Williams}, {Bullock},
  {Dickinson}, {Finkelstein}, {Natarajan}, {Richard}, {Robertson}, {Tumlinson},
  {Zitrin}, {Flanagan}, {Sembach}, {Soifer}, \& {Mountain}}]{Lotz_2017}
{Lotz}, J.~M., {Koekemoer}, A., {Coe}, D., {et~al.} 2017, \apj, 837, 97,
  \dodoi{10.3847/1538-4357/837/1/97}

\bibitem[{{Lovell} {et~al.}(2021){Lovell}, {Vijayan}, {Thomas}, {Wilkins},
  {Barnes}, {Irodotou}, \& {Roper}}]{Lovell_2021}
{Lovell}, C.~C., {Vijayan}, A.~P., {Thomas}, P.~A., {et~al.} 2021, \mnras, 500,
  2127, \dodoi{10.1093/mnras/staa3360}

\bibitem[{{Ly} {et~al.}(2011){Ly}, {Lee}, {Dale}, {Momcheva}, {Salim},
  {Staudaher}, {Moore}, \& {Finn}}]{Ly_2011}
{Ly}, C., {Lee}, J.~C., {Dale}, D.~A., {et~al.} 2011, \apj, 726, 109,
  \dodoi{10.1088/0004-637X/726/2/109}

\bibitem[{{Ly} {et~al.}(2007){Ly}, {Malkan}, {Kashikawa}, {Shimasaku}, {Doi},
  {Nagao}, {Iye}, {Kodama}, {Morokuma}, \& {Motohara}}]{Ly_2007}
{Ly}, C., {Malkan}, M.~A., {Kashikawa}, N., {et~al.} 2007, \apj, 657, 738,
  \dodoi{10.1086/510828}

\bibitem[{{Lyu} {et~al.}(2024){Lyu}, {Alberts}, {Rieke}, {Shivaei},
  {P{\'e}rez-Gonz{\'a}lez}, {Sun}, {Hainline}, {Baum}, {Bonaventura}, {Bunker},
  {Egami}, {Eisenstein}, {Florian}, {Ji}, {Johnson}, {Morrison}, {Rieke},
  {Robertson}, {Rujopakarn}, {Tacchella}, {Scholtz}, \& {Willmer}}]{Lyu_2024}
{Lyu}, J., {Alberts}, S., {Rieke}, G.~H., {et~al.} 2024, \apj, 966, 229,
  \dodoi{10.3847/1538-4357/ad3643}

\bibitem[{{Madau} \& {Dickinson}(2014)}]{Madau_2014}
{Madau}, P., \& {Dickinson}, M. 2014, \araa, 52, 415,
  \dodoi{10.1146/annurev-astro-081811-125615}

\bibitem[{Maseda {et~al.}(2020)Maseda, Bacon, Lam, Matthee, Brinchmann, Schaye,
  Labbe, Schmidt, Boogaard, Bouwens, Cantalupo, Franx, Hashimoto, Inami,
  Kusakabe, Mahler, Nanayakkara, Richard, \& Wisotzki}]{Maseda_2020}
Maseda, M.~V., Bacon, R., Lam, D., {et~al.} 2020, Monthly Notices of the Royal
  Astronomical Society, 493, 5120, \dodoi{10.1093/mnras/staa622}

\bibitem[{{Matharu} {et~al.}(2024){Matharu}, {Nelson}, {Brammer}, {Oesch},
  {Allen}, {Shivaei}, {Naidu}, {Chisholm}, {Covelo-Paz}, {Fudamoto},
  {Giovinazzo}, {Herard-Demanche}, {Kerutt}, {Kramarenko}, {Marchesini},
  {Meyer}, {Prieto-Lyon}, {Reddy}, {Shuntov}, {Weibel}, {Wuyts}, \&
  {Xiao}}]{Matharu_2024}
{Matharu}, J., {Nelson}, E.~J., {Brammer}, G., {et~al.} 2024, arXiv e-prints,
  arXiv:2404.17629, \dodoi{10.48550/arXiv.2404.17629}

\bibitem[{Matthee {et~al.}(2016)Matthee, Sobral, Best, Khostovan, Oteo,
  Bouwens, \& R{\"o}ttgering}]{Matthee_2016}
Matthee, J., Sobral, D., Best, P., {et~al.} 2016, Monthly Notices of the Royal
  Astronomical Society, 465, 3637, \dodoi{10.1093/mnras/stw2973}

\bibitem[{{Matthee} {et~al.}(2024){Matthee}, {Naidu}, {Brammer}, {Chisholm},
  {Eilers}, {Goulding}, {Greene}, {Kashino}, {Labbe}, {Lilly}, {Mackenzie},
  {Oesch}, {Weibel}, {Wuyts}, {Xiao}, {Bordoloi}, {Bouwens}, {van Dokkum},
  {Illingworth}, {Kramarenko}, {Maseda}, {Mason}, {Meyer}, {Nelson}, {Reddy},
  {Shivaei}, {Simcoe}, \& {Yue}}]{Matthee_2024}
{Matthee}, J., {Naidu}, R.~P., {Brammer}, G., {et~al.} 2024, \apj, 963, 129,
  \dodoi{10.3847/1538-4357/ad2345}

\bibitem[{{Mehta} {et~al.}(2017){Mehta}, {Scarlata}, {Rafelski}, {Gburek},
  {Teplitz}, {Alavi}, {Boylan-Kolchin}, {Finkelstein}, {Gardner}, {Grogin},
  {Koekemoer}, {Kurczynski}, {Siana}, {Codoreanu}, {de Mello}, {Lee}, \&
  {Soto}}]{Mehta_2017}
{Mehta}, V., {Scarlata}, C., {Rafelski}, M., {et~al.} 2017, \apj, 838, 29,
  \dodoi{10.3847/1538-4357/aa6259}

\bibitem[{{Moster} {et~al.}(2011){Moster}, {Somerville}, {Newman}, \&
  {Rix}}]{Moster_2011}
{Moster}, B.~P., {Somerville}, R.~S., {Newman}, J.~A., \& {Rix}, H.-W. 2011,
  \apj, 731, 113, \dodoi{10.1088/0004-637X/731/2/113}

\bibitem[{{Muzzin} {et~al.}(2023){Muzzin}, {Abraham}, {Asada}, {Bradac},
  {Brammer}, {Desprez}, {Harshan}, {Iyer}, {Marchesini}, {Martis}, {Mowla},
  {Ravindranath}, {Sarrouh}, {Sawicki}, {Strait}, {Willott}, \&
  {Withers}}]{Muzzin_2023}
{Muzzin}, A., {Abraham}, R.~G., {Asada}, Y., {et~al.} 2023, {JWST in
  Technicolor: Finding and Mapping the Most Extreme Star Forming Galaxies in
  the Epoch of Reionization with Medium and Narrow Bands}, JWST Proposal. Cycle
  2, ID. \#3362

\bibitem[{{Naidu} {et~al.}(2024){Naidu}, {Matthee}, {Kramarenko}, {Weibel},
  {Brammer}, {Oesch}, {Lechner}, {Furtak}, {Di Cesare}, {Torralba}, {Kotiwale},
  {Bezanson}, {Bouwens}, {Chandra}, {Claeyssens}, {Danhaive}, {Frebel}, {de
  Graaff}, {Greene}, {Heintz}, {Ji}, {Kashino}, {Katz}, {Labbe}, {Leja}, {Li},
  {Maseda}, {Richard}, {Shivaei}, {Simcoe}, {Sobral}, {Suess}, {Tacchella}, \&
  {Williams}}]{Naidu_2024}
{Naidu}, R.~P., {Matthee}, J., {Kramarenko}, I., {et~al.} 2024, arXiv e-prints,
  arXiv:2410.01874, \dodoi{10.48550/arXiv.2410.01874}

\bibitem[{{Niemiec} {et~al.}(2023){Niemiec}, {Jauzac}, {Eckert}, {Lagattuta},
  {Sharon}, {Koekemoer}, {Umetsu}, {Acebron}, {Diego}, {Harvey}, {Jullo},
  {Kokorev}, {Limousin}, {Mahler}, {Natarajan}, {Nonino}, {Steinhardt}, {Tam},
  \& {Zitrin}}]{Niemiec_2023}
{Niemiec}, A., {Jauzac}, M., {Eckert}, D., {et~al.} 2023, \mnras, 524, 2883,
  \dodoi{10.1093/mnras/stad1999}

\bibitem[{Novak {et~al.}(2017)Novak, Smol{\v c}i{\'c}, Delhaize, Delvecchio,
  Zamorani, Baran, Bondi, Capak, Carilli, Ciliegi, Civano, Ilbert, Karim,
  Laigle, Le~F{\`e}vre, Marchesi, McCracken, Miettinen, Salvato, Sargent,
  Schinnerer, \& Tasca}]{novak_vla-cosmos_2017}
Novak, M., Smol{\v c}i{\'c}, V., Delhaize, J., {et~al.} 2017, Astronomy \&
  Astrophysics, 602, A5, \dodoi{10.1051/0004-6361/201629436}

\bibitem[{{Oesch} {et~al.}(2023){Oesch}, {Brammer}, {Naidu}, {Bouwens},
  {Chisholm}, {Illingworth}, {Matthee}, {Nelson}, {Qin}, {Reddy}, {Shapley},
  {Shivaei}, {van Dokkum}, {Weibel}, {Whitaker}, {Wuyts}, {Covelo-Paz},
  {Endsley}, {Fudamoto}, {Giovinazzo}, {Herard-Demanche}, {Kerutt},
  {Kramarenko}, {Labbe}, {Leonova}, {Lin}, {Magee}, {Marchesini}, {Maseda},
  {Mason}, {Matharu}, {Meyer}, {Neufeld}, {Prieto Lyon}, {Schaerer}, {Sharma},
  {Shuntov}, {Smit}, {Stefanon}, {Wyithe}, \& {Xiao}}]{Oesch_2023}
{Oesch}, P.~A., {Brammer}, G., {Naidu}, R.~P., {et~al.} 2023, \mnras, 525,
  2864, \dodoi{10.1093/mnras/stad2411}

\bibitem[{{Oguri}(2010)}]{Oguri_2010}
{Oguri}, M. 2010, \pasj, 62, 1017, \dodoi{10.1093/pasj/62.4.1017}

\bibitem[{{Okabe} {et~al.}(2020){Okabe}, {Oguri}, {Peirani}, {Suto}, {Dubois},
  {Pichon}, {Kitayama}, {Sasaki}, \& {Nishimichi}}]{Okabe_2020}
{Okabe}, T., {Oguri}, M., {Peirani}, S., {et~al.} 2020, \mnras, 496, 2591,
  \dodoi{10.1093/mnras/staa1479}

\bibitem[{{Oke} \& {Gunn}(1983)}]{Oke_1983}
{Oke}, J.~B., \& {Gunn}, J.~E. 1983, \apj, 266, 713, \dodoi{10.1086/160817}

\bibitem[{{Pei}(1992)}]{Pei_1992}
{Pei}, Y.~C. 1992, \apj, 395, 130, \dodoi{10.1086/171637}

\bibitem[{Perrin {et~al.}(2014)Perrin, Sivaramakrishnan, Lajoie, Elliott,
  Pueyo, Ravindranath, \& Albert}]{webbpsf}
Perrin, M.~D., Sivaramakrishnan, A., Lajoie, C.-P., {et~al.} 2014, in Space
  Telescopes and Instrumentation 2014: Optical, Infrared, and Millimeter Wave,
  ed. J.~M.~O. Jr., M.~Clampin, G.~G. Fazio, \& H.~A. MacEwen, Vol. 9143,
  International Society for Optics and Photonics (SPIE), 91433X,
  \dodoi{10.1117/12.2056689}

\bibitem[{{Price} {et~al.}(2024){Price}, {Bezanson}, {Labbe}, {Furtak}, {de
  Graaff}, {Greene}, {Kokorev}, {Setton}, {Suess}, {Brammer}, {Cutler}, {Leja},
  {Pan}, {Wang}, {Weaver}, {Whitaker}, {Atek}, {Burgasser}, {Chemerynska},
  {Dayal}, {Feldmann}, {F{\"o}rster Schreiber}, {Fudamoto}, {Fujimoto},
  {Glazebrook}, {Goulding}, {Khullar}, {Kriek}, {Marchesini}, {Maseda},
  {Miller}, {Muzzin}, {Nanayakkara}, {Nelson}, {Oesch}, {Shipley}, {Smit},
  {Taylor}, {van Dokkum}, {Williams}, \& {Zitrin}}]{Price_2024}
{Price}, S.~H., {Bezanson}, R., {Labbe}, I., {et~al.} 2024, arXiv e-prints,
  arXiv:2408.03920, \dodoi{10.48550/arXiv.2408.03920}

\bibitem[{{Qiu} {et~al.}(2018){Qiu}, {Wyithe}, {Oesch}, {Mutch}, {Qin},
  {Labb{\'e}}, {Bouwens}, {Stefanon}, \& {Illingworth}}]{Qiu_2018}
{Qiu}, Y., {Wyithe}, J. S.~B., {Oesch}, P.~A., {et~al.} 2018, \mnras, 481,
  4885, \dodoi{10.1093/mnras/sty2633}

\bibitem[{{Reddy} \& {Steidel}(2009)}]{Reddy_2009}
{Reddy}, N.~A., \& {Steidel}, C.~C. 2009, \apj, 692, 778,
  \dodoi{10.1088/0004-637X/692/1/778}

\bibitem[{{Richard} {et~al.}(2021){Richard}, {Claeyssens}, {Lagattuta},
  {Guaita}, {Bauer}, {Pello}, {Carton}, {Bacon}, {Soucail}, {Lyon}, {Kneib},
  {Mahler}, {Cl{\'e}ment}, {Mercier}, {Variu}, {Tamone}, {Ebeling}, {Schmidt},
  {Nanayakkara}, {Maseda}, {Weilbacher}, {Bouch{\'e}}, {Bouwens}, {Wisotzki},
  {de la Vieuville}, {Martinez}, \& {Patr{\'\i}cio}}]{Richard_2021}
{Richard}, J., {Claeyssens}, A., {Lagattuta}, D., {et~al.} 2021, \aap, 646,
  A83, \dodoi{10.1051/0004-6361/202039462}

\bibitem[{{Richards} {et~al.}(2006){Richards}, {Lacy}, {Storrie-Lombardi},
  {Hall}, {Gallagher}, {Hines}, {Fan}, {Papovich}, {Vanden Berk}, {Trammell},
  {Schneider}, {Vestergaard}, {York}, {Jester}, {Anderson}, {Budav{\'a}ri}, \&
  {Szalay}}]{Richards_2006}
{Richards}, G.~T., {Lacy}, M., {Storrie-Lombardi}, L.~J., {et~al.} 2006, \apjs,
  166, 470, \dodoi{10.1086/506525}

\bibitem[{{Rieke} {et~al.}(2023){Rieke}, {Kelly}, {Misselt}, {Stansberry},
  {Boyer}, {Beatty}, {Egami}, {Florian}, {Greene}, {Hainline}, {Leisenring},
  {Roellig}, {Schlawin}, {Sun}, {Tinnin}, {Williams}, {Willmer}, {Wilson},
  {Clark}, {Rohrbach}, {Brooks}, {Canipe}, {Correnti}, {DiFelice}, {Gennaro},
  {Girard}, {Hartig}, {Hilbert}, {Koekemoer}, {Nikolov}, {Pirzkal}, {Rest},
  {Robberto}, {Sunnquist}, {Telfer}, {Wu}, {Ferry}, {Lewis}, {Baum},
  {Beichman}, {Doyon}, {Dressler}, {Eisenstein}, {Ferrarese}, {Hodapp},
  {Horner}, {Jaffe}, {Johnstone}, {Krist}, {Martin}, {McCarthy}, {Meyer},
  {Rieke}, {Trauger}, \& {Young}}]{Rieke_2023}
{Rieke}, M.~J., {Kelly}, D.~M., {Misselt}, K., {et~al.} 2023, \pasp, 135,
  028001, \dodoi{10.1088/1538-3873/acac53}

\bibitem[{{Rinaldi} {et~al.}(2023){Rinaldi}, {Caputi}, {Costantin}, {Gillman},
  {Iani}, {P{\'e}rez-Gonz{\'a}lez}, {{\"O}stlin}, {Colina}, {Greve},
  {Noorgard-Nielsen}, {Wright}, {Alonso-Herrero}, {{\'A}lvarez-M{\'a}rquez},
  {Eckart}, {Garc{\'\i}a-Mar{\'\i}n}, {Hjorth}, {Ilbert}, {Kendrew}, {Labiano},
  {Le F{\`e}vre}, {Pye}, {Tikkanen}, {Walter}, {van der Werf}, {Ward},
  {Annunziatella}, {Azzollini}, {Bik}, {Boogaard}, {Bosman}, {Crespo
  G{\'o}mez}, {Jermann}, {Langeroodi}, {Melinder}, {Meyer}, {Moutard},
  {Peissker}, {Topinka}, {van Dishoeck}, {G{\"u}del}, {Henning}, {Lagage},
  {Ray}, {Vandenbussche}, {Waelkens}, {Navarro-Carrera}, \&
  {Kokorev}}]{Rinaldi_2023}
{Rinaldi}, P., {Caputi}, K.~I., {Costantin}, L., {et~al.} 2023, \apj, 952, 143,
  \dodoi{10.3847/1538-4357/acdc27}

\bibitem[{{Roberts-Borsani} {et~al.}(2020){Roberts-Borsani}, {Ellis}, \&
  {Laporte}}]{Roberts-Borsani_2020}
{Roberts-Borsani}, G.~W., {Ellis}, R.~S., \& {Laporte}, N. 2020, \mnras, 497,
  3440, \dodoi{10.1093/mnras/staa2085}

\bibitem[{{Salpeter}(1955)}]{Salpeter_1955}
{Salpeter}, E.~E. 1955, \apj, 121, 161, \dodoi{10.1086/145971}

\bibitem[{{Schaerer} \& {de Barros}(2010)}]{Schaerer_2010}
{Schaerer}, D., \& {de Barros}, S. 2010, \aap, 515, A73,
  \dodoi{10.1051/0004-6361/200913946}

\bibitem[{{Schenker} {et~al.}(2013){Schenker}, {Robertson}, {Ellis}, {Ono},
  {McLure}, {Dunlop}, {Koekemoer}, {Bowler}, {Ouchi}, {Curtis-Lake}, {Rogers},
  {Schneider}, {Charlot}, {Stark}, {Furlanetto}, \&
  {Cirasuolo}}]{Schenker_2013}
{Schenker}, M.~A., {Robertson}, B.~E., {Ellis}, R.~S., {et~al.} 2013, \apj,
  768, 196, \dodoi{10.1088/0004-637X/768/2/196}

\bibitem[{{Schmidt}(1968)}]{Schmidt_1968}
{Schmidt}, M. 1968, \apj, 151, 393, \dodoi{10.1086/149446}

\bibitem[{{Scholtz} {et~al.}(2023){Scholtz}, {Maiolino}, {D'Eugenio},
  {Curtis-Lake}, {Carniani}, {Charlot}, {Curti}, {Silcock}, {Arribas}, {Baker},
  {Bhatawdekar}, {Boyett}, {Bunker}, {Chevallard}, {Circosta}, {Eisenstein},
  {Hainline}, {Hausen}, {Ji}, {Ji}, {Johnson}, {Kumari}, {Looser}, {Lyu},
  {Maseda}, {Parlanti}, {Perna}, {Rieke}, {Robertson}, {Rodr{\'\i}guez Del
  Pino}, {Sun}, {Tacchella}, {{\"U}bler}, {Venturi}, {Williams}, {Willmer},
  {Willott}, \& {Witstok}}]{Scholtz_2023}
{Scholtz}, J., {Maiolino}, R., {D'Eugenio}, F., {et~al.} 2023, arXiv e-prints,
  arXiv:2311.18731, \dodoi{10.48550/arXiv.2311.18731}

\bibitem[{Shen {et~al.}(2014)Shen, Madau, Conroy, Governato, \&
  Mayer}]{Shen_baryon_2014}
Shen, S., Madau, P., Conroy, C., Governato, F., \& Mayer, L. 2014, The
  Astrophysical Journal, 792, 99, \dodoi{10.1088/0004-637X/792/2/99}

\bibitem[{{Shen} {et~al.}(2020{\natexlab{a}}){Shen}, {Hopkins},
  {Faucher-Gigu{\`e}re}, {Alexander}, {Richards}, {Ross}, \&
  {Hickox}}]{Shen_2020_quasarLF}
{Shen}, X., {Hopkins}, P.~F., {Faucher-Gigu{\`e}re}, C.-A., {et~al.}
  2020{\natexlab{a}}, \mnras, 495, 3252, \dodoi{10.1093/mnras/staa1381}

\bibitem[{{Shen} {et~al.}(2020{\natexlab{b}}){Shen}, {Vogelsberger}, {Nelson},
  {Pillepich}, {Tacchella}, {Marinacci}, {Torrey}, {Hernquist}, \&
  {Springel}}]{Shen_2020}
{Shen}, X., {Vogelsberger}, M., {Nelson}, D., {et~al.} 2020{\natexlab{b}},
  \mnras, 495, 4747, \dodoi{10.1093/mnras/staa1423}

\bibitem[{{Shen} {et~al.}(2024){Shen}, {Zhuang}, {Burgasser}, {Fan}, {Greene},
  {Li}, {Narayan}, {Shapley}, {Sun}, {Wang}, \& {Yang}}]{Shen_2024jwstprop}
{Shen}, Y., {Zhuang}, M., {Burgasser}, A.~J., {et~al.} 2024, {NEXUS: the North
  ecliptic pole EXtragalactic Unified Survey}, JWST Proposal. Cycle 3, ID.
  \#5105

\bibitem[{{Shioya} {et~al.}(2008){Shioya}, {Taniguchi}, {Sasaki}, {Nagao},
  {Murayama}, {Takahashi}, {Ajiki}, {Ideue}, {Mihara}, {Nakajima}, {Scoville},
  {Mobasher}, {Aussel}, {Giavalisco}, {Guzzo}, {Hasinger}, {Impey}, {Le
  F{\`e}vre}, {Lilly}, {Renzini}, {Rich}, {Sanders}, {Schinnerer}, {Shopbell},
  {Leauthaud}, {Kneib}, {Rhodes}, \& {Massey}}]{Shioya_2008}
{Shioya}, Y., {Taniguchi}, Y., {Sasaki}, S.~S., {et~al.} 2008, \apjs, 175, 128,
  \dodoi{10.1086/523703}

\bibitem[{{Smit} {et~al.}(2016){Smit}, {Bouwens}, {Labb{\'e}}, {Franx},
  {Wilkins}, \& {Oesch}}]{Smit_2016}
{Smit}, R., {Bouwens}, R.~J., {Labb{\'e}}, I., {et~al.} 2016, \apj, 833, 254,
  \dodoi{10.3847/1538-4357/833/2/254}

\bibitem[{{Sobral} {et~al.}(2013){Sobral}, {Smail}, {Best}, {Geach}, {Matsuda},
  {Stott}, {Cirasuolo}, \& {Kurk}}]{Sobral_2013}
{Sobral}, D., {Smail}, I., {Best}, P.~N., {et~al.} 2013, \mnras, 428, 1128,
  \dodoi{10.1093/mnras/sts096}

\bibitem[{{Steidel} {et~al.}(2014){Steidel}, {Rudie}, {Strom}, {Pettini},
  {Reddy}, {Shapley}, {Trainor}, {Erb}, {Turner}, {Konidaris}, {Kulas}, {Mace},
  {Matthews}, \& {McLean}}]{Steidel_2014}
{Steidel}, C.~C., {Rudie}, G.~C., {Strom}, A.~L., {et~al.} 2014, \apj, 795,
  165, \dodoi{10.1088/0004-637X/795/2/165}

\bibitem[{{Steinhardt} {et~al.}(2020){Steinhardt}, {Jauzac}, {Acebron}, {Atek},
  {Capak}, {Davidzon}, {Eckert}, {Harvey}, {Koekemoer}, {Lagos}, {Mahler},
  {Montes}, {Niemiec}, {Nonino}, {Oesch}, {Richard}, {Rodney}, {Schaller},
  {Sharon}, {Strolger}, {Allingham}, {Amara}, {Bah{\'e}}, {B{\oe}hm}, {Bose},
  {Bouwens}, {Bradley}, {Brammer}, {Broadhurst}, {Ca{\~n}as}, {Cen},
  {Cl{\'e}ment}, {Clowe}, {Coe}, {Connor}, {Darvish}, {Diego}, {Ebeling},
  {Edge}, {Egami}, {Ettori}, {Faisst}, {Frye}, {Furtak}, {G{\'o}mez-Guijarro},
  {Remolina Gonz{\'a}lez}, {Gonzalez}, {Graur}, {Gruen}, {Harvey}, {Hensley},
  {Hovis-Afflerbach}, {Jablonka}, {Jha}, {Jullo}, {Kneib}, {Kokorev},
  {Lagattuta}, {Limousin}, {von der Linden}, {Linzer}, {Lopez}, {Magdis},
  {Massey}, {Masters}, {Maturi}, {McCully}, {McGee}, {Meneghetti}, {Mobasher},
  {Moustakas}, {Murphy}, {Natarajan}, {Neyrinck}, {O'Connor}, {Oguri}, {Pagul},
  {Rhodes}, {Rich}, {Robertson}, {Sereno}, {Shan}, {Smith}, {Sneppen},
  {Squires}, {Tam}, {Tchernin}, {Toft}, {Umetsu}, {Weaver}, {van Weeren},
  {Williams}, {Wilson}, {Yan}, \& {Zitrin}}]{Steinhardt_2020}
{Steinhardt}, C.~L., {Jauzac}, M., {Acebron}, A., {et~al.} 2020, \apjs, 247,
  64, \dodoi{10.3847/1538-4365/ab75ed}

\bibitem[{{Stiavelli} {et~al.}(2023){Stiavelli}, {Morishita}, {Chiaberge},
  {Grillo}, {Leethochawalit}, {Rosati}, {Schuldt}, {Trenti}, \&
  {Treu}}]{Stiavelli_2023}
{Stiavelli}, M., {Morishita}, T., {Chiaberge}, M., {et~al.} 2023, \apjl, 957,
  L18, \dodoi{10.3847/2041-8213/ad0159}

\bibitem[{{Stroe} \& {Sobral}(2015)}]{Stroe_2015}
{Stroe}, A., \& {Sobral}, D. 2015, \mnras, 453, 242,
  \dodoi{10.1093/mnras/stv1555}

\bibitem[{{Suess} {et~al.}(2024){Suess}, {Weaver}, {Price}, {Pan}, {Wang},
  {Bezanson}, {Brammer}, {Cutler}, {Labb{\'e}}, {Leja}, {Williams}, {Whitaker},
  {Atek}, {Dayal}, {de Graaff}, {Feldmann}, {Franx}, {Fudamoto}, {Fujimoto},
  {Furtak}, {Goulding}, {Greene}, {Khullar}, {Kokorev}, {Kriek}, {Lorenz},
  {Marchesini}, {Maseda}, {Matthee}, {Miller}, {Mitsuhashi}, {Mowla}, {Muzzin},
  {Naidu}, {Nanayakkara}, {Nelson}, {Oesch}, {Setton}, {Shipley}, {Smit},
  {Spilker}, {van Dokkum}, \& {Zitrin}}]{Suess_2024}
{Suess}, K.~A., {Weaver}, J.~R., {Price}, S.~H., {et~al.} 2024, \apj, 976, 101,
  \dodoi{10.3847/1538-4357/ad75fe}

\bibitem[{{Sun} {et~al.}(2022){Sun}, {Egami}, {Pirzkal}, {Rieke}, {Boyer},
  {Correnti}, {Gennaro}, {Girard}, {Greene}, {Kelly}, {Koekemoer},
  {Leisenring}, {Misselt}, {Nikolov}, {Roellig}, {Stansberry}, {Williams},
  {Willmer}, \& {Members of the JWST/NIRCam Commissioning Team}}]{Sun_2022}
{Sun}, F., {Egami}, E., {Pirzkal}, N., {et~al.} 2022, \apjl, 936, L8,
  \dodoi{10.3847/2041-8213/ac8938}

\bibitem[{{Sun} {et~al.}(2023){Sun}, {Egami}, {Pirzkal}, {Rieke}, {Baum},
  {Boyer}, {Boyett}, {Bunker}, {Cameron}, {Curti}, {Eisenstein}, {Gennaro},
  {Greene}, {Jaffe}, {Kelly}, {Koekemoer}, {Kumari}, {Maiolino}, {Maseda},
  {Perna}, {Rest}, {Robertson}, {Schlawin}, {Smit}, {Stansberry}, {Sunnquist},
  {Tacchella}, {Williams}, \& {Willmer}}]{Sun_2023}
---. 2023, \apj, 953, 53, \dodoi{10.3847/1538-4357/acd53c}

\bibitem[{{Sun} {et~al.}(2025){Sun}, {Wang}, {Yang}, {Champagne}, {Decarli},
  {Fan}, {Ba{\~n}ados}, {Cai}, {Colina}, {Egami}, {Hennawi}, {Jin}, {Jun},
  {Khusanova}, {Li}, {Li}, {Lin}, {Liu}, {Meyer}, {Pudoka}, {Rieke}, {Shen},
  {Tee}, {Venemans}, {Walter}, {Wu}, {Zhang}, \& {Zou}}]{sun_2025}
{Sun}, F., {Wang}, F., {Yang}, J., {et~al.} 2025, \apj, 980, 12,
  \dodoi{10.3847/1538-4357/ad9d0e}

\bibitem[{{Sun} {et~al.}(2024){Sun}, {Ho}, {Zhuang}, {Ma}, {Chen}, \&
  {Li}}]{Sunwen_2024}
{Sun}, W., {Ho}, L.~C., {Zhuang}, M.-Y., {et~al.} 2024, \apj, 960, 104,
  \dodoi{10.3847/1538-4357/acf1f6}

\bibitem[{{Tacchella} {et~al.}(2020){Tacchella}, {Forbes}, \&
  {Caplar}}]{Tacchella_2020a}
{Tacchella}, S., {Forbes}, J.~C., \& {Caplar}, N. 2020, \mnras, 497, 698,
  \dodoi{10.1093/mnras/staa1838}

\bibitem[{{Tacchella} {et~al.}(2022){Tacchella}, {Smith}, {Kannan},
  {Marinacci}, {Hernquist}, {Vogelsberger}, {Torrey}, {Sales}, \&
  {Li}}]{Tacchella_2022a}
{Tacchella}, S., {Smith}, A., {Kannan}, R., {et~al.} 2022, \mnras, 513, 2904,
  \dodoi{10.1093/mnras/stac818}

\bibitem[{{The Astropy Collaboration} {et~al.}(2013){The Astropy
  Collaboration}, {Robitaille, Thomas P.}, {Tollerud, Erik J.}, {Greenfield,
  Perry}, {Droettboom, Michael}, {Bray, Erik}, {Aldcroft, Tom}, {Davis, Matt},
  {Ginsburg, Adam}, {Price-Whelan, Adrian M.}, {Kerzendorf, Wolfgang E.},
  {Conley, Alexander}, {Crighton, Neil}, {Barbary, Kyle}, {Muna, Demitri},
  {Ferguson, Henry}, {Grollier, Fr{\'e}d{\'e}ric}, {Parikh, Madhura M.}, {Nair,
  Prasanth H.}, {G{\"u}nther, Hans M.}, {Deil, Christoph}, {Woillez, Julien},
  {Conseil, Simon}, {Kramer, Roban}, {Turner, James E. H.}, {Singer, Leo},
  {Fox, Ryan}, {Weaver, Benjamin A.}, {Zabalza, Victor}, {Edwards, Zachary I.},
  {Azalee Bostroem, K.}, {Burke, D. J.}, {Casey, Andrew R.}, {Crawford, Steven
  M.}, {Dencheva, Nadia}, {Ely, Justin}, {Jenness, Tim}, {Labrie, Kathleen},
  {Lim, Pey Lian}, {Pierfederici, Francesco}, {Pontzen, Andrew}, {Ptak, Andy},
  {Refsdal, Brian}, {Servillat, Mathieu}, \& {Streicher, Ole}}]{astropy_2013}
{The Astropy Collaboration}, {Robitaille, Thomas P.}, {Tollerud, Erik J.},
  {et~al.} 2013, \aap, 558, A33, \dodoi{10.1051/0004-6361/201322068}

\bibitem[{{Traina} {et~al.}(2024){Traina}, {Gruppioni}, {Delvecchio}, {Calura},
  {Bisigello}, {Feltre}, {Magnelli}, {Schinnerer}, {Liu}, {Adscheid}, {Behiri},
  {Gentile}, {Pozzi}, {Talia}, {Zamorani}, {Algera}, {Gillman}, {Lambrides}, \&
  {Symeonidis}}]{Traina_2024}
{Traina}, A., {Gruppioni}, C., {Delvecchio}, I., {et~al.} 2024, \aap, 681,
  A118, \dodoi{10.1051/0004-6361/202347048}

\bibitem[{{Treu} {et~al.}(2022){Treu}, {Roberts-Borsani}, {Bradac}, {Brammer},
  {Fontana}, {Henry}, {Mason}, {Morishita}, {Pentericci}, {Wang}, {Acebron},
  {Bagley}, {Bergamini}, {Belfiori}, {Bonchi}, {Boyett}, {Boutsia},
  {Calabr{\'o}}, {Caminha}, {Castellano}, {Dressler}, {Glazebrook}, {Grillo},
  {Jacobs}, {Jones}, {Kelly}, {Leethochawalit}, {Malkan}, {Marchesini},
  {Mascia}, {Mercurio}, {Merlin}, {Nanayakkara}, {Nonino}, {Paris},
  {Poggianti}, {Rosati}, {Santini}, {Scarlata}, {Shipley}, {Strait}, {Trenti},
  {Tubthong}, {Vanzella}, {Vulcani}, \& {Yang}}]{Treu_2022}
{Treu}, T., {Roberts-Borsani}, G., {Bradac}, M., {et~al.} 2022, \apj, 935, 110,
  \dodoi{10.3847/1538-4357/ac8158}

\bibitem[{{Trussler} {et~al.}(2025){Trussler}, {Conselice}, {Adams}, {Austin},
  {Caruana}, {Harvey}, {Li}, {Lovell}, {Seeyave}, {Vijayan}, \&
  {Wilkins}}]{Trussler_2025}
{Trussler}, J. A.~A., {Conselice}, C.~J., {Adams}, N., {et~al.} 2025, \mnras,
  \dodoi{10.1093/mnras/staf213}

\bibitem[{{Vijayan} {et~al.}(2021){Vijayan}, {Lovell}, {Wilkins}, {Thomas},
  {Barnes}, {Irodotou}, {Kuusisto}, \& {Roper}}]{Vijayan_2021}
{Vijayan}, A.~P., {Lovell}, C.~C., {Wilkins}, S.~M., {et~al.} 2021, \mnras,
  501, 3289, \dodoi{10.1093/mnras/staa3715}

\bibitem[{{Wang} {et~al.}(2023){Wang}, {Yang}, {Hennawi}, {Fan}, {Sun},
  {Champagne}, {Costa}, {Habouzit}, {Endsley}, {Li}, {Lin}, {Meyer},
  {Schindler}, {Wu}, {Ba{\~n}ados}, {Barth}, {Bhowmick}, {Bieri}, {Blecha},
  {Bosman}, {Cai}, {Colina}, {Connor}, {Davies}, {Decarli}, {De Rosa}, {Drake},
  {Egami}, {Eilers}, {Evans}, {Farina}, {Haiman}, {Jiang}, {Jin}, {Jun},
  {Kakiichi}, {Khusanova}, {Kulkarni}, {Li}, {Liu}, {Loiacono}, {Lupi},
  {Mazzucchelli}, {Onoue}, {Pudoka}, {Rojas-Ruiz}, {Shen}, {Strauss}, {Tee},
  {Trakhtenbrot}, {Trebitsch}, {Venemans}, {Volonteri}, {Walter}, {Xie}, {Yue},
  {Zhang}, {Zhang}, \& {Zou}}]{Wang_ASPIRE_2023}
{Wang}, F., {Yang}, J., {Hennawi}, J.~F., {et~al.} 2023, \apjl, 951, L4,
  \dodoi{10.3847/2041-8213/accd6f}

\bibitem[{{Williams} {et~al.}(2024){Williams}, {Oesch}, {Weibel}, {Brammer},
  {Cloonan}, {Whitaker}, {Barrufet}, {Bezanson}, {Bowler}, {Dayal}, {Franx},
  {Greene}, {Hutter}, {Ji}, {Labb{\'e}}, {Manning}, {Maseda}, \&
  {Xiao}}]{Williams_2024}
{Williams}, C.~C., {Oesch}, P.~A., {Weibel}, A., {et~al.} 2024, arXiv e-prints,
  arXiv:2410.01875, \dodoi{10.48550/arXiv.2410.01875}

\bibitem[{{Willott} {et~al.}(2022){Willott}, {Doyon}, {Albert}, {Brammer},
  {Dixon}, {Muzic}, {Ravindranath}, {Scholz}, {Abraham}, {Artigau},
  {Brada{\v{c}}}, {Goudfrooij}, {Hutchings}, {Iyer}, {Jayawardhana}, {LaMassa},
  {Martis}, {Meyer}, {Morishita}, {Mowla}, {Muzzin}, {Noirot}, {Pacifici},
  {Rowlands}, {Sarrouh}, {Sawicki}, {Taylor}, {Volk}, \& {Zabl}}]{Willott_2022}
{Willott}, C.~J., {Doyon}, R., {Albert}, L., {et~al.} 2022, \pasp, 134, 025002,
  \dodoi{10.1088/1538-3873/ac5158}

\bibitem[{{Willott} {et~al.}(2024){Willott}, {Desprez}, {Asada}, {Sarrouh},
  {Abraham}, {Brada{\v{c}}}, {Brammer}, {Estrada-Carpenter}, {Iyer}, {Martis},
  {Matharu}, {Mowla}, {Muzzin}, {Noirot}, {Sawicki}, {Strait},
  {Rihtar{\v{s}}i{\v{c}}}, \& {Withers}}]{Willott_2024}
{Willott}, C.~J., {Desprez}, G., {Asada}, Y., {et~al.} 2024, \apj, 966, 74,
  \dodoi{10.3847/1538-4357/ad35bc}

\bibitem[{{Windhorst} {et~al.}(2023){Windhorst}, {Cohen}, {Jansen}, {Summers},
  {Tompkins}, {Conselice}, {Driver}, {Yan}, {Coe}, {Frye}, {Grogin},
  {Koekemoer}, {Marshall}, {O'Brien}, {Pirzkal}, {Robotham}, {Ryan}, {Willmer},
  {Carleton}, {Diego}, {Keel}, {Porto}, {Redshaw}, {Scheller}, {Wilkins},
  {Willner}, {Zitrin}, {Adams}, {Austin}, {Arendt}, {Beacom}, {Bhatawdekar},
  {Bradley}, {Broadhurst}, {Cheng}, {Civano}, {Dai}, {Dole}, {D'Silva},
  {Duncan}, {Fazio}, {Ferrami}, {Ferreira}, {Finkelstein}, {Furtak}, {Gim},
  {Griffiths}, {Hammel}, {Harrington}, {Hathi}, {Holwerda}, {Honor}, {Huang},
  {Hyun}, {Im}, {Joshi}, {Kamieneski}, {Kelly}, {Larson}, {Li}, {Lim}, {Ma},
  {Maksym}, {Manzoni}, {Meena}, {Milam}, {Nonino}, {Pascale}, {Petric},
  {Pierel}, {Polletta}, {R{\"o}ttgering}, {Rutkowski}, {Smail}, {Straughn},
  {Strolger}, {Swirbul}, {Trussler}, {Wang}, {Welch}, {B. Wyithe}, {Yun},
  {Zackrisson}, {Zhang}, \& {Zhao}}]{Windhorst_2023}
{Windhorst}, R.~A., {Cohen}, S.~H., {Jansen}, R.~A., {et~al.} 2023, \aj, 165,
  13, \dodoi{10.3847/1538-3881/aca163}

\bibitem[{{Yue} {et~al.}(2018){Yue}, {Castellano}, {Ferrara}, {Fontana},
  {Merlin}, {Amor{\'\i}n}, {Grazian}, {M{\'a}rmol-Queralto}, {Micha{\l}owski},
  {Mortlock}, {Paris}, {Parsa}, {Pilo}, {Santini}, \& {Di
  Criscienzo}}]{YueB_2018}
{Yue}, B., {Castellano}, M., {Ferrara}, A., {et~al.} 2018, \apj, 868, 115,
  \dodoi{10.3847/1538-4357/aae77f}

\end{thebibliography}
\bibliographystyle{aasjournal}

\suppressAffiliationsfalse
\allauthors

\end{document}